\documentclass[twocolumn,prx,amsmath,amssymb,superscriptaddress]{revtex4-1}

\usepackage{graphicx}
\usepackage{bm}
\usepackage{color}

\usepackage[pdftex,dvipsnames]{xcolor}

\usepackage{hyperref}
\usepackage{cleveref}

\Crefname{equation}{Eq.}{Eqs.}

\graphicspath{ {./revised_figs3/} }

\begin{document}

\newcommand\Crefnoabbrv[1]{%
\begingroup
	\Crefname{fig}{Figure}{Figures}\Cref{#1}
\endgroup%
}

\renewcommand{\figureautorefname}{Fig.}
\renewcommand{\equationautorefname}{Eq.}
\renewcommand{\sectionautorefname}{Sec.}

\title{Anomalous Fraunhofer Interference in\\ Epitaxial Superconductor-Semiconductor Josephson Junctions}

\author{H.~J.~Suominen}
\affiliation{Center for Quantum Devices and Station Q Copenhagen, Niels Bohr Institute, University of Copenhagen, Universitetsparken 5, 2100 Copenhagen, Denmark}
\author{J.~Danon}
\affiliation{Center for Quantum Devices and Station Q Copenhagen, Niels Bohr Institute, University of Copenhagen, Universitetsparken 5, 2100 Copenhagen, Denmark}
\affiliation{Department of Physics, Norwegian University of Science and Technology, NO-7491 Trondheim, Norway}
\author{M. Kjaergaard}
\author{K.~Flensberg}
\affiliation{Center for Quantum Devices and Station Q Copenhagen, Niels Bohr Institute, University of Copenhagen, Universitetsparken 5, 2100 Copenhagen, Denmark}

\author{J.~Shabani}\thanks{Now at City College, City University of New York}
\affiliation{California NanoSystems Institute, University of California, Santa Barbara, CA 93106, USA}

\author{C.~J.~Palmstr\o{}m}
\affiliation{California NanoSystems Institute, University of California, Santa Barbara, CA 93106, USA}
\affiliation{Department of Electrical Engineering, University of California, Santa Barbara, CA 93106, USA}
\affiliation{Materials Research Laboratories, University of California, Santa Barbara, CA 93106, USA}
\author{F.~Nichele}
\author{C.~M.~Marcus}
\email{marcus@nbi.ku.dk}
\affiliation{Center for Quantum Devices and Station Q Copenhagen, Niels Bohr Institute, University of Copenhagen, Universitetsparken 5, 2100 Copenhagen, Denmark}

\date{\today}

\begin{abstract}
We investigate patterns of critical current as a function of perpendicular and in-plane magnetic fields in superconductor-semiconductor-superconductor (SNS) junctions based on InAs/InGaAs heterostructures with an epitaxial Al layer. This material system is of interest due to its exceptionally good superconductor-semiconductor coupling, as well as large spin-orbit interaction and $g$-factor in the semiconductor.
Thin epitaxial Al allows the application of large in-plane field without destroying superconductivity. For fields perpendicular to the junction, flux focusing results in aperiodic node spacings in the pattern of critical currents known as Fraunhofer patterns by analogy to the related interference effect in optics. Adding an in-plane field yields two further anomalies in the pattern. First, higher order nodes are systematically strengthened, indicating current flow along the edges of the device, as a result of confinement of Andreev states driven by an induced flux dipole; second, asymmetries in the interference appear that depend on the field direction and magnitude. A model is presented, showing good agreement with experiment, elucidating the roles of flux focusing, Zeeman and spin-orbit coupling, and disorder in producing these effects.\end{abstract}

\maketitle

\section{Introduction}

Materials with strong spin-orbit interaction (SOI) and large Zeeman splitting coupled to superconductors have attracted a great deal of attention in recent years, largely due to the possibility of accessing topological states of matter \cite{Kitaev2001,Sau2010}.
Despite considerable progress on such systems using semiconducting nanowires \cite{Mourik2012,das2012,Deng2012,Churchill2013}, work on two-dimensional platforms, more amenable for quantum computation schemes \cite{alicea2011}, has been limited.

The strong SOI and large Land\'e $g$-factor in InAs \cite{Nitta1997,Heida1998b,Nitta2003} in combination with its natural surface accumulation layer, facilitating coupling to superconductors, make the InAs two-dimensional electron gas (2DEG) a  favorable candidate for creating superconductor-semiconductor hybrids \cite{Kawakami1985,Nguyen1990,Nitta1992}.
Very recently, two-dimensional epitaxial Al/InAs heterostructures have been realized, demonstrating an exceptionally transparent superconductor-semiconductor interface, resulting in a near unity Andreev reflection probability \cite{Shabani2015,Kjaergaard2016,Kjaergaard2016b}.

Despite showing great promise, many properties of these two-dimensional epitaxial structures are not yet well understood.
For instance, basic quantities such as the strength of the SOI in the hybrid system or the orientation of the resulting effective spin-orbit field are not known.
Also, the detailed interplay of superconductivity, SOI, and Zeeman interaction has, to large extent, not been experimentally investigated in two-dimensional systems. Recent investigations of this interplay in the two-dimensional topological insulator HgTe have shown promising results stimulating further studies in more conventional material systems \cite{Hart2015}.
Further, since most envisioned applications of these systems require considerable in-plane magnetic fields, it is important to understand the detailed behavior of the heterostructure under applied magnetic fields with different orientations.

Superconductor-normal-superconductor (SNS) junctions form a well-established platform to study the properties of superconducting hybrid structures.
SNS junctions based on semiconductors with strong SOI have been proposed to study the topological phase transition \cite{SanJose2013,SanJose2014,Hell2016,Pientka2016}, but could also potentially be used to quantify the strength of SOI in the semiconductor \cite{Liu2010}.
For instance, theoretical models have been developed to understand how the detailed SNS current-phase relation depends on SOI in two-dimensional junctions \cite{Bezuglyi2002}, as well as in single-channel junctions \cite{Beri2008}, quantum point contacts \cite{Reynoso2008,Reynoso2012}, and nanowires \cite{Yokoyama2014}.

Many details of the physics occurring in the junction are also encoded in the critical current.
A measurement of the critical current as a function of the out-of-plane magnetic field $B_z$ is paradigmatic in the study of SNS junctions.
For increasing $B_z$, the winding of the superconducting phase by the enclosed flux leads to a characteristic modulation of the critical current $I_c$.
For a rectangular junction with uniform current density
\begin{align}
I_c(B_z) = I_c^{(0)}\left| \frac{\sin(\pi B_z L W/\Phi_0)}{(\pi B_z L W/\Phi_0)} \right|,
\label{eq:frau}
\end{align}
reminiscent of a single-slit Fraunhofer interference pattern in optics \cite{Tinkham2004}.
Here, $L$ and $W$ are the length and width of the normal region, $I_c^{(0)}$ is the zero-field critical current, and $\Phi_0=h/2e$ is the flux quantum.
This behavior has been observed in a wide variety of systems \cite{Rowell1963,Nishino1986} including 2DEGs with strong SOI \cite{Inoue1989}.
Deviations from this Fraunhofer form can yield information about the local magnetic field profile \cite{Miller1985} as well as the supercurrent density in the junction \cite{Dynes1971,Hui2014}.
Recently, such interference mapping has been used to probe edge states arising in two-dimensional topological insulators \cite{Hart2014,Pribiag2015} and graphene \cite{allen2015}.

In this paper, we present an experimental and theoretical study of the magnetic field dependence of the interference pattern of critical currents in epitaxial Al/InAs/Al junctions, with both perpendicular field as well as a separately controlled in-plane field. We identify several interesting effects:
(i) In a purely perpendicular field, we observe a deviation from a simple Fraunhofer pattern (\Cref{eq:frau}), which we interpret as arising from flux focusing due to the Meissner effect in the epitaxial Al leads.
(ii) The interference pattern changes dramatically when an in-plane field is applied.
A crossover is observed in the perpendicular-field interference pattern with increasing in-plane field, from a Fraunhofer-like pattern with rapidly decreasing critical currents with node index, toward one resembling that of a superconducting quantum interference device (SQUID) with critical currents that depend only weakly on node index. We interpret this transition as again resulting from flux focusing: When the \emph{in-plane} flux is excluded from the the leads, an effective \emph{out-of-plane} flux dipole appears in the junction region.
This dipole dephases contributions to the supercurrent in the center of the junction, resulting in coherent transport only near the edges of the sample.
(iii) Application of an in-plane field also induces striking asymmetries (upon reversing perpendicular field) in the interference pattern that depend on the magnitude and direction of the in-plane field, but also vary strongly from lobe to lobe and from sample to sample. Based on these observations, we conclude that flux focusing plays a key role in planar epitaxial devices, particularly in the presence of an in-plane field. Indeed, field modulations due to flux focusing may prove useful, for instance providing magnetic confinement of Andreev states.
In the present devices, observation (iii)---asymmetries in the interference pattern---are dominated by disorder effects, masking related effects due to spin-orbit and Zeeman coupling. 

The paper is organized as follows: \autoref{sec:methods} provides details on device fabrication and magnetotransport measurements. \autoref{sec:outofplane} describes the behavior of the junction with a purely perpendicular magnetic field. \autoref{sec:inplane} describes junction behavior when the applied field is purely in-plane. \autoref{sec:both} reports effects of combined perpendicular and in-plane fields. Conclusions and open questions are discussed in \autoref{sec:discussion}.

\section{Methods}
\label{sec:methods}

The wafer structure, starting at the top surface, consists of $10~\rm{nm}$ Al, $7~\rm{nm}$ InAs, $4~\rm{nm}$ $\rm{In_{0.81}Ga_{0.19}As}$, grown on an InAlAs buffer on an InP substrate by molecular beam epitaxy (see Supplementary Material for more details).  The \emph{in situ} epitaxial growth of the Al layer contrasts with previous approaches to 2D semiconductor-superconductor systems, where the superconductor was deposited in a subsequent processing step \cite{Magnee1995,schapers2001,Hart2014,Pribiag2015}. The clean interface provides high transparency \cite{Shabani2015,Kjaergaard2016b} and a hard induced gap in the semiconductor \cite{Kjaergaard2016}.

Devices are patterned with conventional electron-beam lithography.
In the first processing step, mesas are defined using a wet etch (220:3:3 $\rm{H_2O:H_3PO_4:H_2O_2}$), followed by selective etching of Al using Transene type-D to form the junction. Atomic layer deposition is then used to form an Al$_{2}$O$_{3}$ (40 nm) dielectric, followed by electron-beam deposition of a Ti/Au (5/250 nm) metallic top gate. Ohmic contact is provided directly by the epitaxial Al, which is electrically contacted by wire bonding.

Measurements were carried out in a dilution refrigerator at base temperature $\sim30~\rm{mK}$ using a four-terminal DC+AC current bias with standard lock-in techniques (below 100 Hz), using an AC excitation of $4~\rm{nA}$ or less.

Characterization of the epitaxial Al film yielded a superconducting transition temperature of $T_c=1.5~\rm{K}$, an out-of-plane critical field $B_{z,c}\sim30~\rm{mT}$, and an in-plane critical field $B_{r,c}\sim1.6~\rm{T}$ (see Supplementary Material).
Separate transport measurements of the InAs quantum well (QW) with Al removed demonstrated an electron density of $n_e=3.8\times10^{16}~\rm{m^{-2}}$ and mobility $\mu=0.43~\rm{m^2V^{-1}s^{-1}}$ at zero gate voltage, yielding a mean free path $l_e=140~\rm{nm}$. 
In this density regime, two QW subbands are occupied, as determined by magnetotransport measurements.  Upon partially depleting the 2DEG with the top gate, the single subband limit is reached at gate voltage $V_{\rm{g}}<-2.0~\rm{V}$ with a mobility peak $\mu=0.7~\rm{m^2V^{-1}s^{-1}}$ for $n_e=1.9\times10^{16}~\rm{m^{-2}}$.
The data presented in Secs.~\ref{sec:outofplane} to \ref{sec:both} were all obtained with $V_g = 0$.
Occupation of the second subband appears to play only a minor role in all device characteristics (see Appendix \ref{sec:gating}).
Measurements on similar QWs have demonstrated large SOI, characterized by a spin-orbit length $l_{\rm{so}} \sim 45~\rm{nm}$ \cite{Shabani2015}. The superconducting coherence length is estimated as $\xi=\hbar v_{\rm F}/\Delta^*=1.3~\rm{\mu m}$ \footnote{An effective mass of $m_{\rm{eff}} = 0.05m_e$ is estimated from k.p calculations}, with $v_{\rm F}$ the Fermi velocity and the induced superconducting gap $\Delta^*\sim180~\rm{\mu eV}$ as determined from tunneling measurements (see Ref.~\onlinecite{Kjaergaard2016} and Supplementary Material).

Measurements were performed on six SNS devices, all of which showed qualitatively similar behavior (see Supplementary Material).  The data in \autoref{sec:outofplane} through \autoref{sec:dipole} were characteristic of all devices.
Data similar to those presented in \autoref{sec:twofieldcomponents} were obtained from several samples but with broad quantitative variation, as discussed below.  We focus on data from one SNS junction with contact separation, $L=450~\rm{nm}$, and lateral width, $W=1.5~\rm{\mu m}$ in the regime $l_e<L<\xi$ (see \autoref{fig:setup}(a)). The junction is oriented such that the current flows along the $[011]$ orientation of the underlying crystal structure. 

Throughout, we define the $x$-direction as in the plane of the electron gas and parallel to the average current flow, and the $y$-direction as in plane and perpendicular to average current flow.
The inset in \autoref{fig:setup}(a) shows the corresponding components of the applied magnetic field ${\bf B}$.

To avoid effects of hysteresis as a function of $B_z$ \cite{Courtois2008}, measurements as in Fig.~\ref{fig:setup}(c) were obtained by merging the four quadrants separated by white dashed lines, each taken separately by sweeping current and field away from zero.

\section{Perpendicular magnetic field}
\label{sec:outofplane}

\begin{figure}[t]       
	\includegraphics[]{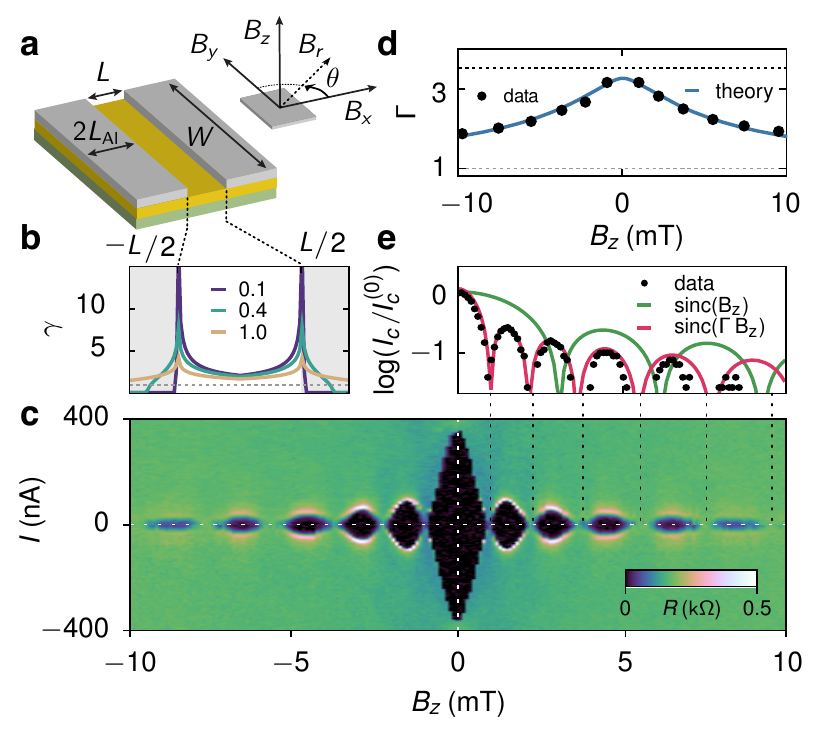}
	\caption{(a) Device schematic illustrating the superconducting Al banks (gray), InAs quantum well (yellow), and InGaAs barrier (green). The coordinate system is illustrated in the inset.
	(b) Local magnetic-field focusing parameter $\gamma$ as a function of position $x$ for three different ratios $\beta=B_z/B_f$, see below.
	(c) Differential resistance $R$, as a function of bias current $I$ and perpendicular magnetic field $B_z$.
	(d) Total magnetic field enhancement in the junction $\Gamma$ as a function of $B_z$, calculated by extraction of the nodes visible in (c,e) (markers), and a fit using \Crefrange{eq:fieldedge}{eq:intgamma} (solid line).
	(e) Critical current $I_c$, plotted logarithmically to highlight periodicity, extracted from (c) (markers).
	Overlaid are the expectation of \Cref{eq:frau} (green) and the modified form taking into account field enhancement due to flux focusing (red).\label{fig:setup}}
\end{figure}

Sweeping the bias current $I$ over a range of perpendicular magnetic fields $B_{z}$ while measuring the differential resistance $R$ results in the interference pattern shown in \autoref{fig:setup}(c).
This pattern deviates from the expected Fraunhofer form predicted by \Cref{eq:frau}.
For instance, from \Cref{eq:frau} we expect equally spaced nodes of the critical current, at values of perpendicular field $B_z =n \Phi_0/(W L)$, where integer numbers of flux quanta penetrate the semiconductor region.
Experimentally, we find a deviation from this uniform node spacing, as can be seen from the vertical dashed lines in \autoref{fig:setup}(c).

In order to investigate this variable node spacing in more detail, we plot in \autoref{fig:setup}(e) the critical current extracted from \autoref{fig:setup}(c) as a function of $B_z$ (markers).
For reference, we also show the expected Fraunhofer pattern (green) using the lithographic device area, for which $\Phi_0/(WL) = 3.1~\rm{mT}$.
From the data, we find a central lobe half-width of $0.97~\rm{mT}$ and a reduced spacing of the subsequent side-lobes, gradually increasing and reaching $1.9~\rm{mT}$ for the fifth side-lobe \footnote{The deviation at high field between our result and the expectation is presumably due to an underestimation of the junction area due to the neglect of the finite penetration depth in the leads \cite{Barone2005}. Utilizing an effective length $L_{\rm{eff}}=L+2\lambda_L$ (with $\lambda_L$ estimated in \autoref{sec:bc1}) yields an expected node spacing of $1.7~\rm{mT}$.}.

To quantify the deviation from the expected uniform spacing, we introduce a dimensionless factor $\Gamma$, the ratio of the expected node position to the observed node position,
\begin{align}
\Gamma(B_z^{(n)}) &= \frac{n\Phi_0}{B_z^{(n)} L W},
\label{eq:Gamma}
\end{align}
where $B_z^{(n)}$ is the perpendicular magnetic field at node number $n$.
A regular Fraunhofer pattern has $\Gamma = 1$ everywhere, as indicated in \autoref{fig:setup}(d).
 At low fields, we find $\Gamma\sim 3$. As $B_z$ increases, $\Gamma$ decreases, approaching unity at high fields.
The black dots in \autoref{fig:setup}(d) show the extracted $\Gamma$ based on the data of \autoref{fig:setup}(c).

The deviation from \autoref{eq:frau} leading to $\Gamma>1$ can be understood as resulting from field-dependent flux focusing from the superconducting contacts.
The qualitative behavior of $\Gamma$ is consistent with the superconducting leads passing from a Meissner state at low field, through a mixed state, towards a fully flux-penetrated state above $10~\rm{mT}$.
In the Meissner state, the contacts completely expel flux, causing the field in the junction region to be enhanced.
When the magnetic field is increased, the thin aluminum banks are slowly pushed into a mixed state as they are penetrated by field lines, leading to a smaller field enhancement in the junction and correspondingly a decreasing $\Gamma$.
At high field the banks are presumably fully penetrated by the incident flux, approaching a negligible field enhancement and $\Gamma \approx 1$.

Previous studies using thick niobium contacts also found large field enhancements in SNS junctions \cite{Harada2002,Paajaste2015}. In those studies, however, the leads remained in a full Meissner regime
for the perpendicular field range studied, resulting in a constant field enhancement.
Because the Al electrodes in the present system are operated close to their critical field $B_c$, the degree of flux focusing depends on field.

To examine the flux-focusing picture more quantitatively, we model the field profile inside the junction following Ref.~\onlinecite{Zeldov1994} (see also \cite{Brandt1993}). The effective field near a single thin superconducting strip of length $2L_{\rm Al}$  and infinite width (see \autoref{fig:setup}(a)), subject to a perpendicular applied field, is given by
\begin{align}
B_{\rm{eff}}(\tilde x) &= B_f\log\left(\frac{|\tilde x|\sqrt{L_{\rm Al}^2-a^2}+L_{\rm Al}\sqrt{\tilde x^2-a^2}}{a\sqrt{\left|\tilde x^2-L_{\rm Al}^2\right|}}\right),
\label{eq:fieldedge}
\end{align}
for $|\tilde x|>a$ and $B_{\rm{eff}}(\tilde x) = 0$ for $|\tilde x| \le a$.
The coordinate $\tilde x$ is the in-plane coordinate perpendicular to the edges of the film, with $\tilde x=0$ corresponding to the center of the film.
The length $2a$ corresponds to the extent of a region centered at $\tilde x=0$ where the field is fully expelled due to Meissner screening; this length is given by $a = L_{\rm Al}/\cosh(B_z/B_f)$, with $B_z$ the applied perpendicular magnetic field \footnote{For $B_z>B_f$, $B_{\rm{eff}}(\tilde x)\propto B_z$. Indeed for large $\tilde x$ we find $B_{\rm{eff}}(\tilde x)=B_f\log\left(\cosh\left(\frac{B_z}{B_f}\right)\left[1+\tanh\left(\frac{B_z}{B_f}\right)\right]\right)=B_z$.} and $B_f$ a characteristic field scale roughly corresponding to the field of first vortex penetration.
To account for the finite width of our junction, we argue that $2L_{\rm Al}$ in this case corresponds not to the physical contact length (on the order of $10~\rm{\mu m}$) but to an effective length over which flux is focused into the junction.
Flux lines further away than $\sim W$ from the junction edge are more likely to be expelled towards the sides rather than into the junction region. We thus use $W$ as a cutoff for the effective contact length and set $L_{\rm Al} = W$.

To account for both contacts in our SNS geometry, we consider for the total effective perpendicular field profile
\begin{align}
B_{\rm{tot}}(x) = B_{\rm{eff}} (L_{\rm Al}+L/2 - |x| ),
\label{eq:beff}
\end{align}
expressed in terms of the $x$-coordinate with $x=0$ corresponding to the center of the SNS junction.
We assumed that the focusing in the junction is dominated by the left(right) contact for negative(positive) $x$.
We then use \Cref{eq:beff,eq:fieldedge} to define a local field enhancement parameter
\begin{align}
\gamma(\beta,x) = B_{\rm{tot}}(x)/B_z,
\label{eq:smallgamma}
\end{align}
which is a function of the ratio $\beta = B_z / B_f$.
In \autoref{fig:setup}(b) we plot $\gamma$ for three different $\beta$, illustrating the inhomogeneous field profile induced by the superconducting leads. The dashed line in \autoref{fig:setup}(b) highlights the expectation in the absence of focusing ($\gamma=1$).
Near zero applied field (blue line), the local enhancement peaks strongly close to the superconducting banks.
Inside of the superconducting contacts, however, $\gamma$ abruptly falls to zero.
When the field is increased (cyan and gold lines) we see a gradual smoothing of the enhancement profile as more of the flux penetrates the superconducting banks.

Integrating \Cref{eq:smallgamma} over the junction length allows us to calculate the total field enhancement,
\begin{align}
\Gamma(B_z) = \frac{1}{L}\int_{-L/2}^{L/2}\gamma(\beta,x)\,dx.
\label{eq:intgamma}
\end{align}
We fit the data using \Cref{eq:intgamma} with $B_f$ as the only free parameter.
The resulting fit is shown as the blue line in \autoref{fig:setup}(d), yielding $B_f=8.2~\rm{mT}$.
This is in good agreement with an estimate for the field of first vortex penetration of the film $B_{c1}=7.7~\rm{mT}$ (see Appendix \ref{sec:bc1}).
Besides, detailed calculations for a finite-width geometry predict a low-field enhancement of $\Gamma = (2W/L)^{2/3} \sim 3.5$ as shown by the black dashed line in \autoref{fig:setup}(d) \cite{Gu1979}.
The good agreement between this low-field prediction and our model further supports our approximation $L_{\rm Al}=W$.
The resulting continuous function $\Gamma(B_z)$ can then be used to plot the full interference pattern of $I_c(B_z)$, corrected for the flux focusing due to the presence of the superconducting contacts.
The resulting $I_c(B_z)$ is plotted in red in \autoref{fig:setup}(e), and shows excellent agreement with the $I_c(B_z)$ extracted from \autoref{fig:setup}(c).

Despite its simplicity, our model captures the observed deviations from a regular Fraunhofer pattern in the interference pattern of critical currents, strongly suggesting that the observed aperiodic node spacings are indeed caused by flux focusing in the mixed state of the superconducting leads where $B_z\sim B_{c1}$.
As a control experiment we have also studied a device of nominally identical dimensions, but with large flux holes located behind the superconducting contacts.
Consistent with our interpretation, negligible field enhancement is observed in this device, independent of the applied field (see Supplementary Material for details).

\section{In-plane magnetic field}
\label{sec:inplane}

\begin{figure}[t]
	\includegraphics[]{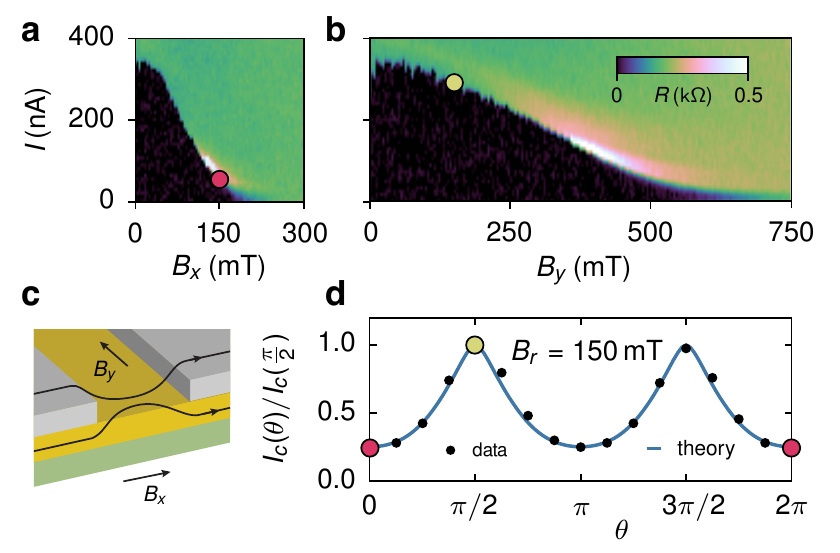}
    \caption{(a) Differential resistance $R$, as a function of bias current $I$ and in-plane magnetic field $B_x$, applied in the $x$-direction (along the direction of current flow).
    (b) As in (a) but with the in-plane field $B_y$ along the $y$-direction.
    (c) Schematic indicating how an in-plane field along $\hat x$ can result in an effective flux dipole in the normal region.
    (d) Normalized critical current $I_c$ as a function of the angle $\theta$ between the in-plane field and $\hat x$; the field has a fixed magnitude of $B_r=150~\rm{mT}$.
    The dots represent the experimental data, the solid line is a theory curve based on a one-parameter fit of $\alpha$ at $\theta = \pi$, using the model based on Eq.~\ref{eq:ephi} (see below).
    The red and yellow markers highlight the correspondence with panels (a) and (b) respectively.}
    \label{fig:inplane}
\end{figure}

We next examine the effects of in-plane magnetic field on the SNS junction, initially without perpendicular field, $B_z=0$.  Differential resistance as a function of bias current and field magnitude is shown in \autoref{fig:inplane} for two field orientations: field parallel to the current ($x$-direction, \autoref{fig:inplane}(a)) and field perpendicular to the current ($y$-direction, \autoref{fig:inplane}(b)).
We see that the critical current exhibits a strong anisotropy.
The critical field (where the supercurrent becomes fully suppressed) changes from $\sim 200~\rm{mT}$ for ${\bf B} \parallel \hat x$ to $\sim 650~\rm{mT}$ for ${\bf B} \parallel \hat y$.
In \autoref{fig:inplane}(d) we show the full dependence of $I_c$ on the direction of the in-plane field, where we fixed the magnitude of the field to $B_r=150~\rm{mT}$ and $\theta$ denotes the angle between ${\bf B}$ and the $x$-direction.

We propose to interpret this anisotropy again in terms of flux focusing due to the Meissner effect.
Indeed, also an in-plane field could give rise to flux focusing, since the thickness of the Al layer ($d\sim10~\rm{nm}$) is comparable to the London penetration depth of Al, $\lambda_{\rm L}=16~\rm{nm}$ \cite{khukhareva1963}.

One consequence of the in-plane Meissner effect would be that the density of flux lines just below the aluminum contacts increases, leading to local enhancements of the effective field inside the QW.
However, this focusing effect is not expected to depend strongly on the direction of the in-plane field.
Another possible effect is that the bending of the field lines around the edges of the contacts may induce a flux dipole in the junction, as shown schematically in \autoref{fig:inplane}(c).
Assuming that ${\bf B} \parallel \hat x$, we see that close to the left contact there is a small component of flux inside the well in the positive $z$-direction, and close to the right contact there is a comparable component in the opposite direction.
This flux dipole couples to the in-plane motion of the electrons and can therefore have a strong effect on the interference pattern of $I_c$.
Furthermore, the effect is proportional to $B_x$ only, and can thus lead to an anisotropy of $I_c$ in the in-plane field direction.

For ${\bf B} \parallel \hat y$ the suppression of the critical current with field appears to be fully accounted for by Zeeman effects only. 
An estimate of the magnitude of the effective $g$-factor in the InAs QW from the critical field $B_{y,c}$ yields $|g^*| = 2\Delta^*/\mu_B B_{y,c} \sim10$, which is in good agreement with previous measurements \cite{Shabani2015}.

As soon as we let the in-plane field deviate from the $y$-direction, a flux dipole will be induced in the N region.
The effect of this dipole is most easily understood within a semiclassical picture, where supercurrent arises from coherent transport of Cooper pairs between S regions along well-defined trajectories through the N region.
A finite flux dipole makes the phase picked up along a trajectory depend explicitly on the angle $\vartheta$ between the trajectory and the $x$-axis.
The dipole will therefore lead to a dephasing of contributions to the current arising from trajectories with different $\vartheta$, and will thus suppress the supercurrent.

We develop a simple but quantitative model of supercurrent through an SNS junction in the presence of a flux dipole by assuming that the junction is ballistic and we can use a semiclassical approximation (where the Fermi wavelength is the smallest length scale in the problem).
In the absence of a perpendicular field (or for finite but small $B_z$) we can associate the Andreev bound states in the normal region with straight trajectories connecting the two proximitized regions in the QW.
For the energy of such a bound state as a function of $\vartheta$ and the average $y$-coordinate $y_0$ one finds in the limit of $W,L \ll \xi$
\begin{align}
E(y_0,\vartheta) = \pm \Delta^* \cos \left(
\frac{\varphi}{2} - \pi \frac{\Phi}{\Phi_0}\frac{y_0}{W} - \pi \alpha \tan \vartheta
\right),
\label{eq:ephi}
\end{align}
where $\varphi$ is the phase difference between the two proximitized regions, $\Phi$ is the homogeneous flux associated with $B_z$, and $\alpha  = \alpha_0 \cos \theta$ depends on $B_x$ and parametrizes the effect of flux focusing~\footnote{We note that this model neglects the effect of SOI. We have verified that spin-orbit effects, calculated along the lines of Ref.~\cite{Bezuglyi2002}, yield changes on the order of a few percent while the experimental anisotropy is of the order 1.}.
The contribution of all Andreev bound states to the free energy $F$ of the junction is found by summing (\ref{eq:ephi}) over all allowed $y_0$ and $\vartheta$, weighted by a Fermi function.
The supercurrent then follows as $I_s(\varphi) = (2e/\hbar) \partial F/\partial \varphi$ and the critical current is simply $I_c = {\rm max}_\varphi\, I_s(\varphi)$.

We convert the sums over $y_0$ and $\vartheta$ into integrals and, assuming for simplicity zero temperature and fully absorbing sides at $y = \pm W/2$, we numerically compute the critical current for $\Phi = 0$ as a function of the in-plane field direction $\theta$.
Comparing the resulting $I_c(\theta)/I_c(\pi/2)$ with the data shown in \autoref{fig:inplane}(d) results in a single-parameter fit yielding $\alpha_0 = 0.32 \pm 0.01$.
The resulting fit is shown as the solid blue line in the figure and shows excellent agreement with the data.
We can also try to connect this numerical value for $\alpha_0$ to our device geometry.
A rough estimate for $\alpha_0$ in terms of device parameters is $\alpha_0 = \eta B_r L d_f / \Phi_0$, where $d_f$ is the width of the strips close to the proximitized regions where flux focusing is significant and $\eta$ is the fraction of $B_x$ that locally contributes to magnetic flux oriented along $\pm \hat z$.
(For instance, $\eta = 1/\sqrt 2$ would correspond to a situation where the flux lines make on average an angle of $45^\circ$ with the plane of the junction within two strips of width $d_f$.)
If we estimate $d_f = d = 10~$nm we find for $B_r = 150~$mT and $\alpha_0 = 0.32$ that $\eta = 0.29$, corresponding to an average local out-of-plane angle of $\sim 20^\circ$.

\section{Combined perpendicular and in-plane magnetic fields}
\label{sec:both}

Sweeping $B_z$ while still applying an in-plane field we observe two new and striking effects, as shown in \autoref{fig:twofields}(a,b).
First, in the presence of an in-plane field, the critical current develops a pronounced asymmetry between positive and negative $B_z$; we observe this for all directions of in-plane field.
Second, increasing the in-plane field when directed along $\hat x$ results in
(i) a decrease of the zero-perpendicular-field critical current, $I_c^{(0)}$;
(ii) a relative enhancement of all side-lobe maxima as compared to the central one, approaching a situation where all observable maxima are roughly equal; and
(iii) a gradual decrease of the width of the central lobe.
We initially focus on the latter effects, associated with $B_x$, and discuss the asymmetries in \autoref{sec:twofieldcomponents}.

\subsection{SNS-to-SQUID transition}
\label{sec:dipole}

\begin{figure}[t]
	\includegraphics[]{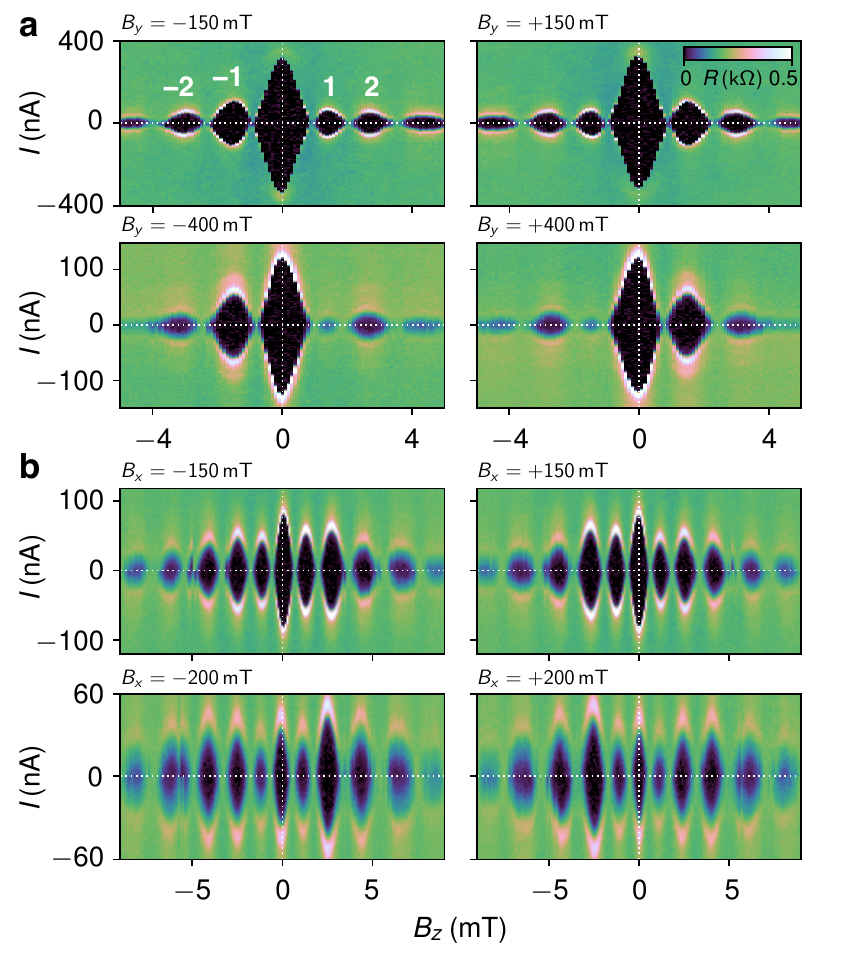}
    \caption{(a) Differential resistance $R$ as a function of bias current $I$ and $B_z$, measured for different values of fixed $B_y$: $B_y = \pm 150~\rm{mT}$ (upper row) and $B_y = \pm 400~\rm{mT}$ (bottom row).
    The white numbers in the upper left panel indicate the lobe indices.
    (b) As (a), for an in-plane magnetic field applied along $\hat x$, using $B_x= \pm 150~\rm{mT}$ (upper row) and $B_x= \pm 200~\rm{mT}$ (bottom row).}
    \label{fig:twofields}
\end{figure}

Both the narrowing of the central lobe and the gradual equalizing of lobe maxima with increasing $B_x$ can be understood as resulting from the flux-focusing mechanism discussed in the previous section.
As argued above, a large $B_x$ could lead to a situation where the supercurrent in the center of the junction is suppressed and most transport takes place along the edges of the normal region, making the system more like a SQUID, with conduction only along sample edges, instead of a planar SNS junction with uniform current flow.
In the pure-SQUID limit, one expects for the critical current $I_c(\Phi) \propto | \cos ( \pi \Phi / \Phi_0 )|$ instead of a Fraunhofer-like pattern, i.e., all lobes will have the same maximum value and the same width $\Phi_0$.
This is qualitatively consistent with the trend we observe in \autoref{fig:twofields}(b).

To further examine the picture of a focusing-induced flux dipole leading to SQUID-like current flow, we use the model from \autoref{sec:inplane} to calculate the critical current as a function of $\Phi = B_zLW$ for different focusing parameters $\alpha$, and compare the resulting theoretical interference patterns $I_c(B_z)$ with experimental data \footnote{In this section we concentrate largely on qualitative features and thus for simplicity neglect the effect of out-of-plane focusing as discussed in \autoref{sec:outofplane}}.

\begin{figure}[t]
	\includegraphics[]{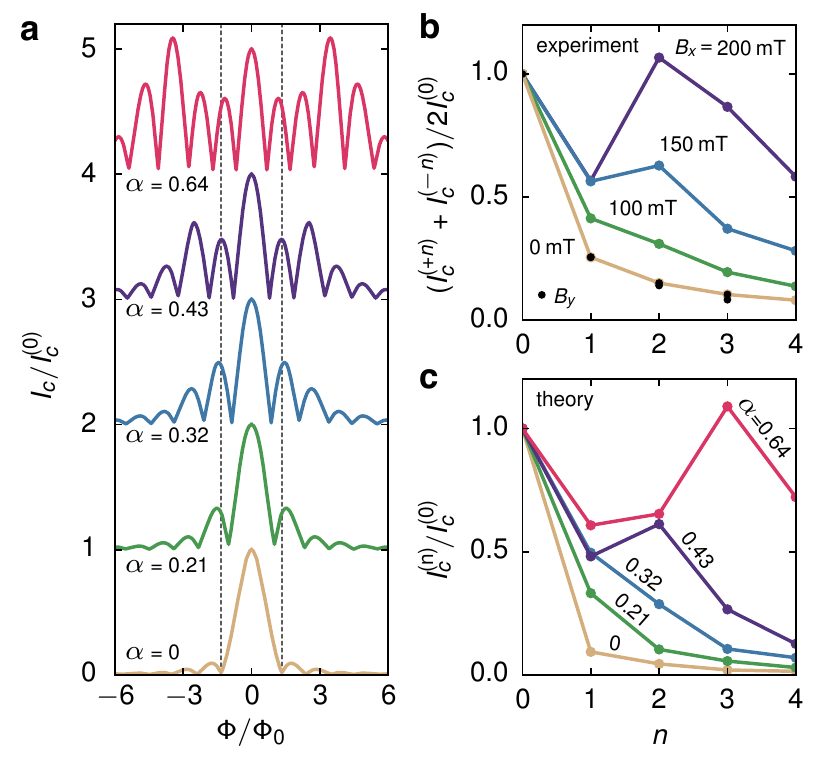}
    \caption{(a) Numerically calculated critical current as a function of $\Phi$, normalized by $I_c^{(0)}$.
    The in-plane field is assumed along $\hat x$ and the different curves correspond to $\alpha = 0$, 0.21, 0.32, 0.43, and 0.64, from bottom to top (each offset by 1).
    (b) Symmetrized side-lobe maxima extracted from experimental data, for different in-plane fields.
    The field magnitudes indicated in the plot refer to ${\bf B} \parallel \hat x$; all data points for ${\bf B} \parallel \hat y$ (black dots) fall on top of the set marked 0~mT.
    (c) Side-lobe maxima obtained from the numerical data shown in (a).}
    \label{fig:SNStoSQUID}
\end{figure}

In \autoref{fig:SNStoSQUID}(a), the calculated $I_c(B_z)$ is plotted for five values of $\alpha$, corresponding to $B_x = 0$, 100, 150, 200, and 300$~{\rm mT}$ (assuming for simplicity a linear relation between $\alpha$ and $B_x$, and setting $\alpha = 0.32$ for $B_x = 150~{\rm mT}$).
These numerical results reproduce the two main features discussed above:
(i) As highlighted by the vertical gray dashed lines, the width of the central lobe decreases with increasing $B_x$.
For $B_x = 0$ we find a width of roughly $2.6 \Phi_0$ (slightly larger than the $2\Phi_0$, corresponding to a regular Fraunhofer pattern, presumably due to finite size effects \footnote{Close to the edges of the junction, where $y_0 \approx \pm W/2$, there are fewer angles $\vartheta$ available to construct Andreev bound states with. Consequently, the flux penetrating the N region close to the edges has less influence on the total average supercurrent through the junction than the flux penetrating the center of the region. To achieve the first full suppression of the supercurrent by perfect destructive interference of all trajectories, one thus needs to go to slightly higher fields than $B_z = \Phi_0/(WL)$.}), and for large $B_x$ it approaches $\Phi_0$, the SQUID limit.
(ii) The heights of all side-lobes in \autoref{fig:SNStoSQUID}(a) increase relative to the central lobe when increasing $B_x$, approaching a situation where all lobes are of comparable height.
Both these trends are qualitatively consistent with the experimental observations and support our interpretation in terms of a focusing-induced flux dipole.

We next examine the behavior of the sequence of side-lobe maxima for different $B_x$ in more detail.
In \autoref{fig:SNStoSQUID}(b) we show the experimentally obtained maxima for four different $B_x$, where we removed the complicating asymmetry in $\pm B_z$ (considered in detail below) by symmetrizing and normalizing the data, $(I_{c}^{(+n)}+I_{c}^{(-n)})/2I_c^{(0)}$, using side-lobe numbers $n$ as indicated in the top left pane of \autoref{fig:twofields}(a).
When $B_x$ is increased we see that (i) the side-lobe maxima are enhanced relative to the central peak, and (ii) the sequence of maxima $I^{(n)}_{c}$ becomes non-monotonic, even yielding side-lobes that exceed the central lobe in magnitude at the highest field ($B_x = 200~{\rm mT}$).
We can extract the same data from the numerical results presented in \autoref{fig:SNStoSQUID}(a), and show in \autoref{fig:SNStoSQUID}(c) the resulting lobe maxima $I_{c}^{(n)}$, normalized by $I_c^{(0)}$.
Comparing with the experimental data, we see that the model not only reproduces the gradual enhancement of the side-lobe maxima for increasing $B_x$, but also captures the more detailed behavior of the series of side-lobes:
Whereas at small $B_x$ the maxima $I_{c}^{(n)}$ monotonically drop for increasing $|n|$, at larger $B_x$ the series becomes non-monotonic, ultimately even producing interference patterns where side-lobes exceed the central maximum in height.

The black dots in \autoref{fig:SNStoSQUID}(b), all falling on top of the yellow curve corresponding to $B_x=0$, represent two data sets with the side-lobe maxima for $B_y = 150$ and 300$~{\rm mT}$ (all at $B_x = 0$), where we removed the asymmetry by symmetrizing $I_c$ in $\pm B_z$.
The fact that all these data are equal to the data without in-plane field, within experimental accuracy, confirms that the qualitative change of the interference pattern that we attribute to an SNS-to-SQUID transition only depends on $B_x$.
It also suggests that the asymmetry in $\pm B_z$ has a physical origin which is distinct from the focusing effects discussed in this section.

In conclusion, the model presented in \autoref{sec:inplane}, that assumes a simple flux dipole in the normal region proportional to $B_x$, appears to capture many aspects of the qualitative behavior of $I_c(B_z)$ as a function of in-plane field.
All global trends we observe in the data are reproduced by our numerical calculations, indicating a transition from Fraunhofer-like interference at zero in-plane field to SQUID-like behavior in the presence of sufficiently strong $B_x$.
A flux dipole in the normal region, induced by flux focusing of the $x$-component of the in-plane field thus appears to provide the likely explanation for our observations.
However, we emphasize that the model used in this section is {\em not} capable of generating the striking asymmetries in $\pm B_z$.

\subsection{Asymmetries in the interference patterns}
\label{sec:twofieldcomponents}

Finally, we turn our attention to the surprising asymmetries observed in the interference patterns of \autoref{fig:twofields}(a,b).
To quantify the asymmetry, we define an asymmetry parameter ${\cal A}_n$ for each side-lobe pair $\{n,-n\}$ as
\begin{align}
\mathcal{A}_n = \frac{I_{c}^{(-n)}-I_{c}^{(n)}}{I_{c}^{(-n)}+I_{c}^{(n)}},
\label{eq:asym}
\end{align}
which yields the relative difference in the side-lobe maxima for $\pm B_z$.
In this section, we will investigate systematic dependences of ${\cal A}_n$ on the magnitude $B_r$ and direction $\theta$ of the in-plane field.

\begin{figure}[t]
	\includegraphics[]{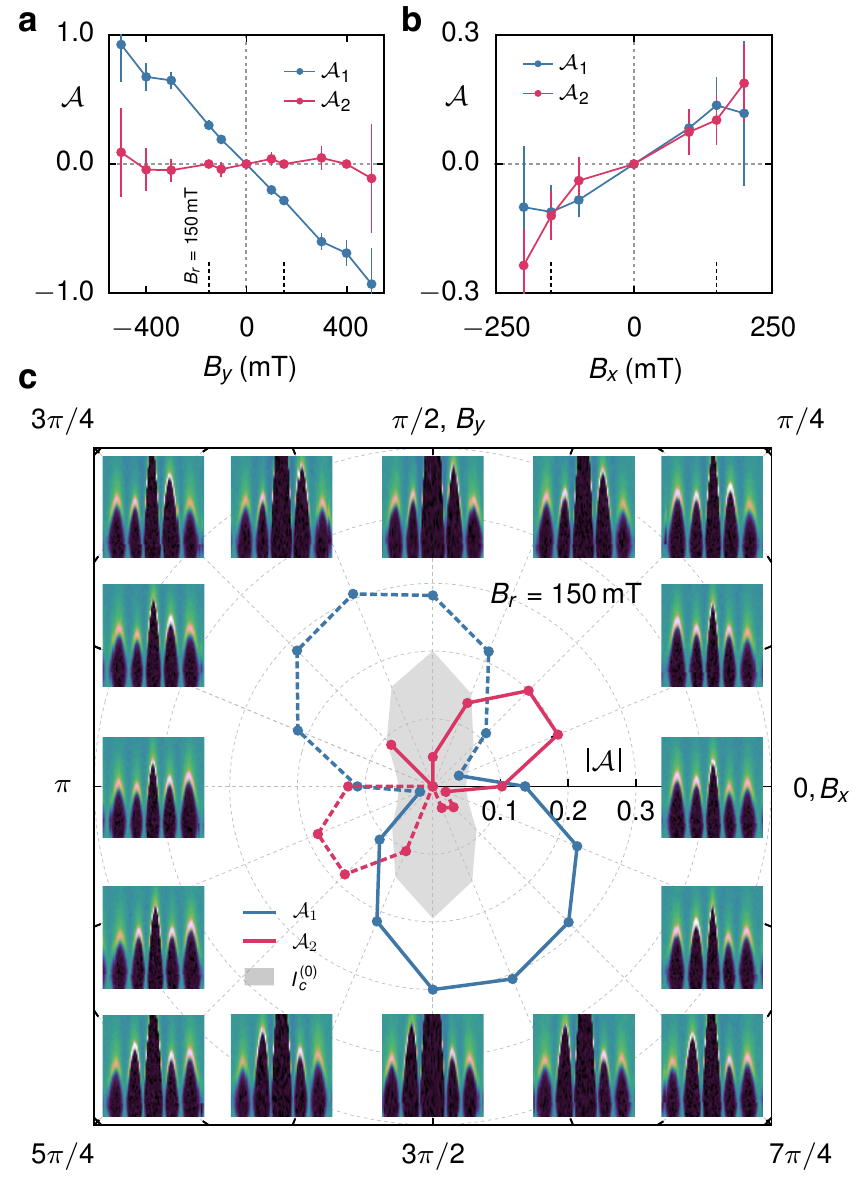}
    \caption{
    (a) Normalized asymmetry $\mathcal{A}$ in the lobe maxima as a function of $B_y$, for the first two side-lobes (shown in blue and red respectively).
    (b) As (a), for magnetic fields oriented along $\hat x$.
    (c) Magnitude of the asymmetry parameter $|\mathcal{A}|$ as a function of in-plane field angle $\theta$.
    Solid(dashed) lines connect points indicating where ${\cal A}_n$ are positive(negative).
    The in-plane field is fixed to $B_r=150~\rm{mT}$.
    To emphasize the deviation of this angular dependence from the anisotropy observed for $B_z = 0$ (see \autoref{sec:inplane}), we include in gray the height of the central lobe $I_c^{(0)}$ as a function of $\theta$ (arbitrary units).
    The panels along the edges show the differential-resistance data from which the asymmetries are extracted.}
    \label{fig:rotation}
\end{figure}

In \autoref{fig:rotation}(a), we plot $\mathcal{A}_1$ (blue) and $\mathcal{A}_2$ (red) as functions of $B_r$ with the field applied along $\hat y$.
The asymmetry of the first node ${\cal A}_1$ is seen to scale roughly linearly with $B_r$, reaching $\sim 100\%$ at the highest fields, while the asymmetry of the second node ${\cal A}_2$ remains zero within experimental uncertainty.
In \autoref{fig:rotation}(b), for in-plane field now along $\hat x$, we now see that both ${\cal A}_1$ and ${\cal A}_2$ increase proportionally to $B_r$, both reaching $\sim 25\%$ at $250~\rm{mT}$, just before $I_c$ gets fully suppressed (see \autoref{fig:inplane}).
All asymmetries thus seem to scale linearly with the magnitude of the applied in-plane field.
The slope of ${\cal A}_n(B_r)$, however, varies considerably: from positive, to zero, to negative for different $n$ and $\theta$.
From these two angles ($\theta = 0$ and $\theta = \pi/2$) no systematics are evident.

The dependence of the ${\cal A}_n$ on the direction of the in-plane field is shown in \autoref{fig:rotation}(c). We plot the measured absolute asymmetries $|\mathcal{A}_1|$ and $|{\cal A}_2|$ for 16 angles at a fixed field magnitude $B_r=150~\rm{mT}$
(we use solid and dashed connectors to indicate where the obtained ${\cal A}_n$ are positive and negative, respectively).
As a reference, we include the anisotropic angular dependence of $I_c^{(0)}$ (filled gray area, plotted in arbitrary units), which we analyzed in terms of a Meissner-induced flux dipole in \autoref{sec:inplane}.
The observed evolution of the asymmetry as a function of $\theta$ in the present sample has a number of interesting characteristics:
(i) The asymmetry of the first side-lobe is maximal for $\theta \sim 5\pi/8$ and minimal in the perpendicular direction $\theta \sim \pi/8$.
(ii) The maximal and minimal asymmetries of the second lobe are roughly perpendicular to those of the first lobe.
(iii) Consistent with the mirroring in $B_z$ observed upon inversion of $B_x$ or $B_y$ (see \autoref{fig:twofields}), both asymmetries have a well defined node at zero about which the behavior of ${\cal A}_n$ are antisymmetric in $\theta$ (or equivalently $B_{r}$).

Separate samples have demonstrated similar behavior, including a linear scaling of the ${\cal A}_n$ in field magnitude and a continuous angular evolution of the asymmetry antisymmetric upon $\pi$ rotation \footnote{See the Supplementary Material for more details and data from different samples.}.
Many of the details, however, are very different from sample to sample:
The observed magnitudes of ${\cal A}_1$ and ${\cal A}_2$ for given $B_r$ fluctuate up to $100\%$, and also the angular alignment of their minima and maxima varies across different samples (also the roughly perpendicular orientation of the maxima of ${\cal A}_1$ and ${\cal A}_2$ observed in \autoref{fig:rotation} is not a consistently observed feature).
The variation of all these details does not display a clear trend following any of the controllable device parameters, such as $W$, $L$, or the orientation of the junction with respect to the crystallographic axes of the InAs wafer.
This suggests that the asymmetries are the result of an intricate interplay of many device-dependent factors, most likely including SOI, disorder, local details of the coupling between the InAs and the Al, and the microscopic shape of the sample.

Although it thus seems difficult to pinpoint the physical mechanism responsible for the asymmetries, we can try to develop a qualitative picture by carrying out a general analysis along the lines of Ref.\ \cite{Rasmussen2015}.
We construct a model (Bogoliubov-de Gennes) Hamiltonian, treating the electrons in the junction as a two-dimensional free electron gas in the presence of a vector potential due to the applied magnetic field (including the flux dipole proportional to $B_x$).
We add to this Hamiltonian finite superconducting pairing potentials of equal magnitude under the left and right contacts, and terms accounting for Rashba and Dresselhaus SOI, Zeeman splitting, and an arbitrary disorder potential $V(x,y)$.
We can then investigate under what circumstances the symmetries of the total Hamiltonian dictate the critical current to be symmetric in $B_z$ and when this symmetry is broken (see the Supplementary Material for details).

The most important conclusion is that if $V(x,y)=0$ the symmetry $I_c(+B_z) = I_c(-B_z)$ is protected, and the model will produce a symmetric interference pattern for a symmetrically shaped sample, no matter how all other parameters are tuned. Disorder or other spatial asymmetries in the junction are thus a necessary ingredient for obtaining an asymmetric critical current.
More specifically, we find:
(i) In the presence of an in-plane field oriented along $\hat x$, only one of the mirror asymmetries $V(x,y) \neq V(-x,y)$ or $V(x,y) \neq V(x,-y)$ has to be present to allow for an asymmetric interference pattern.
(ii) If the in-plane field is along $\hat y$, a direction along which we observe a strong asymmetry (see \autoref{fig:rotation}), only $V(x,y) \neq V(x,-y)$ breaks the symmetry.

As a side note, we mention that some combinations of symmetry-breaking ingredients only affect the higher Fourier components of the current-phase relation $I_s(\varphi)$.
For instance, in order to have $I_c(+B_z) \neq I_c(-B_z)$ in combination with a purely sinusoidal $I_s(\varphi)$, it is required to have (in addition to disorder): (i) a finite $B_x$ or (ii) a finite $B_y$ \emph{and} SOI.
In this case, the degree of asymmetry left at $\theta = \pi/2$ could thus present a measure for the strength of SOI in the junction.
In our experiment, however, current was controlled rather than phase, so we do not know to what degree the current-phase relation is nonsinusoidal.  In general, one expects junctions with weak NS-coupling to have a nearly sinusoidal $I_s(\varphi)$ \cite{Golubov2004}. Engineering a barrier between the normal and proximitized regions in the QW could thus present a way to obtain more detailed knowledge about the SOI in the sample.

Our qualitative analysis thus clearly supports the idea that a key role is played by structural asymmetries in the device, already suggested by the strong sample-to-sample variation observed in the data.
As to the mechanisms that can break spatial symmetries in our samples, we identify three:
(i) spatial variation in the couplings to the superconducting contacts, (ii) imperfections in the microscopic shape of the junction, or (iii) a random disorder potential.
Owing to the epitaxial growth of Al and the small size of the junction, we expect the couplings to the contacts to be relatively homogeneous.
Further, measurements of the asymmetry as a function of gate voltage, presented in Appendix \ref{sec:gating}, show that the asymmetries in $I_c$ are robust to gating in both magnitude and angular dependence.
This weak gate dependence could indicate that the dominant spatial symmetry breaking mechanism is stable, which also suggests that it is either the specific shape of the junction or a fixed disorder potential induced by ionized impurities in the QW.
To further support this picture, we also performed tight-binding numerical simulations of the supercurrent through a two-dimensional disordered SNS junction focusing on the asymmetry parameters ${\cal A}_n$; the results are presented in the Supplementary Material.
We find patterns that look similar to those extracted from the experimental data and also display a strong variation from device to device (i.e.~when we change the disorder configuration).
This also supports our speculation that disorder plays a crucial role in the underlying mechanisms responsible for the asymmetries.

An alternative explanation of the asymmetries one could propose is in terms of Abrikosov vortices near the junction; the presence of such vortices is known to induce asymmetries in the critical current upon inversion of $B_z$.
In the limit of single vortices the behavior is well understood and studies have successfully mapped the position of vortices from the modification of interference patterns \cite{Miller1985,Golod2010}.
For large numbers of vortices, experimental and theoretical investigations exist in the limit of disordered vortex arrays \cite{Fistul1989,Itzler1996}, yielding seemingly random interference patterns.
Theoretical work on ordered vortex arrays predicts symmetric interference patterns described by minor modifications to \Cref{eq:frau} \cite{Fistul1995}.

While we expect flux penetration of the leads in a perpendicular field, and thus vortices to be present, we observe no indication of quantized vortex entrance events, i.e., sudden switches in the critical current \cite{Golod2010}.
Furthermore, we do not observe asymmetries without the application of an in-plane field, which seems to be incompatible with vortices as the origin of the asymmetry.
Furthermore, the mirror symmetry in $B_z$ of the observed asymmetry upon reversing the sign of the in-plane field would require an almost perfect reversal of the vortex configuration, which is highly unlikely.

To conclude, we believe that in the mechanisms underlying the asymmetries we explored in this section, an important role is being played by structural disorder in the samples.
Given the complexity of the system and the randomness of what appears to be the most important symmetry-breaking ingredient, it is currently unclear whether measurements of the asymmetry could be used to quantify the strengths of SOI and Zeeman coupling in these devices.
SNS junctions designed with a well-defined built-in dominant asymmetry might allow for disentangling these effects; this warrants further work.

\section{Conclusion}
\label{sec:discussion}

We report a systematic experimental study of the behavior of two-dimensional epitaxial Al/InAs/Al SNS junctions under the application of out-of-plane as well as in-plane magnetic fields.
Our system is of great interest since it combines strong spin-orbit interaction with exceptionally good semiconductor-superconductor coupling and, due to the epitaxially grown superconductor, it can withstand large in-plane magnetic fields.
Measuring the critical current as a function of the magnitude and direction of the applied magnetic field, we discover a strong influence on the properties of the junction of flux focusing from the superconducting contacts, both for perpendicular and in-plane magnetic fields.
For in-plane fields applied along the direction of average current flow, flux focusing results in an effective flux dipole in the normal region, causing transport to be localized towards the edges of the sample.
We thus find that the in-plane field may act as a novel control knob allowing for magnetic confinement of Andreev states in such hybrid superconductor-semiconductor systems.
We further observe striking asymmetries in the interference pattern $I_c(\pm B_z)$ when an in-plane field is applied.
Although most qualitative properties of these asymmetries remain unexplained, we argue that the microscopic structure of the device plays an crucial role, potentially masking the influences of spin-orbit and Zeeman coupling.

\begin{acknowledgments}
We thank A.~Rasmussen and L.~Levitov for valuable discussions. Research was supported by Microsoft Project Q, the Danish National Research Foundation and the NSF through the National Nanotechnology Infrastructure Network. F.N. acknowledges support from a Marie Curie Fellowship (No. 659653). C.M.M. acknowledges support from the Villum Foundation.
\end{acknowledgments}

\appendix

\section{Estimating $\bf {B_{c1}}$}
\label{sec:bc1}
In order to determine $B_{c1}$ we need to estimate the parameter $\kappa = \lambda / \xi$.
We use values for bulk Al from the literature \cite{Merservey1969}: $\xi_{\rm{bulk}} = 1.6~\rm{\mu m}$ and $T_{c,\rm{bulk}}=1.2~\rm{K}$.
From our measurements we have an accurate value for $T_c$ (see Supplementary Information) and we know from \cite{Tinkham2004} that
\begin{align}
\Delta(0) = 1.76\,k_{\rm B}T_c \;\; {\rm{and}} \;\; \xi = \frac{hv_{\rm F}}{\pi\Delta}.
\end{align}
These expressions allow us to determine the coherence length in the thin film limit as a function of known parameters, yielding
\begin{align}
\xi_{\rm{thin}} = \xi_{\rm{bulk}}\frac{T_{c,\rm{bulk}}}{T_{c,\rm{thin}}},
\end{align}
the same method is e.g.~used in Ref.\ \onlinecite{Hauser1972}.
Substituting the known values of $T_{c,\rm{bulk}}$, $\xi_{\rm{bulk}}$ and the $T_c=1.5~\rm{K}$ measured gives $\xi = 1.28~\rm{\mu m}$ for the superconducting film.
We may also estimate the penetration depth from known quantities \cite{Gubin2005,Tinkham2004}
\begin{align}
\lambda=\lambda_{\rm L}(0)\sqrt{1+\frac{\xi}{d}}.
\end{align}
Using the value for $\lambda_L = \lambda_{{\rm L},\rm{bulk}} = 16~\rm{nm}$ from the literature and using the modified $\xi$ calculated above, we obtain $\lambda = 180~\rm{nm}$ for a film thickness of $d=10~\rm{nm}$.

Finally we can estimate $B_{c1}$.
For type-II superconductors the field of first vortex penetration (assuming a magnetic field perpendicular to the film) is given by \cite{Tinkham2004}
\begin{align}
B_{c1} \approx \frac{\Phi_0}{4\pi\lambda^2}\log\kappa = \frac{B_c}{\sqrt{2}\kappa}\log\kappa.
\end{align}
Importantly this formula assumes that $\kappa > 1/\sqrt{2}$.
For our values $\kappa \approx 0.2 \times (1/\sqrt{2})$, clearly in the type-I regime.
However, in the thin-film limit the penetration depth is renormalized such that $\kappa = \Lambda/\xi$ \cite{Pearl1964,Gladilin2015}, where $\Lambda \sim \lambda^2 / d$. Using this renormalization we obtain $\kappa \sim 2.5$, which lies in the type-II regime. Using these numbers together with $B_{c,z} \sim 30~\rm{mT}$ yields $B_{c1} = 7.7~\rm{mT}$. It is worth noting that the first vortex may penetrate before $B_{c1}$ is reached \cite{Gutierrez2013}.
\begin{figure}[t]
	\includegraphics[]{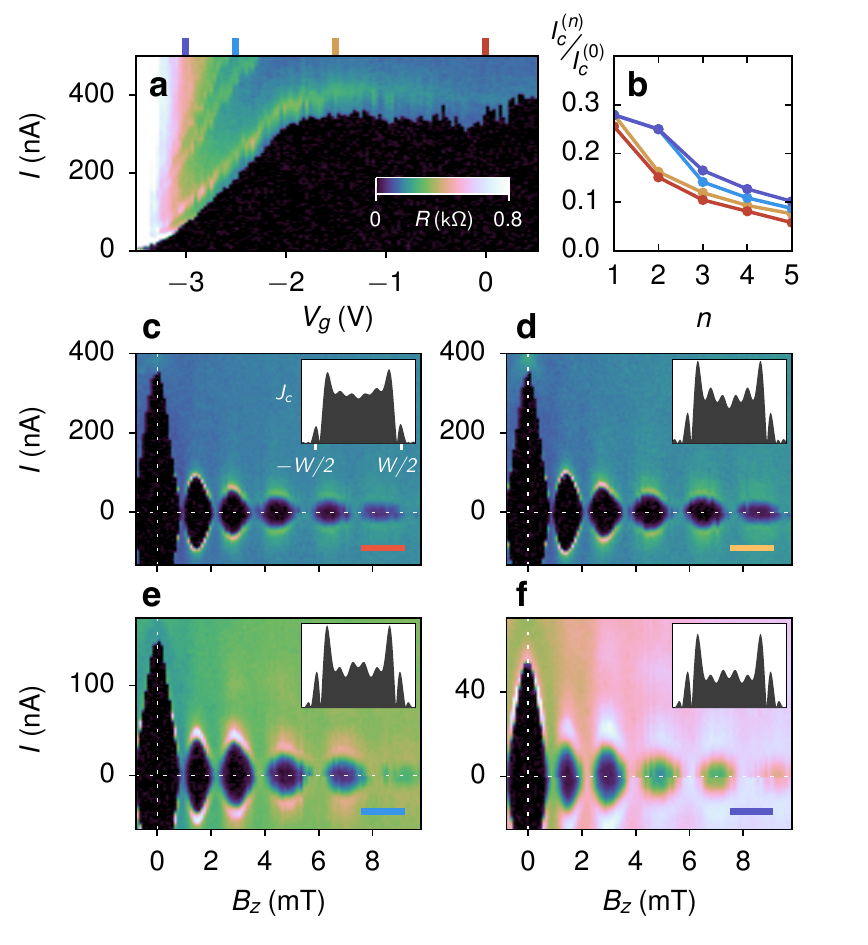}
    \caption{(a) Differential resistance $R$, as a function of gate voltage $V_{\rm{g}}$ and bias current $I$. (b) Normalized critical current as a function of side-lobe index $n$, for varying gate voltages, denoted by the colored markers in (a). (c--f) Differential resistance $R$, as a function of bias current $I$ and out-of-plane magnetic field $B_z$, for the different values of gate voltage marked in (a). Insets show the extracted supercurrent density $J_c(y)$. (c) is based on the same dataset as shown in \autoref{fig:setup}(c)}
    \label{fig:gating}
\end{figure}

\begin{figure}[bt]
	\includegraphics{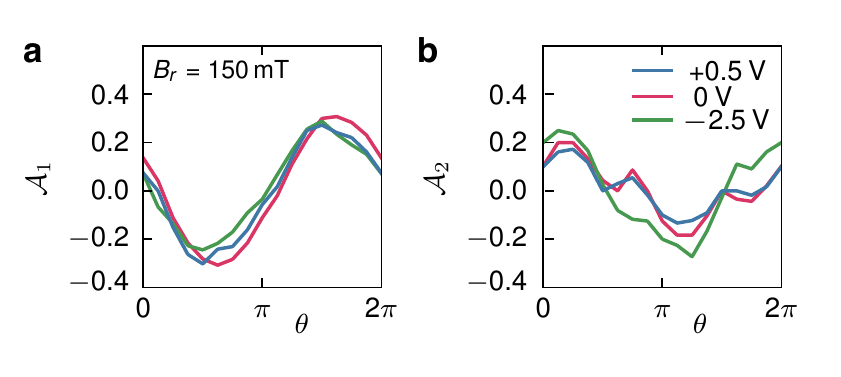}
    \caption{(a) Measured magnetic field asymmetry $\mathcal{A}_1$ for the first side-lobe as a function of in-plane field angle $\theta$. (b) as (a) for the second side-lobe $\mathcal{A}_2$. Curves are colored according to gate voltage. Note that the zero gate voltage data is the same as that displayed in \autoref{fig:rotation}.}
    \label{fig:gateasym}
\end{figure}

\section{Gate dependence}
\label{sec:gating}
The QW used for the experiment hosts two subbands at $V_{\rm{g}}=0$. Based on Hall measurements, we know that the transition to the single subband limit is achieved at $V_{\rm{g}}\sim-2~\rm{V}$. 

\Crefnoabbrv{fig:gating}(a) shows the measured differential resistance $R$, as a function of gate voltage $V_{\rm{g}}$ and bias current $I$.
The interference patterns obtained at four different values of $V_{\rm{g}}$ are shown in \autoref{fig:gating}(c--f).
From these data we can extract the field-dependent critical current $I_c(B_z)$, which we correct for the flux focusing parameter $\Gamma$ (see \autoref{sec:outofplane}). The resulting $I_c(B_z)$ can be used to calculate the supercurrent density $J_c(y)$ using the Dynes and Fulton method \cite{Dynes1971}, the results are shown in the insets in \autoref{fig:gating}(c--f).
All curves show a supercurrent density accumulation towards the lateral edges of the SNS, the effect being more accentuated at negative gate voltages.
This effect is also captured in \autoref{fig:gating}(b), where we plot the normalized side-lobe maxima.
Compatibly with the accumulation of $J_c$ at the edges, the side-lobe maxima are gradually lifted upon depletion of the 2DEG.
For $V_{\rm{g}}<-2~\rm{V}$, an anomalous lifting of the $n=2$ side-lobe is observed, similar to \autoref{fig:SNStoSQUID}(b) where in an in-plane field is applied.
We interpret the gate-voltage-induced enhancement of the critical current density at the mesa edges with band bending.
InAs is well known to host a surface accumulation layer due to the breaking of the translational symmetry of the crystal \cite{Tsui1970,Noguchi1991}.
Due to the presumably high initial electron density at the edges we expect these features to dominate as the 2DEG is depleted.

Finally we investigate the effect of the gate on the asymmetries in the interference pattern.
In \autoref{fig:gateasym}(a,b) we plot the asymmetry of the first two lobes, ${\cal A}_1$ and ${\cal A}_2$ respectively, as a function of in-plane field angle at a fixed magnitude of $B_r=150~\rm{mT}$.
The asymmetry of the two lobes is largely independent of gate voltage, both in amplitude and angular alignment.
These results highlight how the asymmetries are robust against variation of carrier density and subband occupation of the system.

\bibliography{jabref_database}

\begin{thebibliography}{68}%
\makeatletter
\providecommand \@ifxundefined [1]{%
 \@ifx{#1\undefined}
}%
\providecommand \@ifnum [1]{%
 \ifnum #1\expandafter \@firstoftwo
 \else \expandafter \@secondoftwo
 \fi
}%
\providecommand \@ifx [1]{%
 \ifx #1\expandafter \@firstoftwo
 \else \expandafter \@secondoftwo
 \fi
}%
\providecommand \natexlab [1]{#1}%
\providecommand \enquote  [1]{``#1''}%
\providecommand \bibnamefont  [1]{#1}%
\providecommand \bibfnamefont [1]{#1}%
\providecommand \citenamefont [1]{#1}%
\providecommand \href@noop [0]{\@secondoftwo}%
\providecommand \href [0]{\begingroup \@sanitize@url \@href}%
\providecommand \@href[1]{\@@startlink{#1}\@@href}%
\providecommand \@@href[1]{\endgroup#1\@@endlink}%
\providecommand \@sanitize@url [0]{\catcode `\\12\catcode `\$12\catcode
  `\&12\catcode `\#12\catcode `\^12\catcode `\_12\catcode `\%12\relax}%
\providecommand \@@startlink[1]{}%
\providecommand \@@endlink[0]{}%
\providecommand \url  [0]{\begingroup\@sanitize@url \@url }%
\providecommand \@url [1]{\endgroup\@href {#1}{\urlprefix }}%
\providecommand \urlprefix  [0]{URL }%
\providecommand \Eprint [0]{\href }%
\providecommand \doibase [0]{http://dx.doi.org/}%
\providecommand \selectlanguage [0]{\@gobble}%
\providecommand \bibinfo  [0]{\@secondoftwo}%
\providecommand \bibfield  [0]{\@secondoftwo}%
\providecommand \translation [1]{[#1]}%
\providecommand \BibitemOpen [0]{}%
\providecommand \bibitemStop [0]{}%
\providecommand \bibitemNoStop [0]{.\EOS\space}%
\providecommand \EOS [0]{\spacefactor3000\relax}%
\providecommand \BibitemShut  [1]{\csname bibitem#1\endcsname}%
\let\auto@bib@innerbib\@empty
\bibitem [{\citenamefont {Kitaev}(2001)}]{Kitaev2001}%
  \BibitemOpen
  \bibfield  {author} {\bibinfo {author} {\bibfnamefont {A.~Y.}\ \bibnamefont
  {Kitaev}},\ }\href {http://stacks.iop.org/1063-7869/44/i=10S/a=S29}
  {\bibfield  {journal} {\bibinfo  {journal} {Physics-Uspekhi}\ }\textbf
  {\bibinfo {volume} {44}},\ \bibinfo {pages} {131} (\bibinfo {year}
  {2001})}\BibitemShut {NoStop}%
\bibitem [{\citenamefont {Sau}\ \emph {et~al.}(2010)\citenamefont {Sau},
  \citenamefont {Lutchyn}, \citenamefont {Tewari},\ and\ \citenamefont
  {Das~Sarma}}]{Sau2010}%
  \BibitemOpen
  \bibfield  {author} {\bibinfo {author} {\bibfnamefont {J.~D.}\ \bibnamefont
  {Sau}}, \bibinfo {author} {\bibfnamefont {R.~M.}\ \bibnamefont {Lutchyn}},
  \bibinfo {author} {\bibfnamefont {S.}~\bibnamefont {Tewari}}, \ and\ \bibinfo
  {author} {\bibfnamefont {S.}~\bibnamefont {Das~Sarma}},\ }\href {\doibase
  10.1103/PhysRevLett.104.040502} {\bibfield  {journal} {\bibinfo  {journal}
  {Phys. Rev. Lett.}\ }\textbf {\bibinfo {volume} {104}},\ \bibinfo {pages}
  {040502} (\bibinfo {year} {2010})}\BibitemShut {NoStop}%
\bibitem [{\citenamefont {Mourik}\ \emph {et~al.}(2012)\citenamefont {Mourik},
  \citenamefont {Zuo}, \citenamefont {Frolov}, \citenamefont {Plissard},
  \citenamefont {Bakkers},\ and\ \citenamefont {Kouwenhoven}}]{Mourik2012}%
  \BibitemOpen
  \bibfield  {author} {\bibinfo {author} {\bibfnamefont {V.}~\bibnamefont
  {Mourik}}, \bibinfo {author} {\bibfnamefont {K.}~\bibnamefont {Zuo}},
  \bibinfo {author} {\bibfnamefont {S.~M.}\ \bibnamefont {Frolov}}, \bibinfo
  {author} {\bibfnamefont {S.~R.}\ \bibnamefont {Plissard}}, \bibinfo {author}
  {\bibfnamefont {E.~P. A.~M.}\ \bibnamefont {Bakkers}}, \ and\ \bibinfo
  {author} {\bibfnamefont {L.~P.}\ \bibnamefont {Kouwenhoven}},\ }\href
  {\doibase 10.1126/science.1222360} {\bibfield  {journal} {\bibinfo  {journal}
  {Science}\ }\textbf {\bibinfo {volume} {336}},\ \bibinfo {pages} {1003}
  (\bibinfo {year} {2012})}\BibitemShut {NoStop}%
\bibitem [{\citenamefont {Das}\ \emph {et~al.}(2012)\citenamefont {Das},
  \citenamefont {Ronen}, \citenamefont {Most}, \citenamefont {Oreg},
  \citenamefont {Heiblum},\ and\ \citenamefont {Shtrikman}}]{das2012}%
  \BibitemOpen
  \bibfield  {author} {\bibinfo {author} {\bibfnamefont {A.}~\bibnamefont
  {Das}}, \bibinfo {author} {\bibfnamefont {Y.}~\bibnamefont {Ronen}}, \bibinfo
  {author} {\bibfnamefont {Y.}~\bibnamefont {Most}}, \bibinfo {author}
  {\bibfnamefont {Y.}~\bibnamefont {Oreg}}, \bibinfo {author} {\bibfnamefont
  {M.}~\bibnamefont {Heiblum}}, \ and\ \bibinfo {author} {\bibfnamefont
  {H.}~\bibnamefont {Shtrikman}},\ }\href@noop {} {\bibfield  {journal}
  {\bibinfo  {journal} {Nat. Phys.}\ }\textbf {\bibinfo {volume} {8}},\
  \bibinfo {pages} {887} (\bibinfo {year} {2012})}\BibitemShut {NoStop}%
\bibitem [{\citenamefont {Deng}\ \emph {et~al.}(2012)\citenamefont {Deng},
  \citenamefont {Yu}, \citenamefont {Huang}, \citenamefont {Larsson},
  \citenamefont {Caroff},\ and\ \citenamefont {Xu}}]{Deng2012}%
  \BibitemOpen
  \bibfield  {author} {\bibinfo {author} {\bibfnamefont {M.~T.}\ \bibnamefont
  {Deng}}, \bibinfo {author} {\bibfnamefont {C.~L.}\ \bibnamefont {Yu}},
  \bibinfo {author} {\bibfnamefont {G.~Y.}\ \bibnamefont {Huang}}, \bibinfo
  {author} {\bibfnamefont {M.}~\bibnamefont {Larsson}}, \bibinfo {author}
  {\bibfnamefont {P.}~\bibnamefont {Caroff}}, \ and\ \bibinfo {author}
  {\bibfnamefont {H.~Q.}\ \bibnamefont {Xu}},\ }\href {\doibase
  10.1021/nl303758w} {\bibfield  {journal} {\bibinfo  {journal} {Nano Lett.}\
  }\textbf {\bibinfo {volume} {12}},\ \bibinfo {pages} {6414} (\bibinfo {year}
  {2012})}\BibitemShut {NoStop}%
\bibitem [{\citenamefont {Churchill}\ \emph {et~al.}(2013)\citenamefont
  {Churchill}, \citenamefont {Fatemi}, \citenamefont {Grove-Rasmussen},
  \citenamefont {Deng}, \citenamefont {Caroff}, \citenamefont {Xu},\ and\
  \citenamefont {Marcus}}]{Churchill2013}%
  \BibitemOpen
  \bibfield  {author} {\bibinfo {author} {\bibfnamefont {H.~O.~H.}\
  \bibnamefont {Churchill}}, \bibinfo {author} {\bibfnamefont {V.}~\bibnamefont
  {Fatemi}}, \bibinfo {author} {\bibfnamefont {K.}~\bibnamefont
  {Grove-Rasmussen}}, \bibinfo {author} {\bibfnamefont {M.~T.}\ \bibnamefont
  {Deng}}, \bibinfo {author} {\bibfnamefont {P.}~\bibnamefont {Caroff}},
  \bibinfo {author} {\bibfnamefont {H.~Q.}\ \bibnamefont {Xu}}, \ and\ \bibinfo
  {author} {\bibfnamefont {C.~M.}\ \bibnamefont {Marcus}},\ }\href {\doibase
  10.1103/PhysRevB.87.241401} {\bibfield  {journal} {\bibinfo  {journal} {Phys.
  Rev. B}\ }\textbf {\bibinfo {volume} {87}},\ \bibinfo {pages} {241401}
  (\bibinfo {year} {2013})}\BibitemShut {NoStop}%
\bibitem [{\citenamefont {Alicea}\ \emph {et~al.}(2011)\citenamefont {Alicea},
  \citenamefont {Oreg}, \citenamefont {Refael}, \citenamefont {von Oppen},\
  and\ \citenamefont {Fisher}}]{alicea2011}%
  \BibitemOpen
  \bibfield  {author} {\bibinfo {author} {\bibfnamefont {J.}~\bibnamefont
  {Alicea}}, \bibinfo {author} {\bibfnamefont {Y.}~\bibnamefont {Oreg}},
  \bibinfo {author} {\bibfnamefont {G.}~\bibnamefont {Refael}}, \bibinfo
  {author} {\bibfnamefont {F.}~\bibnamefont {von Oppen}}, \ and\ \bibinfo
  {author} {\bibfnamefont {M.~P.}\ \bibnamefont {Fisher}},\ }\href@noop {}
  {\bibfield  {journal} {\bibinfo  {journal} {Nat. Phys.}\ }\textbf {\bibinfo
  {volume} {7}},\ \bibinfo {pages} {412} (\bibinfo {year} {2011})}\BibitemShut
  {NoStop}%
\bibitem [{\citenamefont {Nitta}\ \emph {et~al.}(1997)\citenamefont {Nitta},
  \citenamefont {Akazaki}, \citenamefont {Takayanagi},\ and\ \citenamefont
  {Enoki}}]{Nitta1997}%
  \BibitemOpen
  \bibfield  {author} {\bibinfo {author} {\bibfnamefont {J.}~\bibnamefont
  {Nitta}}, \bibinfo {author} {\bibfnamefont {T.}~\bibnamefont {Akazaki}},
  \bibinfo {author} {\bibfnamefont {H.}~\bibnamefont {Takayanagi}}, \ and\
  \bibinfo {author} {\bibfnamefont {T.}~\bibnamefont {Enoki}},\ }\href
  {\doibase 10.1103/PhysRevLett.78.1335} {\bibfield  {journal} {\bibinfo
  {journal} {Phys. Rev. Lett.}\ }\textbf {\bibinfo {volume} {78}},\ \bibinfo
  {pages} {1335} (\bibinfo {year} {1997})}\BibitemShut {NoStop}%
\bibitem [{\citenamefont {Heida}\ \emph {et~al.}(1998)\citenamefont {Heida},
  \citenamefont {van Wees}, \citenamefont {Kuipers}, \citenamefont {Klapwijk},\
  and\ \citenamefont {Borghs}}]{Heida1998b}%
  \BibitemOpen
  \bibfield  {author} {\bibinfo {author} {\bibfnamefont {J.~P.}\ \bibnamefont
  {Heida}}, \bibinfo {author} {\bibfnamefont {B.~J.}\ \bibnamefont {van Wees}},
  \bibinfo {author} {\bibfnamefont {J.~J.}\ \bibnamefont {Kuipers}}, \bibinfo
  {author} {\bibfnamefont {T.~M.}\ \bibnamefont {Klapwijk}}, \ and\ \bibinfo
  {author} {\bibfnamefont {G.}~\bibnamefont {Borghs}},\ }\href {\doibase
  10.1103/PhysRevB.57.11911} {\bibfield  {journal} {\bibinfo  {journal} {Phys.
  Rev. B}\ }\textbf {\bibinfo {volume} {57}},\ \bibinfo {pages} {11911}
  (\bibinfo {year} {1998})}\BibitemShut {NoStop}%
\bibitem [{\citenamefont {Nitta}\ \emph {et~al.}(2003)\citenamefont {Nitta},
  \citenamefont {Lin}, \citenamefont {Akazaki},\ and\ \citenamefont
  {Koga}}]{Nitta2003}%
  \BibitemOpen
  \bibfield  {author} {\bibinfo {author} {\bibfnamefont {J.}~\bibnamefont
  {Nitta}}, \bibinfo {author} {\bibfnamefont {Y.}~\bibnamefont {Lin}}, \bibinfo
  {author} {\bibfnamefont {T.}~\bibnamefont {Akazaki}}, \ and\ \bibinfo
  {author} {\bibfnamefont {T.}~\bibnamefont {Koga}},\ }\href {\doibase
  http://dx.doi.org/10.1063/1.1631082} {\bibfield  {journal} {\bibinfo
  {journal} {Appl. Phys. Lett.}\ }\textbf {\bibinfo {volume} {83}},\ \bibinfo
  {pages} {4565} (\bibinfo {year} {2003})}\BibitemShut {NoStop}%
\bibitem [{\citenamefont {Kawakami}\ and\ \citenamefont
  {Takayanagi}(1985)}]{Kawakami1985}%
  \BibitemOpen
  \bibfield  {author} {\bibinfo {author} {\bibfnamefont {T.}~\bibnamefont
  {Kawakami}}\ and\ \bibinfo {author} {\bibfnamefont {H.}~\bibnamefont
  {Takayanagi}},\ }\href {\doibase http://dx.doi.org/10.1063/1.95809}
  {\bibfield  {journal} {\bibinfo  {journal} {Appl. Phys. Lett.}\ }\textbf
  {\bibinfo {volume} {46}},\ \bibinfo {pages} {92} (\bibinfo {year}
  {1985})}\BibitemShut {NoStop}%
\bibitem [{\citenamefont {Nguyen}\ \emph {et~al.}(1990)\citenamefont {Nguyen},
  \citenamefont {Werking}, \citenamefont {Kroemer},\ and\ \citenamefont
  {Hu}}]{Nguyen1990}%
  \BibitemOpen
  \bibfield  {author} {\bibinfo {author} {\bibfnamefont {C.}~\bibnamefont
  {Nguyen}}, \bibinfo {author} {\bibfnamefont {J.}~\bibnamefont {Werking}},
  \bibinfo {author} {\bibfnamefont {H.}~\bibnamefont {Kroemer}}, \ and\
  \bibinfo {author} {\bibfnamefont {E.~L.}\ \bibnamefont {Hu}},\ }\href
  {\doibase http://dx.doi.org/10.1063/1.103546} {\bibfield  {journal} {\bibinfo
   {journal} {Appl. Phys. Lett.}\ }\textbf {\bibinfo {volume} {57}},\ \bibinfo
  {pages} {87} (\bibinfo {year} {1990})}\BibitemShut {NoStop}%
\bibitem [{\citenamefont {Nitta}\ \emph {et~al.}(1992)\citenamefont {Nitta},
  \citenamefont {Akazaki}, \citenamefont {Takayanagi},\ and\ \citenamefont
  {Arai}}]{Nitta1992}%
  \BibitemOpen
  \bibfield  {author} {\bibinfo {author} {\bibfnamefont {J.}~\bibnamefont
  {Nitta}}, \bibinfo {author} {\bibfnamefont {T.}~\bibnamefont {Akazaki}},
  \bibinfo {author} {\bibfnamefont {H.}~\bibnamefont {Takayanagi}}, \ and\
  \bibinfo {author} {\bibfnamefont {K.}~\bibnamefont {Arai}},\ }\href {\doibase
  10.1103/PhysRevB.46.14286} {\bibfield  {journal} {\bibinfo  {journal} {Phys.
  Rev. B}\ }\textbf {\bibinfo {volume} {46}},\ \bibinfo {pages} {14286}
  (\bibinfo {year} {1992})}\BibitemShut {NoStop}%
\bibitem [{\citenamefont {Shabani}\ \emph {et~al.}(2016)\citenamefont
  {Shabani}, \citenamefont {Kjaergaard}, \citenamefont {Suominen},
  \citenamefont {Kim}, \citenamefont {Nichele}, \citenamefont {Pakrouski},
  \citenamefont {Stankevic}, \citenamefont {Lutchyn}, \citenamefont
  {Krogstrup}, \citenamefont {Feidenhans'l}, \citenamefont {Kraemer},
  \citenamefont {Nayak}, \citenamefont {Troyer}, \citenamefont {Marcus},\ and\
  \citenamefont {Palmstr\o{}m}}]{Shabani2015}%
  \BibitemOpen
  \bibfield  {author} {\bibinfo {author} {\bibfnamefont {J.}~\bibnamefont
  {Shabani}}, \bibinfo {author} {\bibfnamefont {M.}~\bibnamefont {Kjaergaard}},
  \bibinfo {author} {\bibfnamefont {H.~J.}\ \bibnamefont {Suominen}}, \bibinfo
  {author} {\bibfnamefont {Y.}~\bibnamefont {Kim}}, \bibinfo {author}
  {\bibfnamefont {F.}~\bibnamefont {Nichele}}, \bibinfo {author} {\bibfnamefont
  {K.}~\bibnamefont {Pakrouski}}, \bibinfo {author} {\bibfnamefont
  {T.}~\bibnamefont {Stankevic}}, \bibinfo {author} {\bibfnamefont {R.~M.}\
  \bibnamefont {Lutchyn}}, \bibinfo {author} {\bibfnamefont {P.}~\bibnamefont
  {Krogstrup}}, \bibinfo {author} {\bibfnamefont {R.}~\bibnamefont
  {Feidenhans'l}}, \bibinfo {author} {\bibfnamefont {S.}~\bibnamefont
  {Kraemer}}, \bibinfo {author} {\bibfnamefont {C.}~\bibnamefont {Nayak}},
  \bibinfo {author} {\bibfnamefont {M.}~\bibnamefont {Troyer}}, \bibinfo
  {author} {\bibfnamefont {C.~M.}\ \bibnamefont {Marcus}}, \ and\ \bibinfo
  {author} {\bibfnamefont {C.~J.}\ \bibnamefont {Palmstr\o{}m}},\ }\href
  {\doibase 10.1103/PhysRevB.93.155402} {\bibfield  {journal} {\bibinfo
  {journal} {Phys. Rev. B}\ }\textbf {\bibinfo {volume} {93}},\ \bibinfo
  {pages} {155402} (\bibinfo {year} {2016})}\BibitemShut {NoStop}%
\bibitem [{\citenamefont {Kjaergaard}\ \emph {et~al.}(2016)\citenamefont
  {Kjaergaard}, \citenamefont {Nichele}, \citenamefont {Suominen},
  \citenamefont {Nowak}, \citenamefont {Wimmer}, \citenamefont {Akhmerov},
  \citenamefont {Folk}, \citenamefont {Flensberg}, \citenamefont {Shabani},
  \citenamefont {Palmstr{\o}m},\ and\ \citenamefont {Marcus}}]{Kjaergaard2016}%
  \BibitemOpen
  \bibfield  {author} {\bibinfo {author} {\bibfnamefont {M.}~\bibnamefont
  {Kjaergaard}}, \bibinfo {author} {\bibfnamefont {F.}~\bibnamefont {Nichele}},
  \bibinfo {author} {\bibfnamefont {H.~J.}\ \bibnamefont {Suominen}}, \bibinfo
  {author} {\bibfnamefont {M.~P.}\ \bibnamefont {Nowak}}, \bibinfo {author}
  {\bibfnamefont {M.}~\bibnamefont {Wimmer}}, \bibinfo {author} {\bibfnamefont
  {A.~R.}\ \bibnamefont {Akhmerov}}, \bibinfo {author} {\bibfnamefont {J.~A.}\
  \bibnamefont {Folk}}, \bibinfo {author} {\bibfnamefont {K.}~\bibnamefont
  {Flensberg}}, \bibinfo {author} {\bibfnamefont {J.}~\bibnamefont {Shabani}},
  \bibinfo {author} {\bibfnamefont {C.~J.}\ \bibnamefont {Palmstr{\o}m}}, \
  and\ \bibinfo {author} {\bibfnamefont {C.~M.}\ \bibnamefont {Marcus}},\
  }\href {\doibase 10.1038/ncomms12841} {\bibfield  {journal} {\bibinfo
  {journal} {Nat. Commun.}\ }\textbf {\bibinfo {volume} {7}},\ \bibinfo {pages}
  {12841} (\bibinfo {year} {2016})}\BibitemShut {NoStop}%
\bibitem [{\citenamefont {{Kjaergaard}}\ \emph {et~al.}(2016)\citenamefont
  {{Kjaergaard}}, \citenamefont {{Suominen}}, \citenamefont {{Nowak}},
  \citenamefont {{Akhmerov}}, \citenamefont {{Shabani}}, \citenamefont
  {{Palmstr{\o}m}}, \citenamefont {{Nichele}},\ and\ \citenamefont
  {{Marcus}}}]{Kjaergaard2016b}%
  \BibitemOpen
  \bibfield  {author} {\bibinfo {author} {\bibfnamefont {M.}~\bibnamefont
  {{Kjaergaard}}}, \bibinfo {author} {\bibfnamefont {H.~J.}\ \bibnamefont
  {{Suominen}}}, \bibinfo {author} {\bibfnamefont {M.~P.}\ \bibnamefont
  {{Nowak}}}, \bibinfo {author} {\bibfnamefont {A.~R.}\ \bibnamefont
  {{Akhmerov}}}, \bibinfo {author} {\bibfnamefont {J.}~\bibnamefont
  {{Shabani}}}, \bibinfo {author} {\bibfnamefont {C.~J.}\ \bibnamefont
  {{Palmstr{\o}m}}}, \bibinfo {author} {\bibfnamefont {F.}~\bibnamefont
  {{Nichele}}}, \ and\ \bibinfo {author} {\bibfnamefont {C.~M.}\ \bibnamefont
  {{Marcus}}},\ }\href@noop {} {\  (\bibinfo {year} {2016})},\ \Eprint
  {http://arxiv.org/abs/1607.04164} {arXiv:1607.04164} \BibitemShut {NoStop}%
\bibitem [{\citenamefont {Hart}\ \emph {et~al.}(2016)\citenamefont {Hart},
  \citenamefont {Ren}, \citenamefont {Kosowsky}, \citenamefont {Ben-Shach},
  \citenamefont {Leubner}, \citenamefont {Br\"{u}ne}, \citenamefont {Buhmann},
  \citenamefont {Molenkamp}, \citenamefont {Halperin},\ and\ \citenamefont
  {Yacoby}}]{Hart2015}%
  \BibitemOpen
  \bibfield  {author} {\bibinfo {author} {\bibfnamefont {S.}~\bibnamefont
  {Hart}}, \bibinfo {author} {\bibfnamefont {H.}~\bibnamefont {Ren}}, \bibinfo
  {author} {\bibfnamefont {M.}~\bibnamefont {Kosowsky}}, \bibinfo {author}
  {\bibfnamefont {G.}~\bibnamefont {Ben-Shach}}, \bibinfo {author}
  {\bibfnamefont {P.}~\bibnamefont {Leubner}}, \bibinfo {author} {\bibfnamefont
  {C.}~\bibnamefont {Br\"{u}ne}}, \bibinfo {author} {\bibfnamefont
  {H.}~\bibnamefont {Buhmann}}, \bibinfo {author} {\bibfnamefont {L.~W.}\
  \bibnamefont {Molenkamp}}, \bibinfo {author} {\bibfnamefont {B.~I.}\
  \bibnamefont {Halperin}}, \ and\ \bibinfo {author} {\bibfnamefont
  {A.}~\bibnamefont {Yacoby}},\ }\href {http://dx.doi.org/10.1038/nphys3877}
  {\bibfield  {journal} {\bibinfo  {journal} {Nat. Phys.}\ } (\bibinfo {year}
  {2016})}\BibitemShut {NoStop}%
\bibitem [{\citenamefont {San-Jose}\ \emph {et~al.}(2013)\citenamefont
  {San-Jose}, \citenamefont {Cayao}, \citenamefont {Prada},\ and\ \citenamefont
  {Aguado}}]{SanJose2013}%
  \BibitemOpen
  \bibfield  {author} {\bibinfo {author} {\bibfnamefont {P.}~\bibnamefont
  {San-Jose}}, \bibinfo {author} {\bibfnamefont {J.}~\bibnamefont {Cayao}},
  \bibinfo {author} {\bibfnamefont {E.}~\bibnamefont {Prada}}, \ and\ \bibinfo
  {author} {\bibfnamefont {R.}~\bibnamefont {Aguado}},\ }\href
  {http://stacks.iop.org/1367-2630/15/i=7/a=075019} {\bibfield  {journal}
  {\bibinfo  {journal} {New J. Phys.}\ }\textbf {\bibinfo {volume} {15}},\
  \bibinfo {pages} {075019} (\bibinfo {year} {2013})}\BibitemShut {NoStop}%
\bibitem [{\citenamefont {San-Jose}\ \emph {et~al.}(2014)\citenamefont
  {San-Jose}, \citenamefont {Prada},\ and\ \citenamefont
  {Aguado}}]{SanJose2014}%
  \BibitemOpen
  \bibfield  {author} {\bibinfo {author} {\bibfnamefont {P.}~\bibnamefont
  {San-Jose}}, \bibinfo {author} {\bibfnamefont {E.}~\bibnamefont {Prada}}, \
  and\ \bibinfo {author} {\bibfnamefont {R.}~\bibnamefont {Aguado}},\ }\href
  {\doibase 10.1103/PhysRevLett.112.137001} {\bibfield  {journal} {\bibinfo
  {journal} {Phys. Rev. Lett.}\ }\textbf {\bibinfo {volume} {112}},\ \bibinfo
  {pages} {137001} (\bibinfo {year} {2014})}\BibitemShut {NoStop}%
\bibitem [{\citenamefont {{Hell}}\ \emph {et~al.}(2016)\citenamefont {{Hell}},
  \citenamefont {{Leijnse}},\ and\ \citenamefont {{Flensberg}}}]{Hell2016}%
  \BibitemOpen
  \bibfield  {author} {\bibinfo {author} {\bibfnamefont {M.}~\bibnamefont
  {{Hell}}}, \bibinfo {author} {\bibfnamefont {M.}~\bibnamefont {{Leijnse}}}, \
  and\ \bibinfo {author} {\bibfnamefont {K.}~\bibnamefont {{Flensberg}}},\
  }\href@noop {} {\  (\bibinfo {year} {2016})},\ \Eprint
  {http://arxiv.org/abs/1608.08769} {arXiv:1608.08769} \BibitemShut {NoStop}%
\bibitem [{\citenamefont {{Pientka}}\ \emph {et~al.}(2016)\citenamefont
  {{Pientka}}, \citenamefont {{Keselman}}, \citenamefont {{Berg}},
  \citenamefont {{Yacoby}}, \citenamefont {{Stern}},\ and\ \citenamefont
  {{Halperin}}}]{Pientka2016}%
  \BibitemOpen
  \bibfield  {author} {\bibinfo {author} {\bibfnamefont {F.}~\bibnamefont
  {{Pientka}}}, \bibinfo {author} {\bibfnamefont {A.}~\bibnamefont
  {{Keselman}}}, \bibinfo {author} {\bibfnamefont {E.}~\bibnamefont {{Berg}}},
  \bibinfo {author} {\bibfnamefont {A.}~\bibnamefont {{Yacoby}}}, \bibinfo
  {author} {\bibfnamefont {A.}~\bibnamefont {{Stern}}}, \ and\ \bibinfo
  {author} {\bibfnamefont {B.~I.}\ \bibnamefont {{Halperin}}},\ }\href@noop {}
  {\  (\bibinfo {year} {2016})},\ \Eprint {http://arxiv.org/abs/1609.09482}
  {arXiv:1609.09482} \BibitemShut {NoStop}%
\bibitem [{\citenamefont {Liu}\ and\ \citenamefont {Chan}(2010)}]{Liu2010}%
  \BibitemOpen
  \bibfield  {author} {\bibinfo {author} {\bibfnamefont {J.-F.}\ \bibnamefont
  {Liu}}\ and\ \bibinfo {author} {\bibfnamefont {K.~S.}\ \bibnamefont {Chan}},\
  }\href {\doibase 10.1103/PhysRevB.82.125305} {\bibfield  {journal} {\bibinfo
  {journal} {Phys. Rev. B}\ }\textbf {\bibinfo {volume} {82}},\ \bibinfo
  {pages} {125305} (\bibinfo {year} {2010})}\BibitemShut {NoStop}%
\bibitem [{\citenamefont {Bezuglyi}\ \emph {et~al.}(2002)\citenamefont
  {Bezuglyi}, \citenamefont {Rozhavsky}, \citenamefont {Vagner},\ and\
  \citenamefont {Wyder}}]{Bezuglyi2002}%
  \BibitemOpen
  \bibfield  {author} {\bibinfo {author} {\bibfnamefont {E.~V.}\ \bibnamefont
  {Bezuglyi}}, \bibinfo {author} {\bibfnamefont {A.~S.}\ \bibnamefont
  {Rozhavsky}}, \bibinfo {author} {\bibfnamefont {I.~D.}\ \bibnamefont
  {Vagner}}, \ and\ \bibinfo {author} {\bibfnamefont {P.}~\bibnamefont
  {Wyder}},\ }\href {\doibase 10.1103/PhysRevB.66.052508} {\bibfield  {journal}
  {\bibinfo  {journal} {Phys. Rev. B}\ }\textbf {\bibinfo {volume} {66}},\
  \bibinfo {pages} {052508} (\bibinfo {year} {2002})}\BibitemShut {NoStop}%
\bibitem [{\citenamefont {B\'eri}\ \emph {et~al.}(2008)\citenamefont {B\'eri},
  \citenamefont {Bardarson},\ and\ \citenamefont {Beenakker}}]{Beri2008}%
  \BibitemOpen
  \bibfield  {author} {\bibinfo {author} {\bibfnamefont {B.}~\bibnamefont
  {B\'eri}}, \bibinfo {author} {\bibfnamefont {J.~H.}\ \bibnamefont
  {Bardarson}}, \ and\ \bibinfo {author} {\bibfnamefont {C.~W.~J.}\
  \bibnamefont {Beenakker}},\ }\href {\doibase 10.1103/PhysRevB.77.045311}
  {\bibfield  {journal} {\bibinfo  {journal} {Phys. Rev. B}\ }\textbf {\bibinfo
  {volume} {77}},\ \bibinfo {pages} {045311} (\bibinfo {year}
  {2008})}\BibitemShut {NoStop}%
\bibitem [{\citenamefont {Reynoso}\ \emph {et~al.}(2008)\citenamefont
  {Reynoso}, \citenamefont {Usaj}, \citenamefont {Balseiro}, \citenamefont
  {Feinberg},\ and\ \citenamefont {Avignon}}]{Reynoso2008}%
  \BibitemOpen
  \bibfield  {author} {\bibinfo {author} {\bibfnamefont {A.~A.}\ \bibnamefont
  {Reynoso}}, \bibinfo {author} {\bibfnamefont {G.}~\bibnamefont {Usaj}},
  \bibinfo {author} {\bibfnamefont {C.~A.}\ \bibnamefont {Balseiro}}, \bibinfo
  {author} {\bibfnamefont {D.}~\bibnamefont {Feinberg}}, \ and\ \bibinfo
  {author} {\bibfnamefont {M.}~\bibnamefont {Avignon}},\ }\href {\doibase
  10.1103/physrevlett.101.107001} {\bibfield  {journal} {\bibinfo  {journal}
  {Phys. Rev. Lett.}\ }\textbf {\bibinfo {volume} {101}},\ \bibinfo {pages}
  {107001} (\bibinfo {year} {2008})}\BibitemShut {NoStop}%
\bibitem [{\citenamefont {Reynoso}\ \emph {et~al.}(2012)\citenamefont
  {Reynoso}, \citenamefont {Usaj}, \citenamefont {Balseiro}, \citenamefont
  {Feinberg},\ and\ \citenamefont {Avignon}}]{Reynoso2012}%
  \BibitemOpen
  \bibfield  {author} {\bibinfo {author} {\bibfnamefont {A.~A.}\ \bibnamefont
  {Reynoso}}, \bibinfo {author} {\bibfnamefont {G.}~\bibnamefont {Usaj}},
  \bibinfo {author} {\bibfnamefont {C.~A.}\ \bibnamefont {Balseiro}}, \bibinfo
  {author} {\bibfnamefont {D.}~\bibnamefont {Feinberg}}, \ and\ \bibinfo
  {author} {\bibfnamefont {M.}~\bibnamefont {Avignon}},\ }\href {\doibase
  10.1103/PhysRevB.86.214519} {\bibfield  {journal} {\bibinfo  {journal} {Phys.
  Rev. B}\ }\textbf {\bibinfo {volume} {86}},\ \bibinfo {pages} {214519}
  (\bibinfo {year} {2012})}\BibitemShut {NoStop}%
\bibitem [{\citenamefont {Yokoyama}\ \emph {et~al.}(2014)\citenamefont
  {Yokoyama}, \citenamefont {Eto},\ and\ \citenamefont
  {Nazarov}}]{Yokoyama2014}%
  \BibitemOpen
  \bibfield  {author} {\bibinfo {author} {\bibfnamefont {T.}~\bibnamefont
  {Yokoyama}}, \bibinfo {author} {\bibfnamefont {M.}~\bibnamefont {Eto}}, \
  and\ \bibinfo {author} {\bibfnamefont {Y.~V.}\ \bibnamefont {Nazarov}},\
  }\href {\doibase 10.1103/physrevb.89.195407} {\bibfield  {journal} {\bibinfo
  {journal} {Phys. Rev. B}\ }\textbf {\bibinfo {volume} {89}},\ \bibinfo
  {pages} {195407} (\bibinfo {year} {2014})}\BibitemShut {NoStop}%
\bibitem [{\citenamefont {Tinkham}(2004)}]{Tinkham2004}%
  \BibitemOpen
  \bibfield  {author} {\bibinfo {author} {\bibfnamefont {M.}~\bibnamefont
  {Tinkham}},\ }\href {https://books.google.dk/books?id=k6AO9nRYbioC} {\emph
  {\bibinfo {title} {Introduction to Superconductivity: Second Edition}}},\
  Dover Books on Physics\ (\bibinfo  {publisher} {Dover Publications},\
  \bibinfo {year} {2004})\BibitemShut {NoStop}%
\bibitem [{\citenamefont {Rowell}(1963)}]{Rowell1963}%
  \BibitemOpen
  \bibfield  {author} {\bibinfo {author} {\bibfnamefont {J.~M.}\ \bibnamefont
  {Rowell}},\ }\href {\doibase 10.1103/PhysRevLett.11.200} {\bibfield
  {journal} {\bibinfo  {journal} {Phys. Rev. Lett.}\ }\textbf {\bibinfo
  {volume} {11}},\ \bibinfo {pages} {200} (\bibinfo {year} {1963})}\BibitemShut
  {NoStop}%
\bibitem [{\citenamefont {Nishino}\ \emph {et~al.}(1986)\citenamefont
  {Nishino}, \citenamefont {Kawabe},\ and\ \citenamefont
  {Yamada}}]{Nishino1986}%
  \BibitemOpen
  \bibfield  {author} {\bibinfo {author} {\bibfnamefont {T.}~\bibnamefont
  {Nishino}}, \bibinfo {author} {\bibfnamefont {U.}~\bibnamefont {Kawabe}}, \
  and\ \bibinfo {author} {\bibfnamefont {E.}~\bibnamefont {Yamada}},\ }\href
  {\doibase 10.1103/PhysRevB.34.4857} {\bibfield  {journal} {\bibinfo
  {journal} {Phys. Rev. B}\ }\textbf {\bibinfo {volume} {34}},\ \bibinfo
  {pages} {4857} (\bibinfo {year} {1986})}\BibitemShut {NoStop}%
\bibitem [{\citenamefont {Inoue}\ and\ \citenamefont
  {Kawakami}(1989)}]{Inoue1989}%
  \BibitemOpen
  \bibfield  {author} {\bibinfo {author} {\bibfnamefont {K.}~\bibnamefont
  {Inoue}}\ and\ \bibinfo {author} {\bibfnamefont {T.}~\bibnamefont
  {Kawakami}},\ }\href {\doibase 10.1063/1.342930} {\bibfield  {journal}
  {\bibinfo  {journal} {J. Appl. Phys.}\ }\textbf {\bibinfo {volume} {65}},\
  \bibinfo {pages} {1631} (\bibinfo {year} {1989})}\BibitemShut {NoStop}%
\bibitem [{\citenamefont {Miller}\ \emph {et~al.}(1985)\citenamefont {Miller},
  \citenamefont {Biagi}, \citenamefont {Clem},\ and\ \citenamefont
  {Finnemore}}]{Miller1985}%
  \BibitemOpen
  \bibfield  {author} {\bibinfo {author} {\bibfnamefont {S.~L.}\ \bibnamefont
  {Miller}}, \bibinfo {author} {\bibfnamefont {K.~R.}\ \bibnamefont {Biagi}},
  \bibinfo {author} {\bibfnamefont {J.~R.}\ \bibnamefont {Clem}}, \ and\
  \bibinfo {author} {\bibfnamefont {D.~K.}\ \bibnamefont {Finnemore}},\ }\href
  {\doibase 10.1103/PhysRevB.31.2684} {\bibfield  {journal} {\bibinfo
  {journal} {Phys. Rev. B}\ }\textbf {\bibinfo {volume} {31}},\ \bibinfo
  {pages} {2684} (\bibinfo {year} {1985})}\BibitemShut {NoStop}%
\bibitem [{\citenamefont {Dynes}\ and\ \citenamefont
  {Fulton}(1971)}]{Dynes1971}%
  \BibitemOpen
  \bibfield  {author} {\bibinfo {author} {\bibfnamefont {R.~C.}\ \bibnamefont
  {Dynes}}\ and\ \bibinfo {author} {\bibfnamefont {T.~A.}\ \bibnamefont
  {Fulton}},\ }\href {\doibase 10.1103/PhysRevB.3.3015} {\bibfield  {journal}
  {\bibinfo  {journal} {Phys. Rev. B}\ }\textbf {\bibinfo {volume} {3}},\
  \bibinfo {pages} {3015} (\bibinfo {year} {1971})}\BibitemShut {NoStop}%
\bibitem [{\citenamefont {Hui}\ \emph {et~al.}(2014)\citenamefont {Hui},
  \citenamefont {Lobos}, \citenamefont {Sau},\ and\ \citenamefont
  {Das~Sarma}}]{Hui2014}%
  \BibitemOpen
  \bibfield  {author} {\bibinfo {author} {\bibfnamefont {H.-Y.}\ \bibnamefont
  {Hui}}, \bibinfo {author} {\bibfnamefont {A.~M.}\ \bibnamefont {Lobos}},
  \bibinfo {author} {\bibfnamefont {J.~D.}\ \bibnamefont {Sau}}, \ and\
  \bibinfo {author} {\bibfnamefont {S.}~\bibnamefont {Das~Sarma}},\ }\href
  {\doibase 10.1103/PhysRevB.90.224517} {\bibfield  {journal} {\bibinfo
  {journal} {Phys. Rev. B}\ }\textbf {\bibinfo {volume} {90}},\ \bibinfo
  {pages} {224517} (\bibinfo {year} {2014})}\BibitemShut {NoStop}%
\bibitem [{\citenamefont {Hart}\ \emph {et~al.}(2014)\citenamefont {Hart},
  \citenamefont {Ren}, \citenamefont {Wagner}, \citenamefont {Leubner},
  \citenamefont {Muhlbauer}, \citenamefont {Brune}, \citenamefont {Buhmann},
  \citenamefont {Molenkamp},\ and\ \citenamefont {Yacoby}}]{Hart2014}%
  \BibitemOpen
  \bibfield  {author} {\bibinfo {author} {\bibfnamefont {S.}~\bibnamefont
  {Hart}}, \bibinfo {author} {\bibfnamefont {H.}~\bibnamefont {Ren}}, \bibinfo
  {author} {\bibfnamefont {T.}~\bibnamefont {Wagner}}, \bibinfo {author}
  {\bibfnamefont {P.}~\bibnamefont {Leubner}}, \bibinfo {author} {\bibfnamefont
  {M.}~\bibnamefont {Muhlbauer}}, \bibinfo {author} {\bibfnamefont
  {C.}~\bibnamefont {Brune}}, \bibinfo {author} {\bibfnamefont
  {H.}~\bibnamefont {Buhmann}}, \bibinfo {author} {\bibfnamefont {L.~W.}\
  \bibnamefont {Molenkamp}}, \ and\ \bibinfo {author} {\bibfnamefont
  {A.}~\bibnamefont {Yacoby}},\ }\href {http://dx.doi.org/10.1038/nphys3036}
  {\bibfield  {journal} {\bibinfo  {journal} {Nat. Phys.}\ }\textbf {\bibinfo
  {volume} {10}},\ \bibinfo {pages} {638} (\bibinfo {year} {2014})}\BibitemShut
  {NoStop}%
\bibitem [{\citenamefont {Pribiag}\ \emph {et~al.}(2015)\citenamefont
  {Pribiag}, \citenamefont {Beukman}, \citenamefont {Qu}, \citenamefont
  {Cassidy}, \citenamefont {Charpentier}, \citenamefont {Wegscheider},\ and\
  \citenamefont {Kouwenhoven}}]{Pribiag2015}%
  \BibitemOpen
  \bibfield  {author} {\bibinfo {author} {\bibfnamefont {V.~S.}\ \bibnamefont
  {Pribiag}}, \bibinfo {author} {\bibfnamefont {A.~J.~A.}\ \bibnamefont
  {Beukman}}, \bibinfo {author} {\bibfnamefont {F.}~\bibnamefont {Qu}},
  \bibinfo {author} {\bibfnamefont {M.~C.}\ \bibnamefont {Cassidy}}, \bibinfo
  {author} {\bibfnamefont {C.}~\bibnamefont {Charpentier}}, \bibinfo {author}
  {\bibfnamefont {W.}~\bibnamefont {Wegscheider}}, \ and\ \bibinfo {author}
  {\bibfnamefont {L.~P.}\ \bibnamefont {Kouwenhoven}},\ }\href
  {http://dx.doi.org/10.1038/nnano.2015.86} {\bibfield  {journal} {\bibinfo
  {journal} {Nat. Nano}\ }\textbf {\bibinfo {volume} {10}},\ \bibinfo {pages}
  {593} (\bibinfo {year} {2015})}\BibitemShut {NoStop}%
\bibitem [{\citenamefont {Allen}\ \emph {et~al.}(2015)\citenamefont {Allen},
  \citenamefont {Shtanko}, \citenamefont {Fulga}, \citenamefont {Akhmerov},
  \citenamefont {Watanabe}, \citenamefont {Taniguchi}, \citenamefont
  {Jarillo-Herrero}, \citenamefont {Levitov},\ and\ \citenamefont
  {Yacoby}}]{allen2015}%
  \BibitemOpen
  \bibfield  {author} {\bibinfo {author} {\bibfnamefont {M.~T.}\ \bibnamefont
  {Allen}}, \bibinfo {author} {\bibfnamefont {O.}~\bibnamefont {Shtanko}},
  \bibinfo {author} {\bibfnamefont {I.~C.}\ \bibnamefont {Fulga}}, \bibinfo
  {author} {\bibfnamefont {A.~R.}\ \bibnamefont {Akhmerov}}, \bibinfo {author}
  {\bibfnamefont {K.}~\bibnamefont {Watanabe}}, \bibinfo {author}
  {\bibfnamefont {T.}~\bibnamefont {Taniguchi}}, \bibinfo {author}
  {\bibfnamefont {P.}~\bibnamefont {Jarillo-Herrero}}, \bibinfo {author}
  {\bibfnamefont {L.~S.}\ \bibnamefont {Levitov}}, \ and\ \bibinfo {author}
  {\bibfnamefont {A.}~\bibnamefont {Yacoby}},\ }\href {\doibase
  10.1038/nphys3534} {\bibfield  {journal} {\bibinfo  {journal} {Nat. Phys.}\
  }\textbf {\bibinfo {volume} {12}},\ \bibinfo {pages} {128} (\bibinfo {year}
  {2015})}\BibitemShut {NoStop}%
\bibitem [{\citenamefont {Magn\'ee}\ \emph {et~al.}(1995)\citenamefont
  {Magn\'ee}, \citenamefont {den Hartog}, \citenamefont {Wees}, \citenamefont
  {Klapwijk}, \citenamefont {van~de Graaf},\ and\ \citenamefont
  {Borghs}}]{Magnee1995}%
  \BibitemOpen
  \bibfield  {author} {\bibinfo {author} {\bibfnamefont {P.~H.~C.}\
  \bibnamefont {Magn\'ee}}, \bibinfo {author} {\bibfnamefont {S.~G.}\
  \bibnamefont {den Hartog}}, \bibinfo {author} {\bibfnamefont {B.~J.~v.}\
  \bibnamefont {Wees}}, \bibinfo {author} {\bibfnamefont {T.~M.}\ \bibnamefont
  {Klapwijk}}, \bibinfo {author} {\bibfnamefont {W.}~\bibnamefont {van~de
  Graaf}}, \ and\ \bibinfo {author} {\bibfnamefont {G.}~\bibnamefont
  {Borghs}},\ }\href {\doibase http://dx.doi.org/10.1063/1.115320} {\bibfield
  {journal} {\bibinfo  {journal} {Appl. Phys. Lett.}\ }\textbf {\bibinfo
  {volume} {67}},\ \bibinfo {pages} {3569} (\bibinfo {year}
  {1995})}\BibitemShut {NoStop}%
\bibitem [{\citenamefont {Sch{\"a}pers}(2001)}]{schapers2001}%
  \BibitemOpen
  \bibfield  {author} {\bibinfo {author} {\bibfnamefont {T.}~\bibnamefont
  {Sch{\"a}pers}},\ }\href@noop {} {\emph {\bibinfo {title}
  {Superconductor/semiconductor junctions}}},\ Vol.\ \bibinfo {volume} {174}\
  (\bibinfo  {publisher} {Springer Science \& Business Media},\ \bibinfo {year}
  {2001})\BibitemShut {NoStop}%
\bibitem [{Note1()}]{Note1}%
  \BibitemOpen
  \bibinfo {note} {An effective mass of $m_{\protect \rm {eff}} = 0.05m_e$ is
  estimated from k.p calculations}\BibitemShut {NoStop}%
\bibitem [{\citenamefont {Courtois}\ \emph {et~al.}(2008)\citenamefont
  {Courtois}, \citenamefont {Meschke}, \citenamefont {Peltonen},\ and\
  \citenamefont {Pekola}}]{Courtois2008}%
  \BibitemOpen
  \bibfield  {author} {\bibinfo {author} {\bibfnamefont {H.}~\bibnamefont
  {Courtois}}, \bibinfo {author} {\bibfnamefont {M.}~\bibnamefont {Meschke}},
  \bibinfo {author} {\bibfnamefont {J.~T.}\ \bibnamefont {Peltonen}}, \ and\
  \bibinfo {author} {\bibfnamefont {J.~P.}\ \bibnamefont {Pekola}},\ }\href
  {\doibase 10.1103/PhysRevLett.101.067002} {\bibfield  {journal} {\bibinfo
  {journal} {Phys. Rev. Lett.}\ }\textbf {\bibinfo {volume} {101}},\ \bibinfo
  {pages} {067002} (\bibinfo {year} {2008})}\BibitemShut {NoStop}%
\bibitem [{Note2()}]{Note2}%
  \BibitemOpen
  \bibinfo {note} {The deviation at high field between our result and the
  expectation is presumably due to an underestimation of the junction area due
  to the neglect of the finite penetration depth in the leads \cite
  {Barone2005}. Utilizing an effective length $L_{\protect \rm
  {eff}}=L+2\lambda _L$ (with $\lambda _L$ estimated in \protect \autoref
  {sec:bc1}) yields an expected node spacing of $1.7~\protect \rm
  {mT}$.}\BibitemShut {Stop}%
\bibitem [{\citenamefont {Harada}\ \emph {et~al.}(2002)\citenamefont {Harada},
  \citenamefont {Jensen}, \citenamefont {Akazaki},\ and\ \citenamefont
  {Takayanagi}}]{Harada2002}%
  \BibitemOpen
  \bibfield  {author} {\bibinfo {author} {\bibfnamefont {Y.}~\bibnamefont
  {Harada}}, \bibinfo {author} {\bibfnamefont {S.}~\bibnamefont {Jensen}},
  \bibinfo {author} {\bibfnamefont {T.}~\bibnamefont {Akazaki}}, \ and\
  \bibinfo {author} {\bibfnamefont {H.}~\bibnamefont {Takayanagi}},\ }\href
  {\doibase http://dx.doi.org/10.1016/S0921-4534(01)01023-1} {\bibfield
  {journal} {\bibinfo  {journal} {Physica C: Superconductivity}\ }\textbf
  {\bibinfo {volume} {367}},\ \bibinfo {pages} {229 } (\bibinfo {year}
  {2002})}\BibitemShut {NoStop}%
\bibitem [{\citenamefont {Paajaste}\ \emph {et~al.}(2015)\citenamefont
  {Paajaste}, \citenamefont {Amado}, \citenamefont {Roddaro}, \citenamefont
  {Bergeret}, \citenamefont {Ercolani}, \citenamefont {Sorba},\ and\
  \citenamefont {Giazotto}}]{Paajaste2015}%
  \BibitemOpen
  \bibfield  {author} {\bibinfo {author} {\bibfnamefont {J.}~\bibnamefont
  {Paajaste}}, \bibinfo {author} {\bibfnamefont {M.}~\bibnamefont {Amado}},
  \bibinfo {author} {\bibfnamefont {S.}~\bibnamefont {Roddaro}}, \bibinfo
  {author} {\bibfnamefont {F.~S.}\ \bibnamefont {Bergeret}}, \bibinfo {author}
  {\bibfnamefont {D.}~\bibnamefont {Ercolani}}, \bibinfo {author}
  {\bibfnamefont {L.}~\bibnamefont {Sorba}}, \ and\ \bibinfo {author}
  {\bibfnamefont {F.}~\bibnamefont {Giazotto}},\ }\href {\doibase
  10.1021/nl504544s} {\bibfield  {journal} {\bibinfo  {journal} {Nano Lett.}\
  }\textbf {\bibinfo {volume} {15}},\ \bibinfo {pages} {1803} (\bibinfo {year}
  {2015})}\BibitemShut {NoStop}%
\bibitem [{\citenamefont {Zeldov}\ \emph {et~al.}(1994)\citenamefont {Zeldov},
  \citenamefont {Clem}, \citenamefont {McElfresh},\ and\ \citenamefont
  {Darwin}}]{Zeldov1994}%
  \BibitemOpen
  \bibfield  {author} {\bibinfo {author} {\bibfnamefont {E.}~\bibnamefont
  {Zeldov}}, \bibinfo {author} {\bibfnamefont {J.~R.}\ \bibnamefont {Clem}},
  \bibinfo {author} {\bibfnamefont {M.}~\bibnamefont {McElfresh}}, \ and\
  \bibinfo {author} {\bibfnamefont {M.}~\bibnamefont {Darwin}},\ }\href
  {\doibase 10.1103/PhysRevB.49.9802} {\bibfield  {journal} {\bibinfo
  {journal} {Phys. Rev. B}\ }\textbf {\bibinfo {volume} {49}},\ \bibinfo
  {pages} {9802} (\bibinfo {year} {1994})}\BibitemShut {NoStop}%
\bibitem [{\citenamefont {Brandt}\ and\ \citenamefont
  {Indenbom}(1993)}]{Brandt1993}%
  \BibitemOpen
  \bibfield  {author} {\bibinfo {author} {\bibfnamefont {E.~H.}\ \bibnamefont
  {Brandt}}\ and\ \bibinfo {author} {\bibfnamefont {M.}~\bibnamefont
  {Indenbom}},\ }\href {\doibase 10.1103/PhysRevB.48.12893} {\bibfield
  {journal} {\bibinfo  {journal} {Phys. Rev. B}\ }\textbf {\bibinfo {volume}
  {48}},\ \bibinfo {pages} {12893} (\bibinfo {year} {1993})}\BibitemShut
  {NoStop}%
\bibitem [{Note3()}]{Note3}%
  \BibitemOpen
  \bibinfo {note} {For $B_z>B_f$, $B_{\protect \rm {eff}}(\protect \mathaccentV
  {tilde}07Ex)\propto B_z$. Indeed for large $\protect \mathaccentV
  {tilde}07Ex$ we find $B_{\protect \rm {eff}}(\protect \mathaccentV
  {tilde}07Ex)=B_f\protect \qopname \relax o{log}\left (\protect \qopname
  \relax o{cosh}\left (\protect \frac {B_z}{B_f}\right )\left [1+\protect
  \qopname \relax o{tanh}\left (\protect \frac {B_z}{B_f}\right )\right ]\right
  )=B_z$.}\BibitemShut {Stop}%
\bibitem [{\citenamefont {Gu}\ \emph {et~al.}(1979)\citenamefont {Gu},
  \citenamefont {Cha}, \citenamefont {Gamo},\ and\ \citenamefont
  {Namba}}]{Gu1979}%
  \BibitemOpen
  \bibfield  {author} {\bibinfo {author} {\bibfnamefont {J.}~\bibnamefont
  {Gu}}, \bibinfo {author} {\bibfnamefont {W.}~\bibnamefont {Cha}}, \bibinfo
  {author} {\bibfnamefont {K.}~\bibnamefont {Gamo}}, \ and\ \bibinfo {author}
  {\bibfnamefont {S.}~\bibnamefont {Namba}},\ }\href {\doibase
  http://dx.doi.org/10.1063/1.325736} {\bibfield  {journal} {\bibinfo
  {journal} {J. Appl. Phys.}\ }\textbf {\bibinfo {volume} {50}},\ \bibinfo
  {pages} {6437} (\bibinfo {year} {1979})}\BibitemShut {NoStop}%
\bibitem [{\citenamefont {Khukhareva}(1963)}]{khukhareva1963}%
  \BibitemOpen
  \bibfield  {author} {\bibinfo {author} {\bibfnamefont {I.~S.}\ \bibnamefont
  {Khukhareva}},\ }\href@noop {} {\bibfield  {journal} {\bibinfo  {journal}
  {JETP Lett.}\ }\textbf {\bibinfo {volume} {16}} (\bibinfo {year}
  {1963})}\BibitemShut {NoStop}%
\bibitem [{Note4()}]{Note4}%
  \BibitemOpen
  \bibinfo {note} {We note that this model neglects the effect of SOI. We have
  verified that spin-orbit effects, calculated along the lines of Ref.~\cite
  {Bezuglyi2002}, yield changes on the order of a few percent while the
  experimental anisotropy is of the order 1.}\BibitemShut {Stop}%
\bibitem [{Note5()}]{Note5}%
  \BibitemOpen
  \bibinfo {note} {In this section we concentrate largely on qualitative
  features and thus for simplicity neglect the effect of out-of-plane focusing
  as discussed in \protect \autoref {sec:outofplane}}\BibitemShut {NoStop}%
\bibitem [{Note6()}]{Note6}%
  \BibitemOpen
  \bibinfo {note} {Close to the edges of the junction, where $y_0 \approx \pm
  W/2$, there are fewer angles $\vartheta $ available to construct Andreev
  bound states with. Consequently, the flux penetrating the N region close to
  the edges has less influence on the total average supercurrent through the
  junction than the flux penetrating the center of the region. To achieve the
  first full suppression of the supercurrent by perfect destructive
  interference of all trajectories, one thus needs to go to slightly higher
  fields than $B_z = \Phi _0/(WL)$.}\BibitemShut {Stop}%
\bibitem [{Note7()}]{Note7}%
  \BibitemOpen
  \bibinfo {note} {See the Supplementary Material for more details and data
  from different samples.}\BibitemShut {Stop}%
\bibitem [{\citenamefont {Rasmussen}\ \emph {et~al.}(2016)\citenamefont
  {Rasmussen}, \citenamefont {Danon}, \citenamefont {Suominen}, \citenamefont
  {Nichele}, \citenamefont {Kjaergaard},\ and\ \citenamefont
  {Flensberg}}]{Rasmussen2015}%
  \BibitemOpen
  \bibfield  {author} {\bibinfo {author} {\bibfnamefont {A.}~\bibnamefont
  {Rasmussen}}, \bibinfo {author} {\bibfnamefont {J.}~\bibnamefont {Danon}},
  \bibinfo {author} {\bibfnamefont {H.}~\bibnamefont {Suominen}}, \bibinfo
  {author} {\bibfnamefont {F.}~\bibnamefont {Nichele}}, \bibinfo {author}
  {\bibfnamefont {M.}~\bibnamefont {Kjaergaard}}, \ and\ \bibinfo {author}
  {\bibfnamefont {K.}~\bibnamefont {Flensberg}},\ }\href {\doibase
  10.1103/PhysRevB.93.155406} {\bibfield  {journal} {\bibinfo  {journal} {Phys.
  Rev. B}\ }\textbf {\bibinfo {volume} {93}},\ \bibinfo {pages} {155406}
  (\bibinfo {year} {2016})}\BibitemShut {NoStop}%
\bibitem [{\citenamefont {Golubov}\ \emph {et~al.}(2004)\citenamefont
  {Golubov}, \citenamefont {Kupriyanov},\ and\ \citenamefont
  {Il'ichev}}]{Golubov2004}%
  \BibitemOpen
  \bibfield  {author} {\bibinfo {author} {\bibfnamefont {A.~A.}\ \bibnamefont
  {Golubov}}, \bibinfo {author} {\bibfnamefont {M.~Y.}\ \bibnamefont
  {Kupriyanov}}, \ and\ \bibinfo {author} {\bibfnamefont {E.}~\bibnamefont
  {Il'ichev}},\ }\href {\doibase 10.1103/RevModPhys.76.411} {\bibfield
  {journal} {\bibinfo  {journal} {Rev. Mod. Phys.}\ }\textbf {\bibinfo {volume}
  {76}},\ \bibinfo {pages} {411} (\bibinfo {year} {2004})}\BibitemShut
  {NoStop}%
\bibitem [{\citenamefont {Golod}\ \emph {et~al.}(2010)\citenamefont {Golod},
  \citenamefont {Rydh},\ and\ \citenamefont {Krasnov}}]{Golod2010}%
  \BibitemOpen
  \bibfield  {author} {\bibinfo {author} {\bibfnamefont {T.}~\bibnamefont
  {Golod}}, \bibinfo {author} {\bibfnamefont {A.}~\bibnamefont {Rydh}}, \ and\
  \bibinfo {author} {\bibfnamefont {V.~M.}\ \bibnamefont {Krasnov}},\ }\href
  {http://dx.doi.org/10.1103/PhysRevLett.104.227003} {\bibfield  {journal}
  {\bibinfo  {journal} {Phys. Rev. Lett.}\ }\textbf {\bibinfo {volume} {104}}
  (\bibinfo {year} {2010})}\BibitemShut {NoStop}%
\bibitem [{\citenamefont {Fistul'}(1989)}]{Fistul1989}%
  \BibitemOpen
  \bibfield  {author} {\bibinfo {author} {\bibfnamefont {M.~V.}\ \bibnamefont
  {Fistul'}},\ }\href@noop {} {\bibfield  {journal} {\bibinfo  {journal} {JETP
  Lett.}\ }\textbf {\bibinfo {volume} {69}},\ \bibinfo {pages} {209} (\bibinfo
  {year} {1989})}\BibitemShut {NoStop}%
\bibitem [{\citenamefont {Itzler}\ and\ \citenamefont
  {Tinkham}(1996)}]{Itzler1996}%
  \BibitemOpen
  \bibfield  {author} {\bibinfo {author} {\bibfnamefont {M.~A.}\ \bibnamefont
  {Itzler}}\ and\ \bibinfo {author} {\bibfnamefont {M.}~\bibnamefont
  {Tinkham}},\ }\href {\doibase 10.1103/PhysRevB.53.R11949} {\bibfield
  {journal} {\bibinfo  {journal} {Phys. Rev. B}\ }\textbf {\bibinfo {volume}
  {53}},\ \bibinfo {pages} {R11949} (\bibinfo {year} {1996})}\BibitemShut
  {NoStop}%
\bibitem [{\citenamefont {Fistul'}\ and\ \citenamefont
  {Giuliani}(1995)}]{Fistul1995}%
  \BibitemOpen
  \bibfield  {author} {\bibinfo {author} {\bibfnamefont {M.~V.}\ \bibnamefont
  {Fistul'}}\ and\ \bibinfo {author} {\bibfnamefont {G.~F.}\ \bibnamefont
  {Giuliani}},\ }\href {\doibase 10.1103/PhysRevB.51.1090} {\bibfield
  {journal} {\bibinfo  {journal} {Phys. Rev. B}\ }\textbf {\bibinfo {volume}
  {51}},\ \bibinfo {pages} {1090} (\bibinfo {year} {1995})}\BibitemShut
  {NoStop}%
\bibitem [{\citenamefont {Merservey}\ and\ \citenamefont
  {Schwartz}(1969)}]{Merservey1969}%
  \BibitemOpen
  \bibfield  {author} {\bibinfo {author} {\bibfnamefont {R.}~\bibnamefont
  {Merservey}}\ and\ \bibinfo {author} {\bibfnamefont {B.~B.}\ \bibnamefont
  {Schwartz}},\ }in\ \href@noop {} {\emph {\bibinfo {booktitle}
  {Superconductivity Vol. 1}}},\ \bibinfo {editor} {edited by\ \bibinfo
  {editor} {\bibfnamefont {R.~D.}\ \bibnamefont {Parks}}}\ (\bibinfo
  {publisher} {Marcel Dekker Inc.},\ \bibinfo {year} {1969})\ p.\ \bibinfo
  {pages} {126}\BibitemShut {NoStop}%
\bibitem [{\citenamefont {Hauser}(1972)}]{Hauser1972}%
  \BibitemOpen
  \bibfield  {author} {\bibinfo {author} {\bibfnamefont {J.}~\bibnamefont
  {Hauser}},\ }\href {\doibase 10.1007/BF00660071} {\bibfield  {journal}
  {\bibinfo  {journal} {J. Low Temp. Phys.}\ }\textbf {\bibinfo {volume} {7}},\
  \bibinfo {pages} {335} (\bibinfo {year} {1972})}\BibitemShut {NoStop}%
\bibitem [{\citenamefont {Gubin}\ \emph {et~al.}(2005)\citenamefont {Gubin},
  \citenamefont {Il'in}, \citenamefont {Vitusevich}, \citenamefont {Siegel},\
  and\ \citenamefont {Klein}}]{Gubin2005}%
  \BibitemOpen
  \bibfield  {author} {\bibinfo {author} {\bibfnamefont {A.~I.}\ \bibnamefont
  {Gubin}}, \bibinfo {author} {\bibfnamefont {K.~S.}\ \bibnamefont {Il'in}},
  \bibinfo {author} {\bibfnamefont {S.~A.}\ \bibnamefont {Vitusevich}},
  \bibinfo {author} {\bibfnamefont {M.}~\bibnamefont {Siegel}}, \ and\ \bibinfo
  {author} {\bibfnamefont {N.}~\bibnamefont {Klein}},\ }\href {\doibase
  10.1103/PhysRevB.72.064503} {\bibfield  {journal} {\bibinfo  {journal} {Phys.
  Rev. B}\ }\textbf {\bibinfo {volume} {72}},\ \bibinfo {pages} {064503}
  (\bibinfo {year} {2005})}\BibitemShut {NoStop}%
\bibitem [{\citenamefont {Pearl}(1964)}]{Pearl1964}%
  \BibitemOpen
  \bibfield  {author} {\bibinfo {author} {\bibfnamefont {J.}~\bibnamefont
  {Pearl}},\ }\href {\doibase http://dx.doi.org/10.1063/1.1754056} {\bibfield
  {journal} {\bibinfo  {journal} {Appl. Phys. Lett.}\ }\textbf {\bibinfo
  {volume} {5}},\ \bibinfo {pages} {65} (\bibinfo {year} {1964})}\BibitemShut
  {NoStop}%
\bibitem [{\citenamefont {Gladilin}\ \emph {et~al.}(2015)\citenamefont
  {Gladilin}, \citenamefont {Ge}, \citenamefont {Gutierrez}, \citenamefont
  {Timmermans}, \citenamefont {de~Vondel}, \citenamefont {Tempere},
  \citenamefont {Devreese},\ and\ \citenamefont {Moshchalkov}}]{Gladilin2015}%
  \BibitemOpen
  \bibfield  {author} {\bibinfo {author} {\bibfnamefont {V.~N.}\ \bibnamefont
  {Gladilin}}, \bibinfo {author} {\bibfnamefont {J.}~\bibnamefont {Ge}},
  \bibinfo {author} {\bibfnamefont {J.}~\bibnamefont {Gutierrez}}, \bibinfo
  {author} {\bibfnamefont {M.}~\bibnamefont {Timmermans}}, \bibinfo {author}
  {\bibfnamefont {J.~V.}\ \bibnamefont {de~Vondel}}, \bibinfo {author}
  {\bibfnamefont {J.}~\bibnamefont {Tempere}}, \bibinfo {author} {\bibfnamefont
  {J.~T.}\ \bibnamefont {Devreese}}, \ and\ \bibinfo {author} {\bibfnamefont
  {V.~V.}\ \bibnamefont {Moshchalkov}},\ }\href
  {http://stacks.iop.org/1367-2630/17/i=6/a=063032} {\bibfield  {journal}
  {\bibinfo  {journal} {New J. Phys.}\ }\textbf {\bibinfo {volume} {17}},\
  \bibinfo {pages} {063032} (\bibinfo {year} {2015})}\BibitemShut {NoStop}%
\bibitem [{\citenamefont {Gutierrez}\ \emph {et~al.}(2013)\citenamefont
  {Gutierrez}, \citenamefont {Raes}, \citenamefont {Van~de Vondel},
  \citenamefont {Silhanek}, \citenamefont {Kramer}, \citenamefont {Ataklti},\
  and\ \citenamefont {Moshchalkov}}]{Gutierrez2013}%
  \BibitemOpen
  \bibfield  {author} {\bibinfo {author} {\bibfnamefont {J.}~\bibnamefont
  {Gutierrez}}, \bibinfo {author} {\bibfnamefont {B.}~\bibnamefont {Raes}},
  \bibinfo {author} {\bibfnamefont {J.}~\bibnamefont {Van~de Vondel}}, \bibinfo
  {author} {\bibfnamefont {A.~V.}\ \bibnamefont {Silhanek}}, \bibinfo {author}
  {\bibfnamefont {R.~B.~G.}\ \bibnamefont {Kramer}}, \bibinfo {author}
  {\bibfnamefont {G.~W.}\ \bibnamefont {Ataklti}}, \ and\ \bibinfo {author}
  {\bibfnamefont {V.~V.}\ \bibnamefont {Moshchalkov}},\ }\href {\doibase
  10.1103/PhysRevB.88.184504} {\bibfield  {journal} {\bibinfo  {journal} {Phys.
  Rev. B}\ }\textbf {\bibinfo {volume} {88}},\ \bibinfo {pages} {184504}
  (\bibinfo {year} {2013})}\BibitemShut {NoStop}%
\bibitem [{\citenamefont {Tsui}(1970)}]{Tsui1970}%
  \BibitemOpen
  \bibfield  {author} {\bibinfo {author} {\bibfnamefont {D.~C.}\ \bibnamefont
  {Tsui}},\ }\href {\doibase 10.1103/PhysRevLett.24.303} {\bibfield  {journal}
  {\bibinfo  {journal} {Phys. Rev. Lett.}\ }\textbf {\bibinfo {volume} {24}},\
  \bibinfo {pages} {303} (\bibinfo {year} {1970})}\BibitemShut {NoStop}%
\bibitem [{\citenamefont {Noguchi}\ \emph {et~al.}(1991)\citenamefont
  {Noguchi}, \citenamefont {Hirakawa},\ and\ \citenamefont
  {Ikoma}}]{Noguchi1991}%
  \BibitemOpen
  \bibfield  {author} {\bibinfo {author} {\bibfnamefont {M.}~\bibnamefont
  {Noguchi}}, \bibinfo {author} {\bibfnamefont {K.}~\bibnamefont {Hirakawa}}, \
  and\ \bibinfo {author} {\bibfnamefont {T.}~\bibnamefont {Ikoma}},\ }\href
  {\doibase 10.1103/PhysRevLett.66.2243} {\bibfield  {journal} {\bibinfo
  {journal} {Phys. Rev. Lett.}\ }\textbf {\bibinfo {volume} {66}},\ \bibinfo
  {pages} {2243} (\bibinfo {year} {1991})}\BibitemShut {NoStop}%
\bibitem [{\citenamefont {Barone}\ and\ \citenamefont
  {Paterno}(1982)}]{Barone2005}%
  \BibitemOpen
  \bibfield  {author} {\bibinfo {author} {\bibfnamefont {A.}~\bibnamefont
  {Barone}}\ and\ \bibinfo {author} {\bibfnamefont {G.}~\bibnamefont
  {Paterno}},\ }\href {http://dx.doi.org/10.1002/352760278X.fmatter} {\emph
  {\bibinfo {title} {Physics and Applications of the Josephson Effect}}}\
  (\bibinfo  {publisher} {Wiley and Sons Inc.},\ \bibinfo {year}
  {1982})\BibitemShut {NoStop}%
\end{thebibliography}%

\clearpage
\pagebreak
\widetext
\begin{center}
\textbf{\large Supplementary Material: Anomalous Fraunhofer Interference in\\ Epitaxial Superconductor-Semiconductor Josephson Junctions}
\end{center}
\setcounter{equation}{0}
\setcounter{figure}{0}
\makeatletter
\renewcommand{\theequation}{S\arabic{equation}}
\renewcommand{\thefigure}{S\arabic{figure}}
\newcommand{\ua}{\uparrow}
\newcommand{\da}{\downarrow}

\begin{center}
This Supplemental Material Section describes the wafer structure, characterization of the superconducting Al film, supplementary devices, symmetry analysis, and tight-binding simulations. It also provides complete datasets for the extracted data presented in the paper.
\end{center}

\twocolumngrid

\section*{Aluminum film characterization}
\begin{figure*}[t!]
\includegraphics{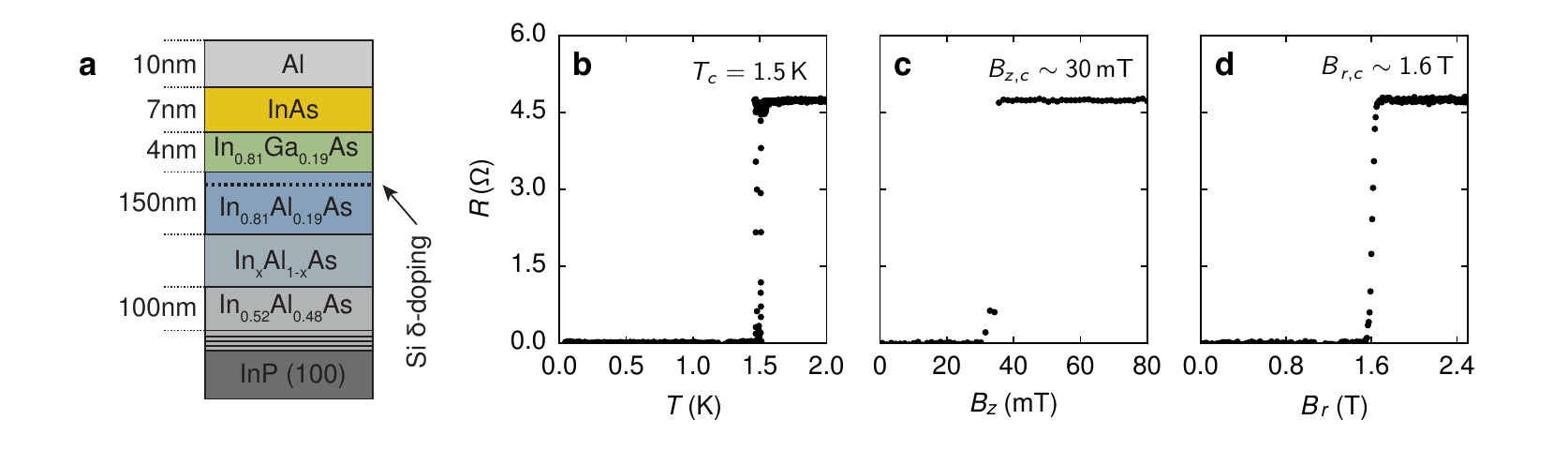}
\caption{(a) Full wafer structure of the Al/InAs heterostructure. (b--d) Resistance in a Hall bar geometry, (b) as a function of temperature of the aluminum film, (c) as a function of perpendicular field, and (d) as a function of in-plane field.}
\label{fig:filmcharac}
\end{figure*}

The full wafer structure used in this study is presented in \autoref{fig:filmcharac}(a). In \autoref{fig:filmcharac}(b--d) we show a characterization of the epitaxial aluminum film measured in a Hall bar geometry as a function of temperature (b), perpendicular magnetic field (c) and in-plane magnetic field (d). Extracted parameters are the superconducting transition temperature $T_c=1.5\,\rm{K}$, perpendicular critical field $B_{z,c} = 30\,\rm{mT}$, and in-plane critical field $B_{r,c} = 1.6\,\rm{T}$.

\section*{Additional devices}

\subsection*{Flux focusing control device}

\begin{figure}[tb!]
\includegraphics[width=\linewidth]{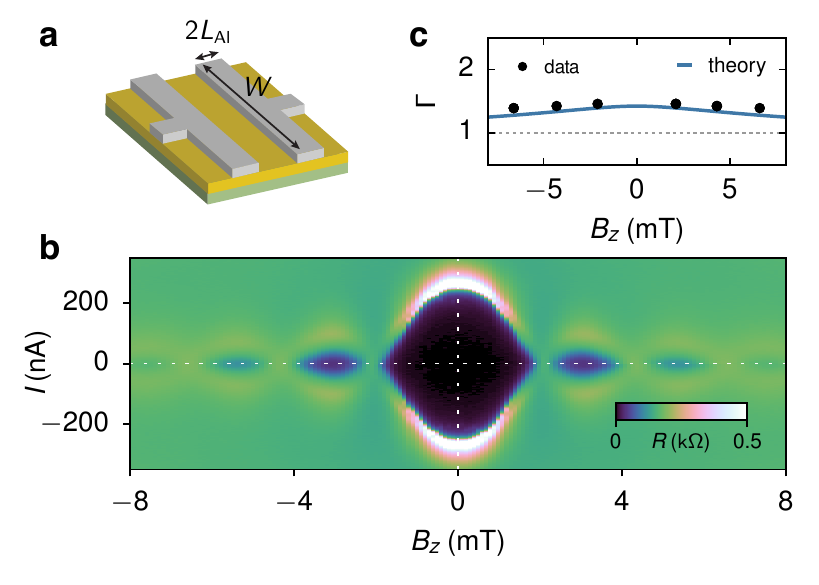}
\caption{(a) Schematic of the flux-focusing control device. (b) Differential resistance $R$ as a function of current $I$ and perpendicular field $B_z$. (c) Extracted field enhancement at the nodes visible in (b) (markers), and fit using Eqs.~(3)--(6) of the main text (solid line).}
\label{fig:focuscontrol}
\end{figure}

The behavior of a control device with large regions of removed aluminum behind the junction, as shown in \autoref{fig:focuscontrol}(a), is demonstrated in \autoref{fig:focuscontrol}(b) and (c). The dimensions of the central semiconducting region are lithographically identical to that of the primary device studied in the main text, cf.~Fig.~1(a). In contrast the device in the main text with Al lead dimensions $W=1.5~\rm{\mu m}$ and $2L_{\rm Al} \sim 10~\rm{\mu m}$, the contacts of the device presented here have $W=1.5~\rm{\mu m}$ and $2L_{\rm Al} = 0.3~\rm{\mu m}$.
The lack of extended aluminum planes atop the leads results in a more uniform magnetic field profile perpendicular to the junction plane, minimizing flux focusing.
\autoref{fig:focuscontrol}(b) shows the measured interference pattern of $I_c(B_z)$ on this device. All figures from here on including colorplots are displayed on a constant colorscale ranging from $R=0$ to $R=0.5~\rm{k\Omega}$. By extracting the positions of the visible node closings we obtain an effective field enhancement as shown in \autoref{fig:focuscontrol}(c), which can be compared to Fig.~1(d).
Whilst a finite enhancement is observed, the value is roughly constant in the field range measured. Applying the model developed in Eqs.~(3)--(6) of the main text to the present flux-minimizing geometry yields the blue curve, in good agreement with the data.

\subsection*{Devices rotated with respect to crystal axes}

\begin{figure*}[tb!]
\includegraphics{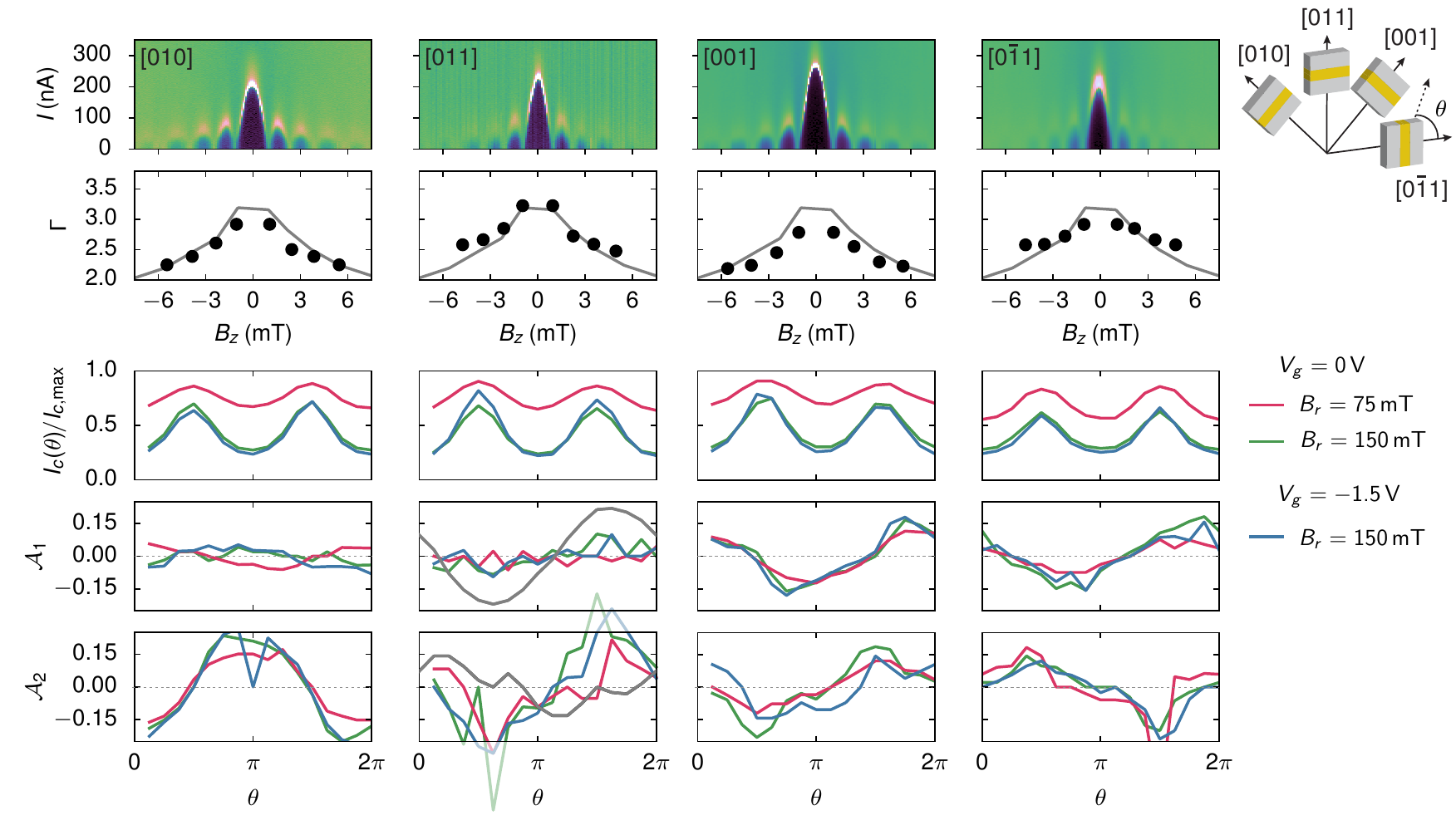} 
\caption{(Top row) Differential resistance as function of applied current and perpendicular magnetic field for four additional devices. (Second row) Extracted field enhancement at the nodes of the interference pattern (markers). The gray line indicates the behavior of the device studied in the main text. (Third row) Behavior of the critical current at $B_z = 0$, as a function of in plane field angle $\theta$.  Different curves correspond to different field magnitudes and gate voltages. Note that for all devices $\theta$ is measured with respect to the direction of current flow as indicated in the top right inset. (Fourth row) Extracted asymmetry of the first side-lobe pair $\mathcal{A}_1$. The gray line for the [011] device indicates the behavior of the nominally identical device studied in the main text. (Fifth row) The same for the second side-lobe pair $\mathcal{A}_2$.}
\label{fig:extradevices}
\end{figure*}

A number of additional samples were investigated where the device design was rotated relative to the crystal, as shown in the top right of \autoref{fig:extradevices}.
These devices are otherwise lithographically identical to the one examined in the main text.

The top row of \autoref{fig:extradevices} shows the interference patterns observed in all devices.
The second row shows the extracted field enhancement parameters of all visible nodes for each device (markers).
For comparison with the main text we also plot the enhancement envelope from Fig.~1(d) (solid gray line).
All devices show aperiodic node spacings, with the effective field enhancement decreasing with increasing applied field.
The variation in $\Gamma(B_z)$ observed across the samples is attributed to small variations in the effective sample dimensions arising during processing.
The third row demonstrates the behavior of the critical current for a purely in-plane field ($B_z = 0$) as a function of field angle $\theta$ (the current is normalized to the maximum value $I_{c,\rm{max}}$ measured at zero field).
Curves are shown for varying in-plane field magnitudes and gate voltages as detailed in the legend.
Overall, we see roughly a factor of two suppression of $I_{c}$ between $B_r=75\,\rm{mT}$ and $150\,\rm{mT}$ when the field is applied along the current ($x$ direction, $\theta=0$).
For fields applied perpendicular to the current ($y$ direction, $\theta=\pi/2$) the suppression is considerably weaker, consistent with our interpretation in terms of flux focusing (see main text).
Negligible differences are observed for different values of gate voltage.
The fourth and fifth rows demonstrate the behavior of the asymmetry parameters ${\cal A}$ of the first and second side-lobe pair respectively.
Concentrating initially on the $[011]$ column, corresponding to a device nominally identical to the one examined in the main text, it is clear that the specific behavior of the asymmetry is not quantitatively reproducible across devices (data from Fig.~5(c) and 7(a) of the main text are shown in solid gray for comparison).
Furthermore, comparing all four junctions we do not observe any systematic dependence on crystal orientation as one might expect for an intrinsic spin-orbit dominated effect.
These results support our suggestion outlined in the main text that disorder plays a key role in determining the precise magnitude and alignment of the asymmetries.
Full data sets including magnetic diffraction patterns for all measured angles and configurations are provided in \Crefrange{fig:dev2}{fig:dev4}.
For completeness we also include in \autoref{fig:focus} the full data set for the control sample with narrow Al contacts shown in \autoref{fig:focuscontrol}.

\section*{Symmetry analysis}

In this section we investigate a simple model Hamiltonian describing the two-dimensional SNS junction, and try to identify which ingredients could be responsible for the striking asymmetries in the interference patterns that are reported in Sec. V.B of the main text.
Following the approach of Ref.~\onlinecite{Rasmussen2015}, we describe the electrons in the junction by a Bogoliubov-de Gennes Hamiltonian $H = \frac{1}{2} \int d{\bf r} \, \Psi^\dagger {\cal H}\Psi$, with $\Psi = [ \psi_\ua({\bf r}), \psi_\da({\bf r}), \psi^\dagger_\da({\bf r}), -\psi^\dagger_\ua({\bf r})]^T$, where $\psi_{\ua(\da)}({\bf r})$ is the electronic annihilation operator for an electron with spin up(down) at position ${\bf r}$. We use
\begin{align}
{\cal H} = \Big\{ \frac{\hat{{\bf p}}^2}{2m} -\mu +V(x,y)\Big\} \tau_z
+{\cal H}_{\rm S}
+\frac{1}{2} g\mu_{\rm B}{\bf B}\cdot \boldsymbol\sigma
+{\cal H}_{\rm SO},
\label{eq:ham}
\end{align}
where the Pauli matrices $\boldsymbol\tau$ and $\boldsymbol\sigma$ act in electron-hole and spin space respectively.
The momentum operator $\hat{\bf p} = -i\hbar\nabla_{\bf r} - e{\bf A}$ includes the effect of a vector potential ${\bf A} = -B_zy\hat x\tau_z$, the term ${\cal H}_{\rm S}$ introduces a superconducting pairing potential $\propto \Delta e^{-i\varphi/2}$($\Delta e^{i\varphi/2}$) under the left(right) contact, the third term describes the Zeeman coupling of the electron spin to the applied magnetic field, and ${\cal H}_{\rm SO}$ accounts for the spin-orbit coupling. Disorder is modeled with a local electrostatic potential $V(x,y)$.
For average current flow aligned with the $[011]$ crystallographic direction, the spin-orbit Hamiltonian reads
\begin{align}
{\cal H}_{\rm SO} = i(\alpha-\beta) \partial_y \tau_z\sigma_x +(\alpha+\beta)\big\{ eB_zy- i \partial_x \tau_z\big \} \sigma_y,
\end{align}
where $\alpha$($\beta$) is the Rashba(Dresselhaus) coefficient.
The flux focusing due to the in-plane field is taken into account by making $B_z$ effectively $x$-dependent, $B_z(x) = B_z + \delta B(x)$, where $\delta B(x)$ is proportional to $B_x$ and antisymmetric under $x$-reflection ${\cal V}_x \delta B(x) = -\delta B(x){\cal V}_x$.

We can now investigate which combinations and orientations of spin-orbit coupling, in-plane field, and asymmetric potential $V(x,y)$ can in principle yield an asymmetric interference pattern.
Without any of these ingredients, we can find four symmetry transformations of the Hamiltonian that guarantee a symmetric interference pattern with $I_c(+B_z) = I_c(-B_z)$: (i) $\sigma_x{\cal V}_y$, (ii) $\sigma_y{\cal V}_y$, (iii) ${\cal P}{\cal T}$, and (iv) $\sigma_z{\cal P}{\cal T}$, where ${\cal V}_y$ is the $y$-reflection operator, ${\cal P} = {\cal V}_x{\cal V}_y$ the spatial inversion operator, and ${\cal T}$ the time-reversal operator.
Adding a finite spin-orbit coupling or an in-plane field with $B_y \neq 0$ breaks symmetries (i) and (iii), an in-plane field with $B_x \neq 0$ breaks symmetries (i)--(iii) due to the presence of the flux dipole, an asymmetric potential with $V(x,y) \neq V(-x,y)$ breaks symmetries (iii) and (iv), and a potential with $V(x,y)\neq V(x,-y)$ breaks all four symmetries.
When all four symmetries are broken, the symmetry $I_c(+B_z) = I_c(-B_z)$ is no longer protected and the interference pattern will in general be asymmetric.

This qualitative analysis suggests that an important role is played by the disorder potential $V(x,y)$.
Indeed, symmetry (iv) is only broken in the presence of an asymmetric potential, making it within this model a necessary ingredient for observing $I_c(+B_z) \neq I_c(-B_z)$.
Moreover, when the in-plane field is oriented along $\hat y$, a direction along which we observe a strong asymmetry in the device studied in the main text, the only ingredient left that can break symmetry (ii) is having $V(x,y) \neq V(x,-y)$.

Finite disorder in principle also allows for asymmetries in $I_c(\pm B_z)$ in the {\it absence} of any in-plane fields, seemingly at odds with the behavior observed in Fig.~1(c) of the main text.
However, careful analysis reveals that even in the presence of disorder the ``inversion'' symmetry between positive and negative critical current $I_{c+}(+B_z) = I_{c-}(-B_z)$ can be protected, by ${\cal T}$ and by $\sigma_z {\cal T}$.
If one of these symmetries is present, one can observe $I_{c}(+B_z) \neq I_{c}(-B_z)$ only if positive and negative critical currents at a fixed $B_z$ are allowed to be different, which requires higher-order Fourier components (i.e.~the non-sinusoidal part) of the Josephson current-phase relation to be significant.
If that would be the case, the asymmetries in the interference pattern could arise due to (i) a finite $B_x$ or (ii) a finite $B_y$ in combination with spin-orbit interaction.
Then, the degree of asymmetry left when the in-plane field is oriented perpendicular to the current could present a measure for the strength of spin-orbit interaction in the junction.
The consistently observed equal positive and negative critical currents in the experimental data suggest, however, that the current-phase relation is more or less sinusoidal.

\section*{Tight binding simulations}

To support the suggestion of Sec.~V.B that the microscopic (disordered) structure of the junction can play a crucial role for the behavior of the asymmetries in the interference pattern, we present numerical simulations of the supercurrent through a two-dimensional disordered SNS-junction.
For the normal region we write the model Hamiltonian
\begin{align}
{\cal H}_{\rm N} =  \Big\{ \frac{\hat{{\bf p}}^2}{2m} -\mu +V(x,y)\Big\} \tau_z
+\frac{1}{2} g\mu_{\rm B}\tilde{\bf B}\cdot \boldsymbol\sigma 
+{\cal H}_{\rm SO},
\label{eq:ham2}
\end{align}
where momentum operator $\hat{\bf p} = -i\hbar\nabla_{\bf r} - e{\bf A}$ again includes the effect of a vector potential ${\bf A} = -\tilde B_z y\hat x\tau_z$.
As before, we include in-plane flux focusing by making the magnetic field position-dependent: the field $\tilde{\bf B}$ is the effective field including the flux focusing, whereas ${\bf B}$ is the actual applied field.
Explicitly, we use
\begin{widetext}
\begin{align}
\tilde{\bf B}(x) = \begin{cases}
(\sqrt{1-f^2} B_x, B_y, B_z + f B_x) & \text{for } {-L/2} \leq x < -L/2 + d_f, \\
(B_x, B_y, B_z) & \text{for } {-L/2} + d_f \leq x < L/2 - d_f, \\
(\sqrt{1-f^2} B_x, B_y, B_z - f B_x) & \text{for } L/2 - d_f \leq x \leq L/2. \\
\end{cases}
\end{align}
\end{widetext}
The $z$-component of the field thus gets shifted by a $\pm fB_x$ in a strip of width $d_f$ next to the contacts.
We further include the chemical potential $\mu$ and a (possibly disordered) electronic potential $V(x,y)$ in the first term of ${\cal H}$.

The calculations that follow are based on a perturbative expansion of the free energy of the central normal region, assuming for ease of calculation weak coupling to the superconductors (see Ref.~\cite{Rasmussen2015} and especially its Supplementary Material for all details of the calculation).
Integrating out the degrees of freedom of the superconducting contacts, one finds to leading order in the semiconductor-superconductor coupling for the supercurrent
\begin{widetext}
\begin{align}
I_s(\varphi) = & -{\rm Im} \bigg[ e^{-i\varphi} \frac{4eT}{\hbar}\sum_{n,\sigma,\sigma'}\int_{-\tfrac{W}{2}}^{\tfrac{W}{2}}dy_1dy_2 \,
\frac{(\kappa W \Delta)^{2}}{\Delta^{2}+\omega_{n}^{2}}
\,\sigma\,\sigma'\, {\cal G}_{\sigma'\sigma}^{RL}(y_2,y_1;i\omega_{n}){\cal G}^{RL}_{\bar \sigma'\bar \sigma}(y_2,y_1;-i\omega_{n}) \bigg],
\label{eq:supcur}
\end{align}
\end{widetext}
where $T$ is the temperature, $\kappa$ parametrizes the strength of the NS-coupling, the fermionic Matsubara frequencies are $\omega_n = (2n\pi +1)T/\hbar$, and $\sigma,\sigma' = \pm 1$ denote the two electronic spin directions.
${\cal G}^{RL}_{\sigma'\sigma}(y_2,y_1;i\omega)$ is the Matsubara Green function for an electron in the normal region with spin $\sigma$ at position $(-L/2,y_1)$ evolving to a spin $\sigma'$ at position $(L/2,y_2)$; it thus describes propagation from the left to the right boundary of the normal region.
We note that this lowest-order expression only produces the first Fourier component of $I_s(\varphi)$, i.e.~the result will always have the form $I_s(\varphi) = I_c \sin(\varphi - \varphi_0)$ leading to equal positive and negative critical currents $I_{c+} = I_{c-}$.

For the numerical simulations we discretize the full Hamiltonian (\ref{eq:ham2}) for the electrons in the normal part.
The required Green functions ${\cal G}^{RL}$ are found by solving for elements of $[i\hbar\omega-H_{\rm N}]^{-1}$, where $H_{\rm N}$ is the discretized lattice version of (\ref{eq:ham2}), and the supercurrent through the junction then follows from (\ref{eq:supcur}), where the integrals over $y_{1,2}$ are replaced by sums over lattice sites.

In our simulations we use a $30\times 120$ lattice with lattice constant $a = 2.5$~nm, resulting in $L = 75$~nm and $W = 300$~nm.
Using an effective electronic mass of $m = 0.026\, m_e$ this yields a hopping matrix element $t = \hbar^2 / 2ma^2 = 234$ meV.
We use a Fermi wavelength of $\lambda_{\rm F} = 20$~nm, which corresponds to $\mu = 0.62\,t$, and a $g$-factor of $g = -10$, yielding a ``Zeeman length'' $l_{\rm Z} = 2\pi \hbar v_{\rm F} / |g|\mu_{\rm B} B \approx 50~\mu$m for $B=200$~mT.
The Rashba and Dresselhaus coefficients are set to $\alpha = 1$~eV\AA{} and $\beta = 0.25$~eV\AA{} respectively, corresponding to spin-orbit lengths $\pi \hbar^2 / m \alpha = 92$~nm and $\pi\hbar^2/ m \beta = 368$~nm.
We further take $\Delta = 0.2$~meV, such that the coherence length $\xi = \hbar v_{\rm F} / \pi \Delta \approx 1.5~\mu$m, in the short-junction limit.
The temperature is set to $T = 100$ mK and we use an NS coupling parameter $\kappa = 3$~meV.
We include disorder by adding an onsite potential $V(x,y)$ with its elements picked from a uniform distribution between $[-U/2, U/2]$, where $U = (48 a/l_e)^{1/2}(\mu/t)^{1/4}t$ with $l_e = 50$~nm being the effective electronic mean free path.
The width of the strips where flux focusing is present is set to $d_f = 15$~nm, with its strength $f$ as well as the in-plane field magnitude $B_\parallel$ varied for different plots, see below.

\begin{figure}[b!]
\includegraphics{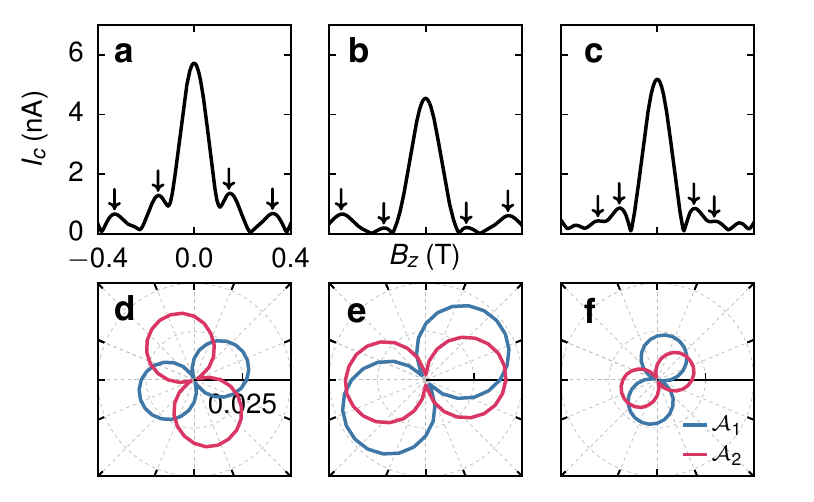} 
\caption{(a--c) Critical current as a function of $B_z$ for three distinct disorder configurations. In all panels an in-plane field of $200~\rm{mT}$ is applied along $B_x$ and other parameters are fixed as detailed in the text. The local maxima corresponding to the first two side-lobes are marked with arrows. (d--f) Lobe asymmetries $\mathcal{A}_{1,2}$, in blue and red respectively, as a function of in-plane field angle, for the disorder configurations in (a--c).}
\label{fig:disorder}
\end{figure}

\begin{figure}[tb!]
\includegraphics{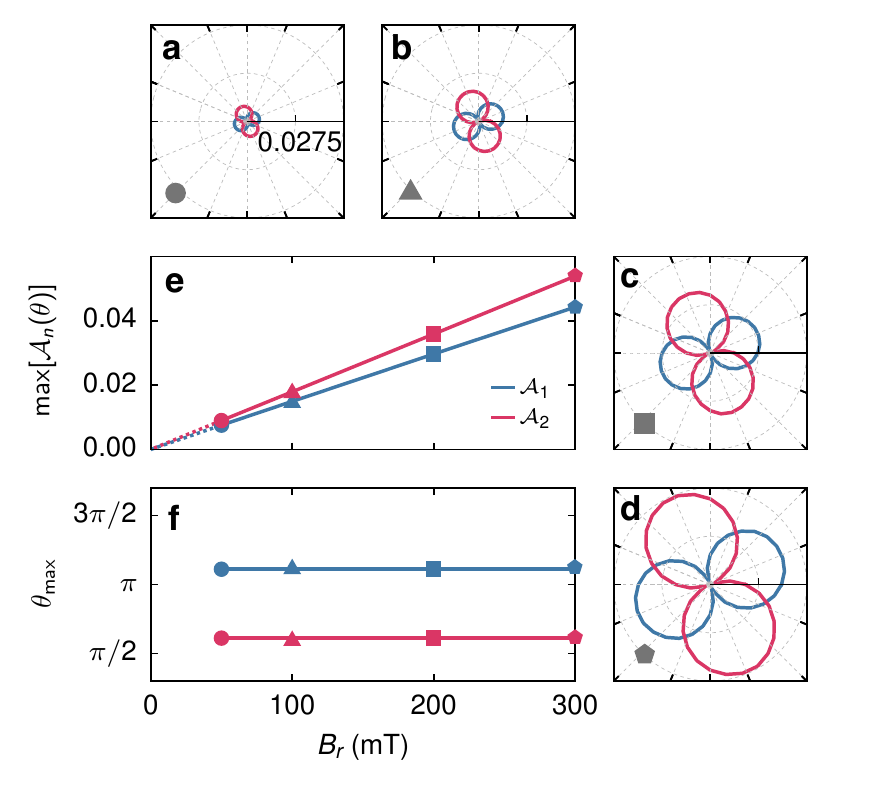} 
\caption{(a--d) Behavior of the side-lobe asymmetries $\mathcal{A}_{1,2}$, in blue and red respectively, as a function of in-plane field angle for increasing $B_r$, denoted by markers indicated in (e,f).
In all plots the same disorder configuration as in \autoref{fig:disorder}(a,d) was used.
(e) The maximum asymmetry $\rm{max}_\theta[\mathcal{A}_{1,2}]$ as a function of $B_r$. (f) The angle $\theta_{\rm{max}}$ where the maximal asymmetries occur, as a function of $B_r$.}
\label{fig:zeeman}
\end{figure}

\begin{figure}[tb!]
\includegraphics{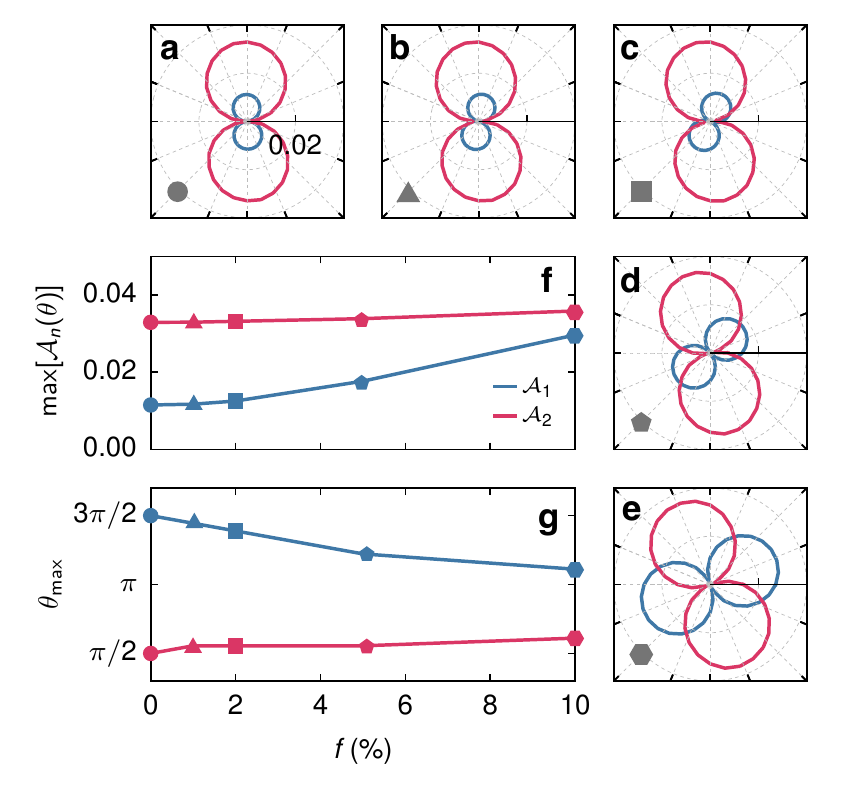} 
\caption{(a--e) Behavior of the side-lobe asymmetries $\mathcal{A}_{1,2}$, in blue and red respectively, as a function of in-plane field angle for increasing dipole strength $f$, denoted by markers indicated in (f,g). (f) The behavior of the maximum asymmetry $\rm{max}_\theta[\mathcal{A}_{1,2}]$ as a function of $f$. (g) The angle $\theta_{\rm{max}}$ where the maximal asymmetries occur, as a function of $f$.}
\label{fig:dipole}
\end{figure}

The results are presented in \Crefrange{fig:disorder}{fig:dipole}.
In \autoref{fig:disorder}(a--c) we show the interference pattern of critical currents $I_{c}(B_z)$ for three different disorder realizations, using $f = 10$\% and an in-plane field of $B_r = 200$~mT oriented along the $x$-direction.
We repeated these calculations, varying the angle $\theta$ between the in-plane field and the $x$-axis from $0$ to $2\pi$ in 36 steps.
For each interference pattern we find the local maxima, which give the $I_c^{(n)}$ as defined in the main text.
The resulting asymmetry of the first two side lobe pairs,
\begin{align}
{\cal A}_n = \frac{ I_c^{(-n)} - I_c^{(n)} }{ I_c^{(-n)} + I_c^{(n)} }
\end{align}
with $n = 1,2$, is then calculated as a function of $\theta$.
In \autoref{fig:disorder}(d--f) we present polar plots of the resulting $|{\cal A}_{1,2}|$ for the three disorder configurations used in \autoref{fig:disorder}.
These results can be qualitatively compared with Fig.~5 of the main text, as well as the experimental data shown in \autoref{fig:extradevices}.
We see that the overall patterns always look similar in shape, but with significant differences in both angular alignment of the maxima, as well as angular separation between the lobes. Furthermore, the overall magnitudes of the asymmetries appear to depend strongly (on the order of $\sim100\%$) on the precise disorder configuration. In general numerical simulations yield consistently smaller asymmetries than those observed in experiment for a wide range of parameters. Furthermore, for comparable disorder strengths to those estimated experimentally (as characterized by the mean free path), the obtained diffraction patterns deviate strongly from the Fraunhofer form and we regularly observe a finite lifting of the nodes as seen in \autoref{fig:disorder}(a), incompatible with experimental observations. The reasons for these discrepancies between experiment and numerical simulations are at present not well understood. 

In \autoref{fig:zeeman} we investigate the effect of the magnitude of the in-plane field.
In (a--d) we again plot the asymmetry parameters $|{\cal A}_{1,2}|$ using the same disorder configuration and other parameters as in \autoref{fig:disorder}(a,d), but now for different in-plane field magnitudes $B_r = 50$, $100$, $200$, and $300$~mT. The maximal asymmetries of the first and second lobe ${\rm max}_\theta[\mathcal{A}_{1,2}]$, in blue and red respectively, are shown in \autoref{fig:zeeman}(e) as function of in-plane field magnitude. The dotted lines intercepting zero are added to emphasize that the model yields effectively zero asymmetry (up to floating point accuracy) in the absence of an in-plane field. Consistent with our experimental findings, the asymmetries of both lobes appear to grow linearly with different slopes. To allow for a fair comparison with Fig.~3 of the main text we track the angular position of the asymmetry maxima $\theta_{\rm max}$ in \autoref{fig:zeeman} and find that changes in in-plane field strength do not affect the angular alignment of the observed asymmetry pattern. This appears to consistent with the data presented in \autoref{fig:extradevices} for the supplementary devices.

In \autoref{fig:dipole}(a--e) we gradually change the strength of the flux focusing, setting $f = 0$\%, 1\%, 2\%, 5\%, and 10\% respectively, using the same disorder configuration as in \autoref{fig:disorder}(a,d) and \autoref{fig:zeeman} and with $B_r = 200$~mT. We find that a change in the effective dipole has a significant effect on the angular alignment of the first lobe asymmetry ${\cal A}_1$ (blue), rotating roughly by $\pi/2$ when the dipole strength is changed from 0\% to 10\%, as shown explicitly in \autoref{fig:dipole}(g):
In the absence of a dipole, the asymmetry is zero for a field parallel to current flow; when the dipole is strong ($f=10$\%) the asymmetry is almost maximal in this direction.
This change may indicate that there are asymmetries of different origins.
The overall increase in magnitude of ${\cal A}_1$ for increasing $f$, as shown in \autoref{fig:dipole}(f), could support this interpretation.
The largely independent behavior of the second lobe asymmetry ${\cal A}_2$ (red) in both magnitude and angular alignment as seen in (f) and (g) is currently not understood.

\section*{Full rotation and magnitude datasets}

\begin{figure*}[tb!]
\includegraphics{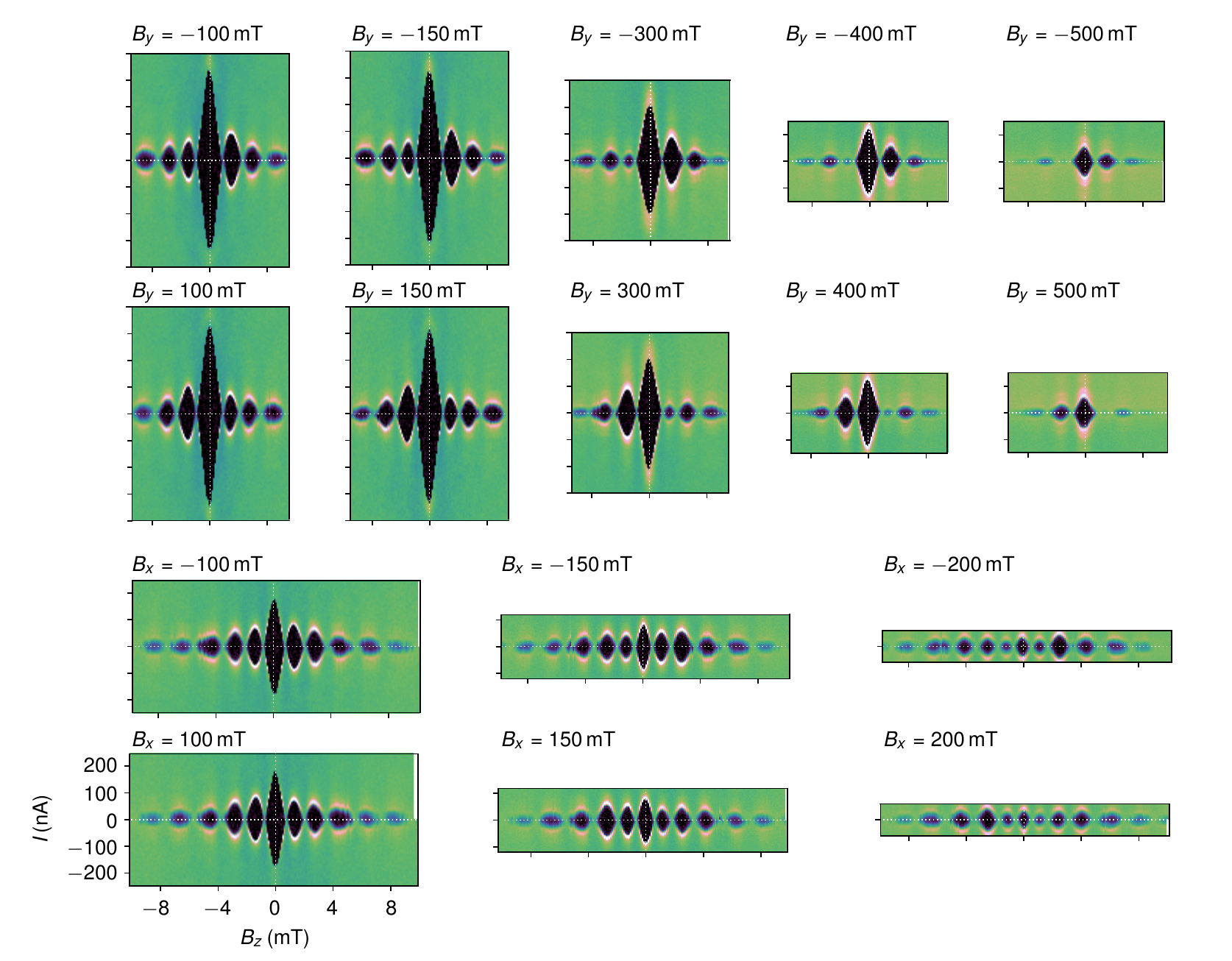} 
\caption{Complete data set of in-plane field magnitude dependence along the principal axes of the junction. To emphasize the anisotropies in field orientation, all data are shown on axes with a fixed scale. The data presented here are used in the extractions presented in Fig.~5(a,b), with a subset shown in Fig.~3.}
\label{fig:fullmagnitude}
\end{figure*}

\begin{figure*}[tb!]
\includegraphics{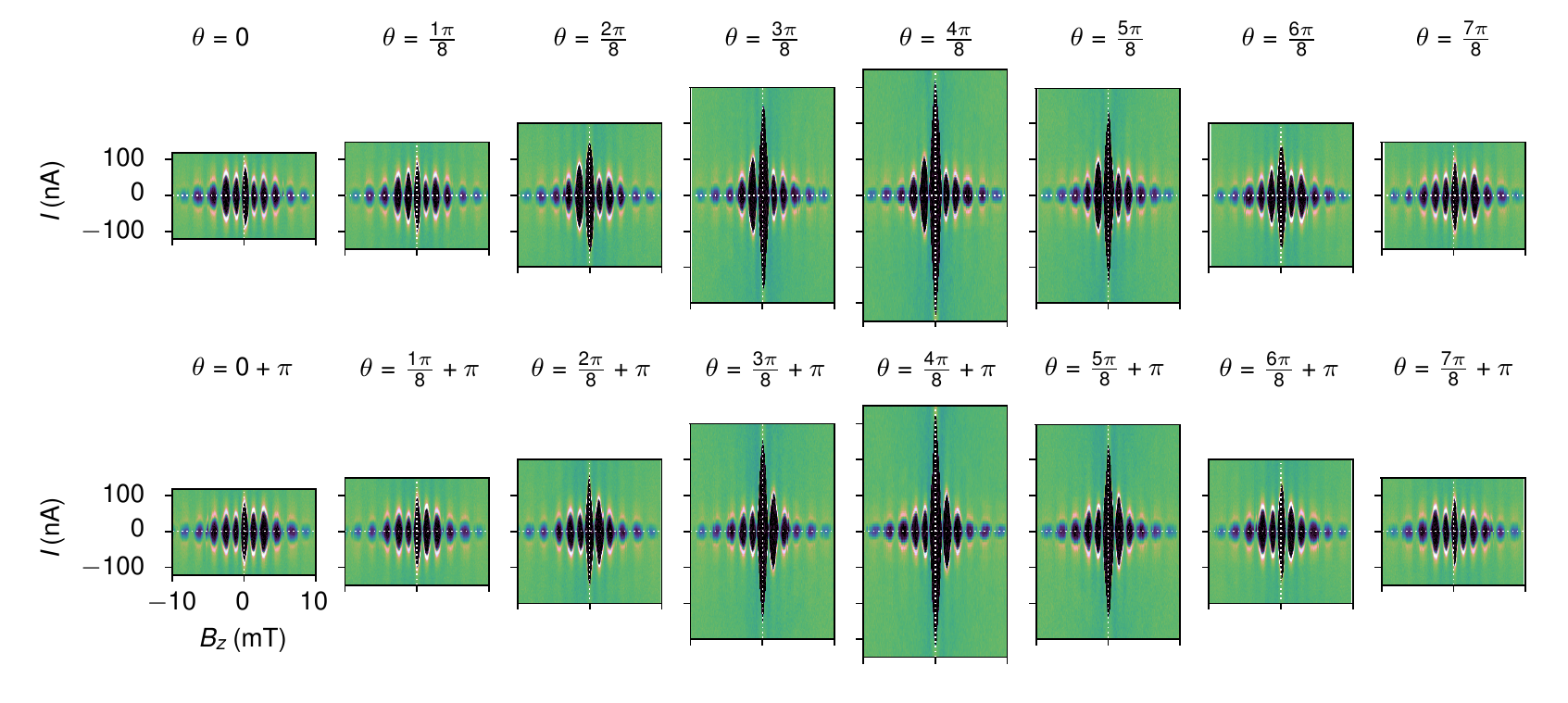} 
\caption{Complete data set of the in-plane angular dependence similar to that shown in Fig.~5 but here shown in full scale. To emphasize the anisotropies in field all data are shown on axes with a fixed scale.}
\label{fig:fullrotation}
\end{figure*}

\begin{figure*}[tb!]
\includegraphics{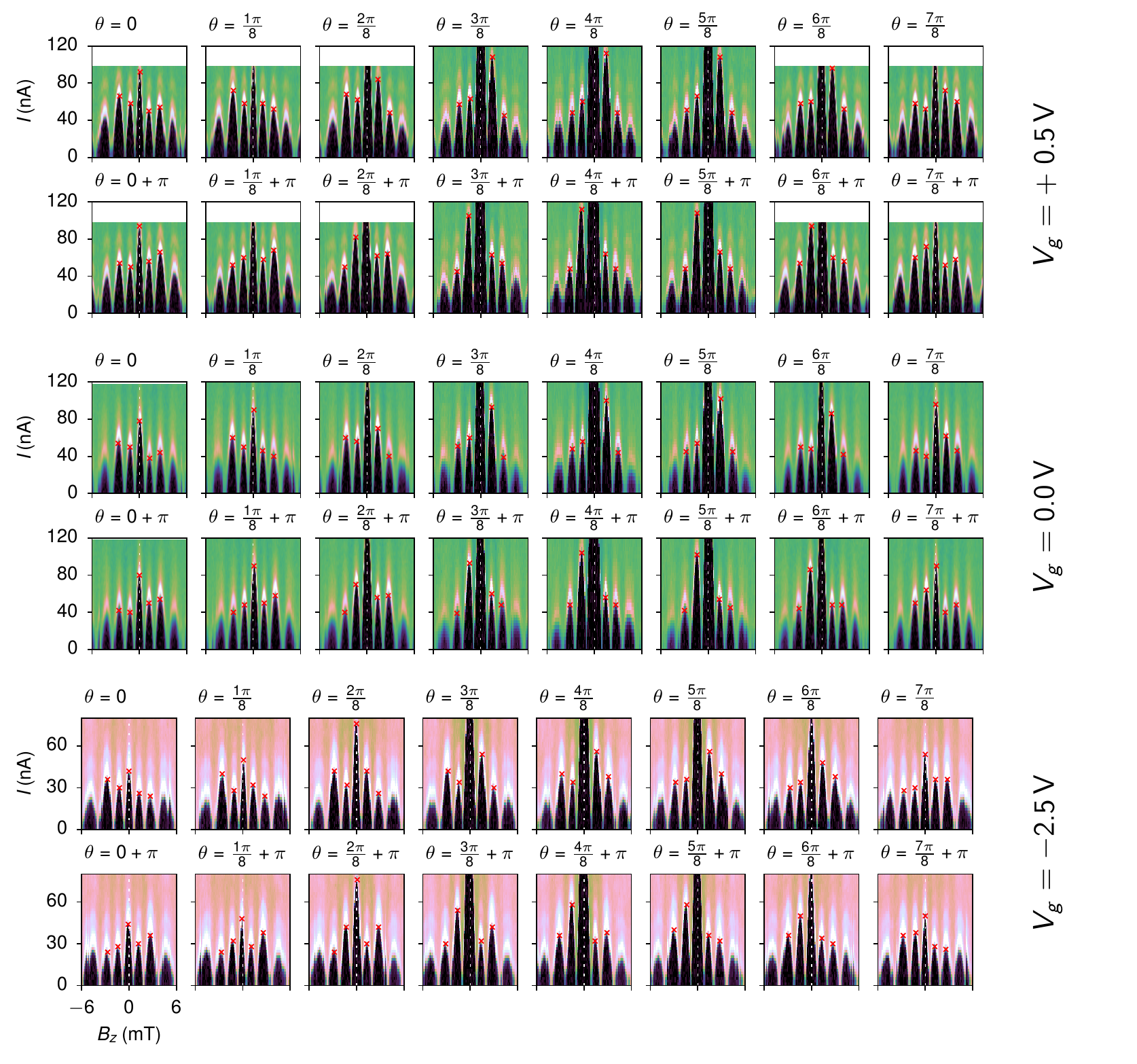} 
\caption{Complete data sets of the in-plane field angular dependence of the device presented in the main text at three different gate voltages, used to extract the data shown in Fig.~5 and 7. Extracted side-lobe maxima are indicated by red crosses.}
\label{fig:asymzoom}
\end{figure*}

\begin{figure*}[tb!]
\includegraphics{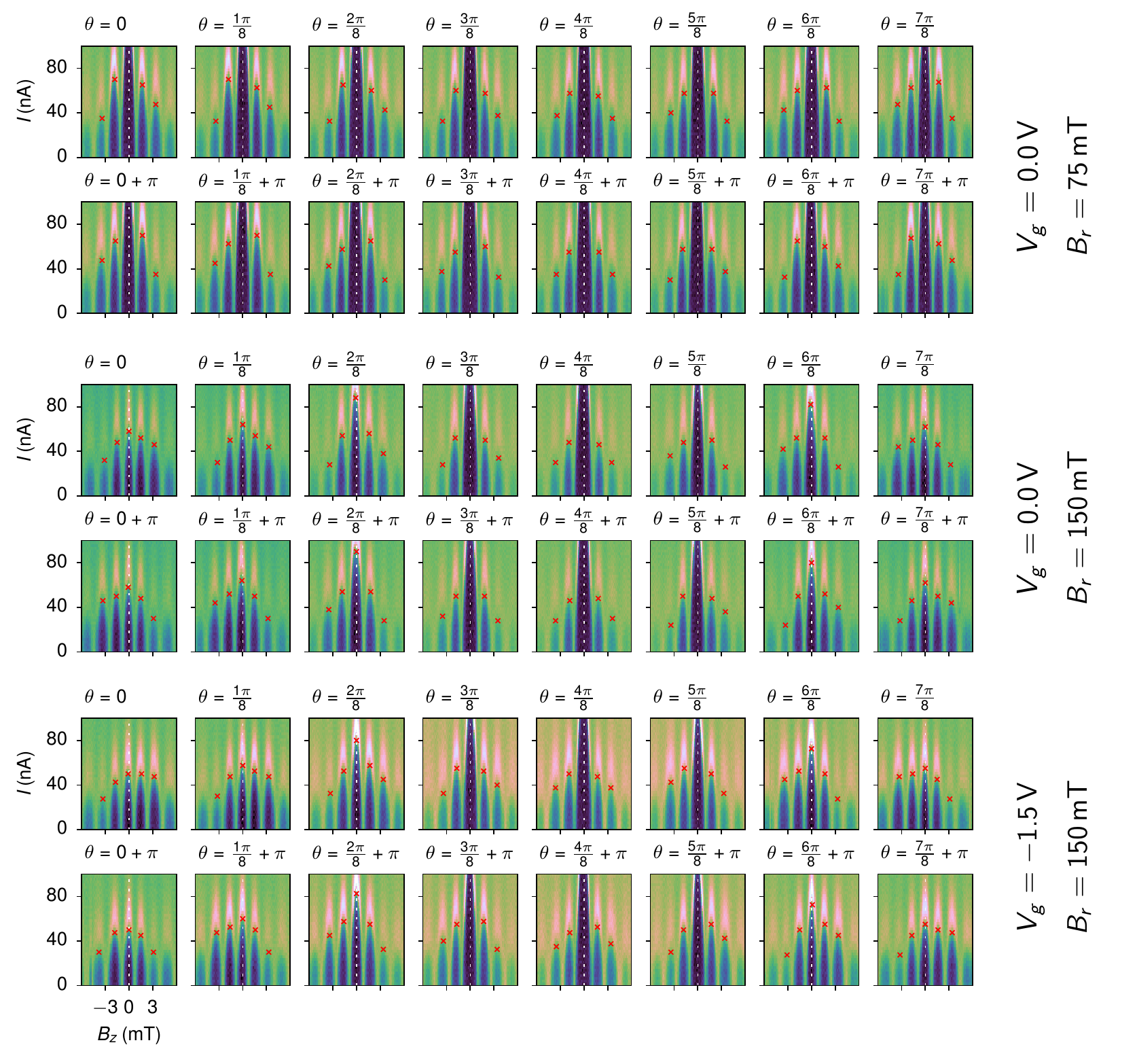} 
\caption{Complete data sets of the in-plane field angular dependence of the supplementary device oriented along $[010]$ in three different configurations used to extract the data shown in \autoref{fig:extradevices}. Extracted side-lobe maxima are indicated by red crosses.}
\label{fig:dev2}
\end{figure*}

\begin{figure*}[tb!]
\includegraphics{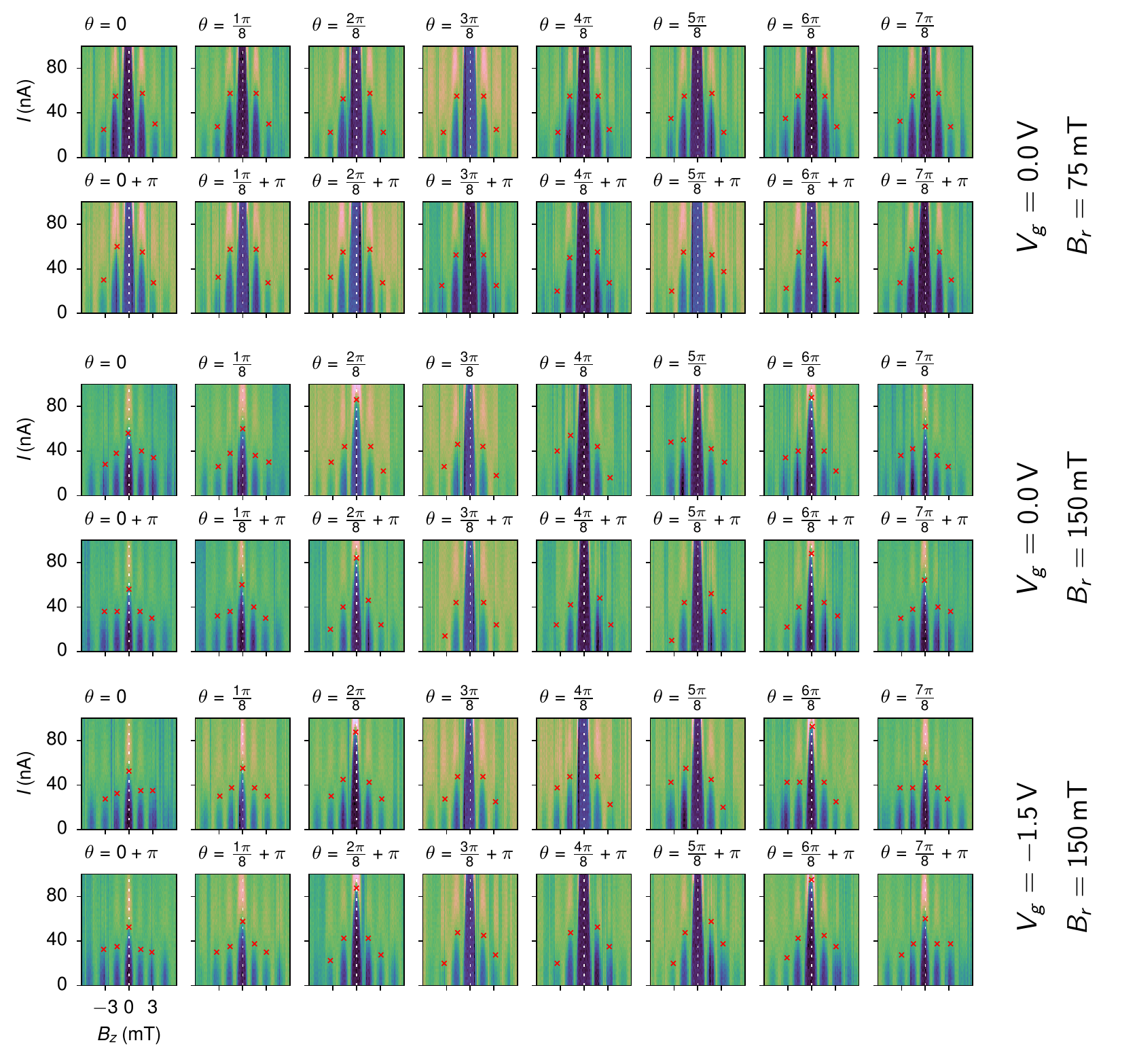} 
\caption{Complete data sets of the in-plane field angular dependence of the supplementary device oriented along $[011]$ in three different configurations used to extract the data shown in \autoref{fig:extradevices}. Extracted side-lobe maxima are indicated by red crosses.}
\label{fig:dev1}
\end{figure*}

\begin{figure*}[tb!]
\includegraphics{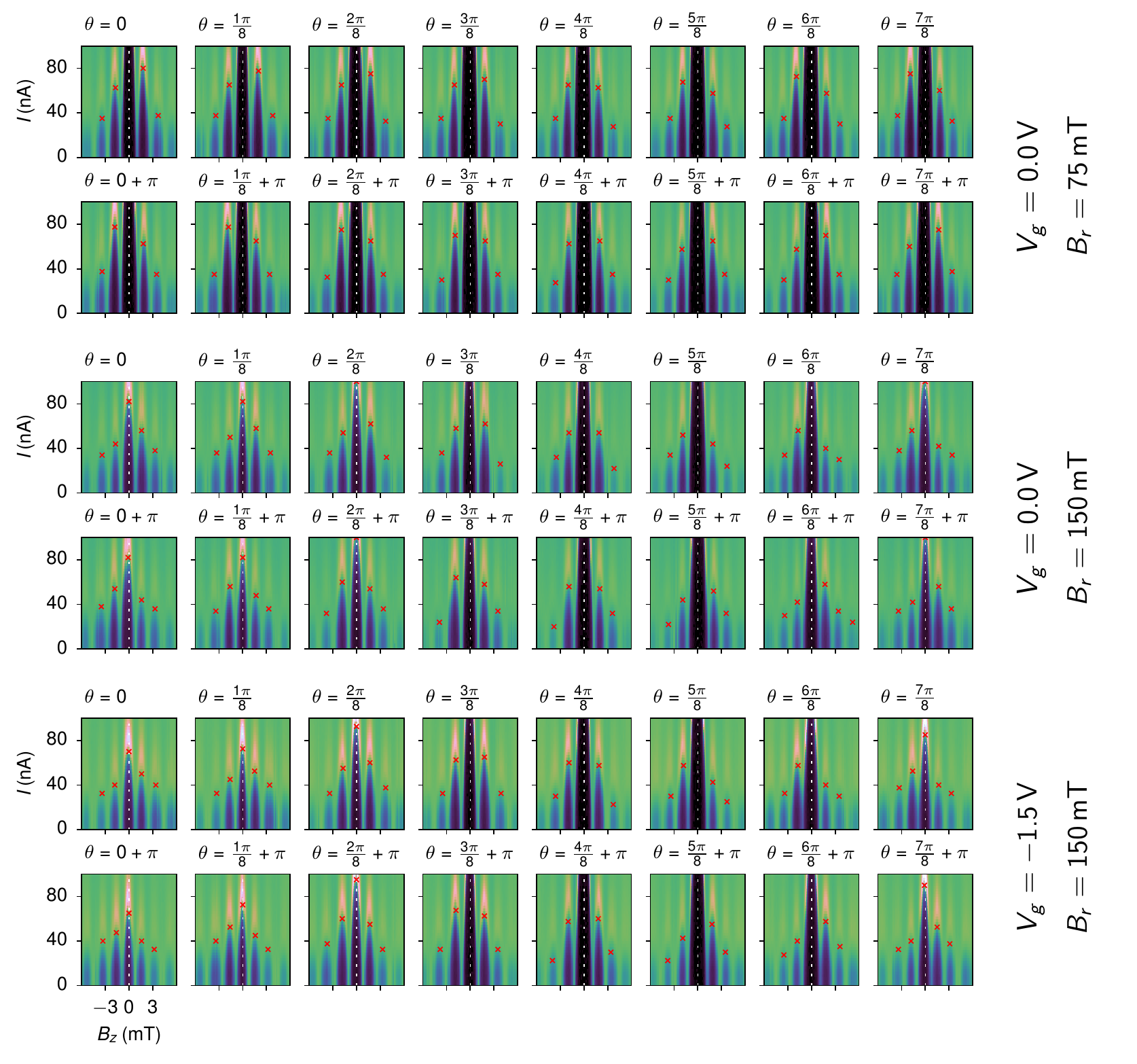} 
\caption{Complete data sets of the in-plane field angular dependence of the supplementary device oriented along $[001]$ in three different configurations used to extract the data shown in \autoref{fig:extradevices}. Extracted side-lobe maxima are indicated by red crosses.}
\label{fig:dev3}
\end{figure*}

\begin{figure*}[tb!]
\includegraphics{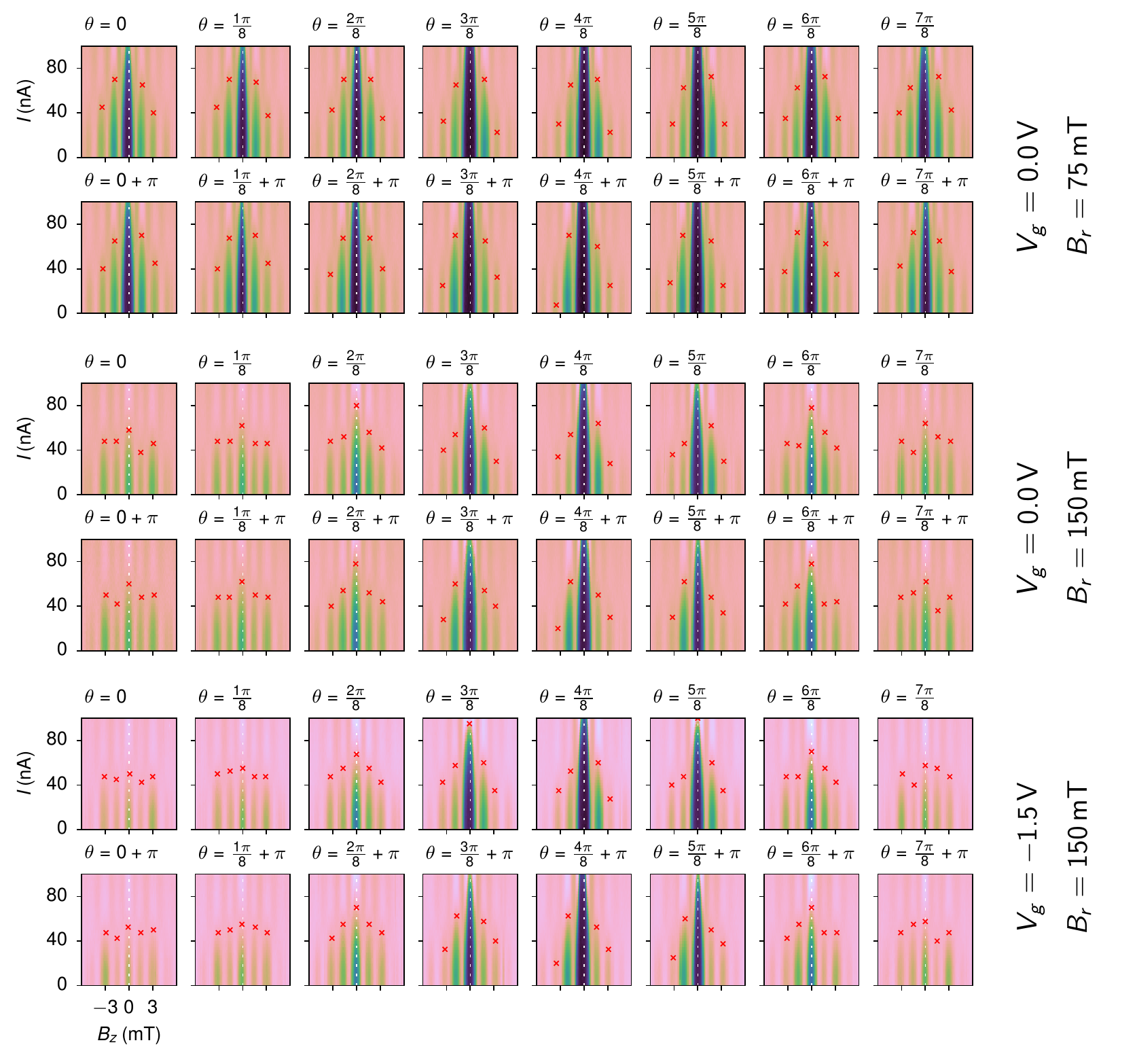} 
\caption{Complete data sets of the in-plane field angular dependence of the supplementary device oriented along $[0\bar{1}1]$ in three different configurations used to extract the data shown in \autoref{fig:extradevices}. Extracted side-lobe maxima are indicated by red crosses.}
\label{fig:dev4}
\end{figure*}

\begin{figure*}[tb!]
\includegraphics{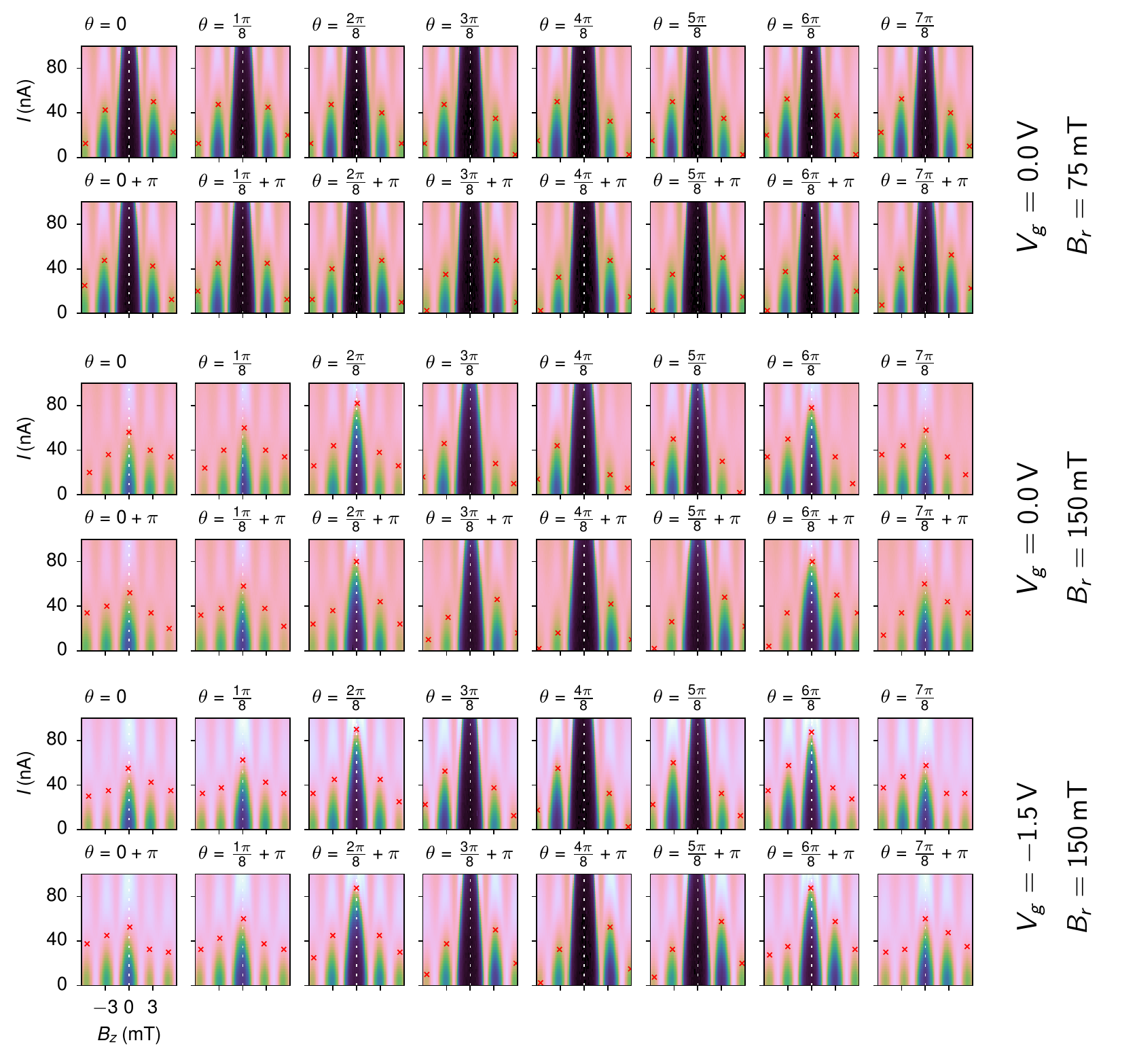} 
\caption{Complete data sets of the in-plane field angular dependence of the supplementary device oriented along [011] in three different configurations used to extract the data shown in \autoref{fig:extradevices}. Extracted side-lobe maxima are indicated by red crosses.}
\label{fig:focus}
\end{figure*}

Complete data sets of the dependence on in-plane field magnitude and in-plane field angle of the device investigated in Fig.~3 are provided in \autoref{fig:fullmagnitude} and \autoref{fig:fullrotation}. Detailed data sets of the asymmetry including markers highlighting extracted side-lobe maxima for all configurations are provided in \autoref{fig:asymzoom}.

\end{document}